\documentclass[11pt, a4paper, logo, twocolumn, copyright, nonumbering]{deepmind}

\usepackage[authoryear, sort&compress, round]{natbib}
\usepackage{ifthen}
\newboolean{showNotes}
\setboolean{showNotes}{true}  
\usepackage{soul}
\usepackage[dvipsnames]{xcolor}
\newcommand{\authornote}[3]{%
    \ifthenelse{\boolean{showNotes}}{%
        \begingroup\sethlcolor{#1}\hl{(#2) #3}\endgroup
    }{}%
}

\usepackage{blindtext}
\usepackage{setspace}

\usepackage{listings}
\usepackage{subfig}

\newlist{SC}{enumerate}{1}
\setlist[SC]{label=SC~\arabic*, start=0, align=left, labelwidth=2em, leftmargin=\labelwidth+\labelsep}
\usepackage{enumitem}
\usepackage{multirow}
\usepackage{longtable}

\usepackage{placeins}

\title{Melting Pot 2.0}

\correspondingauthor{JPA: jagapiou@, AV: vezhnick@, EDG: duenez@, JZL: jzl@; ... @deepmind.com}


\renewcommand{\today}{October 2023}

\author[*,1]{John P. Agapiou}
\author[*,1]{Alexander Sasha Vezhnevets}
\author[*,1]{Edgar A. Du{\'e}{\~n}ez-Guzm{\'a}n}
\author[1]{Jayd Matyas}
\author[1]{\\Yiran Mao}
\author[1]{Peter Sunehag}
\author[1]{Raphael K{\"o}ster}
\author[1]{Udari Madhushani}
\author[1]{Kavya Kopparapu}
\author[1]{Ramona Comanescu}
\author[1]{DJ Strouse}
\author[1]{Michael B. Johanson}
\author[1]{Sukhdeep Singh}
\author[1]{Julia Haas}
\author[2]{Igor Mordatch}
\author[3]{\\Dean Mobbs}
\author[*,1]{Joel Z. Leibo}

\affil[*]{Equal contributions}
\affil[1]{DeepMind}
\affil[2]{Google}
\affil[3]{California Institute of Technology}

\begin{abstract}
Multi-agent artificial intelligence research promises a path to develop intelligent technologies that are more human-like and more human-compatible than those produced by ``solipsistic'' approaches, which do not consider interactions between agents. Melting Pot is a research tool developed to facilitate work on multi-agent artificial intelligence, and provides an evaluation protocol that measures generalization to novel social partners in a set of canonical test scenarios. Each scenario pairs a physical environment (a ``substrate'') with a reference set of co-players (a ``background population''), to create a social situation with substantial interdependence between the individuals involved. For instance, some scenarios were inspired by institutional-economics-based accounts of natural resource management and public-good-provision dilemmas. Others were inspired by considerations from evolutionary biology, game theory, and artificial life. Melting Pot aims to cover a maximally diverse set of interdependencies and incentives. It includes the commonly-studied extreme cases of perfectly-competitive (zero-sum) motivations and perfectly-cooperative (shared-reward) motivations, but does not stop with them. As in real-life, a clear majority of scenarios in Melting Pot have mixed incentives. They are neither purely competitive nor purely cooperative and thus demand successful agents be able to navigate the resulting ambiguity. Here we describe Melting Pot 2.0, which revises and expands on Melting Pot. We also introduce support for scenarios with asymmetric roles, and explain how to integrate them into the evaluation protocol. This report also contains: (1) details of all substrates and scenarios; (2) a complete description of all baseline algorithms and results. Our intention is for it to serve as a reference for researchers using Melting Pot 2.0.
\end{abstract}

\begin{document}

\maketitle

\begin{spacing}{0.5}
\tableofcontents
\end{spacing}

\section{Introduction}

AI research is often divided into one branch that aims to improve the fundamental capabilities of AI systems, and another branch that aims to ensure AI is used in a way that benefits humanity. Research in the fundamental capabilities branch is typically concerned with solipsistic metrics such as performance in perception and cognition challenges~(e.g.~\cite{russakovsky2015imagenet, koehn2017six, crosby2019animal, ott2022mapping}). Research in the second branch is typically concerned with outcomes arising from introducing AI technology in real human societies like perpetuating unfair biases, diminishing privacy, promoting misinformation, and exacerbating inequality~(e.g.~\cite{obermeyer2019dissecting, rahwan2019machine, acemoglu2021harms, mehrabi2021survey, weidinger2021ethical}). This separation can lead to researchers in the capabilities branch to construct and deploy systems without enough consideration of their likely social consequences, and engaging with researchers in the beneficence branch only after the fact. This is not ideal. We think there should be a way to make fundamental capabilities research less solipsistic. 

Of course, it will always be desirable to measure the social consequences of AI deployment in the real world. Nevertheless, we may still benefit from developing challenges that are less solipsistic, while still being usable by fundamental-capabilities researchers engaged in the earlier stages of AI technology development. At core, AI provides ways of simulating agents. So we should be able to use it in a self-reflective way that constructs models of societies where multiple agents interact, and social consequences can be simulated. It may work similarly to agent-based modeling approaches in economics~(e.g.~\cite{tesfatsion2021}), but here using individuals modeled by learning agents. By adopting such an approach, early stage fundamental capabilities research need not be entirely blind to questions concerning how technology can be deployed in a socially optimal way. This is the human compatibility purpose of Melting Pot.

Melting Pot was also constructed to facilitate research on human-like AI, adopting the stance that it is our ultra-sociality that makes human intelligence unique. Along most of the traditionally studied domains of intelligence we do not outshine other mammals: our perception, memory, attention, planning, and decision-making skills are nothing special for a mammal of our size~\citep{henrich2021origins}. Rather we distinguish ourselves primarily along \emph{social}-cognitive dimensions. For instance, we improvisationally achieve cooperation through means such as persuasion and negotiation, reciprocity and reputational concern, alliances and incentive alignment, imitation and teaching, leadership and responsibility, commitments and promises, trust and monitoring, sharing and division of labor, social norms and institutions, and other social-cognitive capacities, representations,  motivations, and mechanisms. 

We propose an approach to building advanced AI systems based on reverse-engineering human intelligence. This reverse engineering approach is common in AI, especially with people who work on the ``classic'' cognitive abilities like perception, attention, and memory. However, we think it has been under explored with regard to the social-cognitive abilities that underlie important skills such as cooperation. In this case, we aim understand the social-cognitive abilities and biases that underlie human cooperation, and build them into our AI systems.

For this multi-agent-first AI research program to progress, it needs to settle two main methodological questions: (1) how to train agents so that they develop social-cognitive abilities on par with those of humans; and (2) how to measure progress toward the goal. The \emph{Melting Pot}  methodology we describe in this report addresses both of these questions.

\noindent \textbf{Melting Pot: the evaluation protocol}

Let's start first by talking about how Melting Pot provides a recipe for measuring progress. The key idea is to always measure generalization, just as it is in other parts of machine learning~\citep{hastie2009elements, cobbe2019quantifying, farebrother2018generalization, zhang2018dissection}. If we don't measure generalization then we run the risk of fooling ourselves into thinking we have made progress despite agents merely becoming increasingly overfit to one another's behavior~\citep{carroll2019utility, vinyals2019grandmaster, vezhnevets2020options}. Such overfitting can be extremely brittle and unlikely to work in the real world. To measure social-cognitive skills and to simulate questions of human compatibility, the key type of generalization we need to look at is social generalization, i.e. generalization to interactions in unfamiliar social situations involving both familiar and unfamiliar individuals~\citep{leibo2021scalable}.

Melting Pot consists of a set of \emph{test scenarios} and a protocol for using them. A scenario is a multi-agent environment that evaluates the ability of a \emph{focal population} of agents to generalize to novel social situations. Each scenario is formed by a \emph{substrate} and a \emph{background population}. The substrate is the physical part of the world: its spatial layout, where the objects are, how they move, the rules of physics, etc. The background population is the part of the simulation that is imbued with agency---excluding the focal population. While the focal population experiences the substrate during training, it has never experienced any of the individuals in the background population. Thus the performance of the focal population in a test scenario measures how well its agents generalize to social situations they were not directly exposed to at training time.

Since Melting Pot seeks to isolate social-environment generalization, it assumes that agents are familiar with the substrate. They have unlimited access to it during training. The evaluation scheme concentrates on assessing the ability of agents (and populations composed thereof) to cope with the presence of unfamiliar individuals in the background population (in a zero-shot way), typically mixed in with familiar individuals from the population under test. If the mixture contains more focal individuals than background individuals we call it a \emph{resident} mode scenario. Whereas, if the mixture contains more background individuals than focal individuals, we call it a \emph{visitor} mode scenario. The name Melting Pot is inspired by viewing this mixing of two distinct populations as analogous to that which occurs in real life when many members of different cultures immigrate to the same destination. 

We use the term \emph{multi-agent population learning algorithm} (MAPLA) to refer to any training process that produces a decentralized population of agents capable of simultaneous interaction with one another. The class of MAPLA algorithms is very broad. Most multi-agent reinforcement learning approaches can be made to produce populations. For instance self-play schemes like those used for AlphaGo~\citep{silver2016mastering, silver2017mastering}, AlphaZero~\citep{silver2018general}, FTW (Capture the Flag)~\citep{jaderberg2019human}, hide and seek~\citep{baker2019emergent}, and AlphaStar~\citep{vinyals2019grandmaster} fit in the class of MAPLAs, as does recent work on DOTA~\citep{berner2019dota} and MOBA~\citep{ye2020towards} games, as well as algorithms like MADDPG~\citep{lowe2017multi}, LOLA~\citep{foerster2018learning}, $\text{PSRO}$~\citep{lanctot2017unified}, $\text{PSRO}_{rN}$~\citep{balduzzi2019open}, and Malthusian reinforcement learning~\citep{leibo2019malthusian}. It includes both ``centralized training and decentralized execution'' algorithms and ``fully decentralized'' (decentralized at both training and test time) algorithms. Melting Pot evaluates MAPLAs by evaluating the generalization of populations they produce.

\noindent \textbf{How to train MAPLAs that will perform well in Melting Pot}

Now lets talk about the training side of Melting Pot. Researchers and the agents they train are allowed unlimited access to the substrate. This means that at test time the agents will usually already be familiar with their physical environment. So the kind of generalization that Melting Pot probes is mainly along social dimensions, not physical dimensions. This is in contrast to most work on generalization in reinforcement learning (e.g.~\cite{cobbe2019quantifying}), which are primarily concerned with generalization along physical dimensions.

In thinking about training for Melting Pot, our guiding principle has been that we shouldn't make decisions that restrict the scope for creativity in how researchers can train their agents. We can get away with few restrictions on training because we have a rigorous evaluation. As long as researchers do not cheat by training their agents on the test set, then we should be able to say of training that ``anything goes''. All agents will be judged according to a common benchmark in the end.

One interpretation of Melting Pot-compatible training processes, which we especially like, is that they give groups of agents the opportunity to form their own artificial civil society, complete with their own conventions, norms, and institutions. Forming a civil society entails the genesis and successful resolution of all kinds of obstacles to cooperation. The specific obstacles that arise, and the ways they can be resolved, depend on properties of the substrate. For instance, free riding in public-good provision is one obstacle to cooperation, and unsustainable resource usage leading to environmental degradation is quite another~\citep{ostrom2009understanding}. The challenge of overcoming these obstacles is what pushes agents to develop their social-cognitive abilities. For instance, if a particular social dilemma can be resolved by inventing a concept of fairness and enforcing its associated norms then that very fact motivates agents capable of representing the fairness concept and its norms to actually do so.

\section{Training and testing}

A \emph{substrate} is an $N$-player partially observable general-sum Markov game~\citep{leibo2021scalable}. Previously, in Melting Pot 1.0, we only fully supported symmetric games so each player received observations from the same observation space $\mathcal{X}$, and took actions from the same action space $\mathcal{A}$. Melting Pot 2.0 now adds full support for asymmetric games. We implement it by adding one new concept, that a player may have a \emph{role}.

\subsection{Roles}

Here, we extend the definition of a substrate to explicitly allow for situations where players do not all have the same role (e.g. for asymmetric games). Now a player's observation and action space may vary according to their \emph{role}. Substrates have a set of allowed roles $\mathcal{R}$. The player in slot $i$ is assigned role $r_i \in \mathcal{R}$, which determines their observation space ($\mathcal{X}_i \equiv \mathcal{X}_{r_i}$) and action space ($\mathcal{A}_i \equiv \mathcal{A}_{r_i}$).

\subsection{Substrate factories}

We define a \emph{substrate factory} $F(\mathbf{r})$, a parameterized function that returns a substrate conditional on a particular \emph{role configuration}, $\mathbf{r} = (r_1, \dots, r_N) \in \mathcal{R}_F^N$, where the number of players supported by the returned substrate is $|\mathbf{r}|$. In principle, $F$ can support any number of players, but in practice it is constrained to $ N_F^\textrm{min} \leq |\mathbf{r}| \leq N_F^\textrm{max}$.

The substrate factory may be used freely to sample a substrate, allowing agents to be trained for each specific role supported by that substrate factory. This results in a \emph{focal population}, $\pi \sim f_F(\cdot|r)$; a distribution of policies conditioned on role $r \in \mathcal{R}_F$. It is this conditional distribution that is submitted for evaluation on test scenarios.

\begin{figure*}[t]
    	\includegraphics[width=\linewidth]{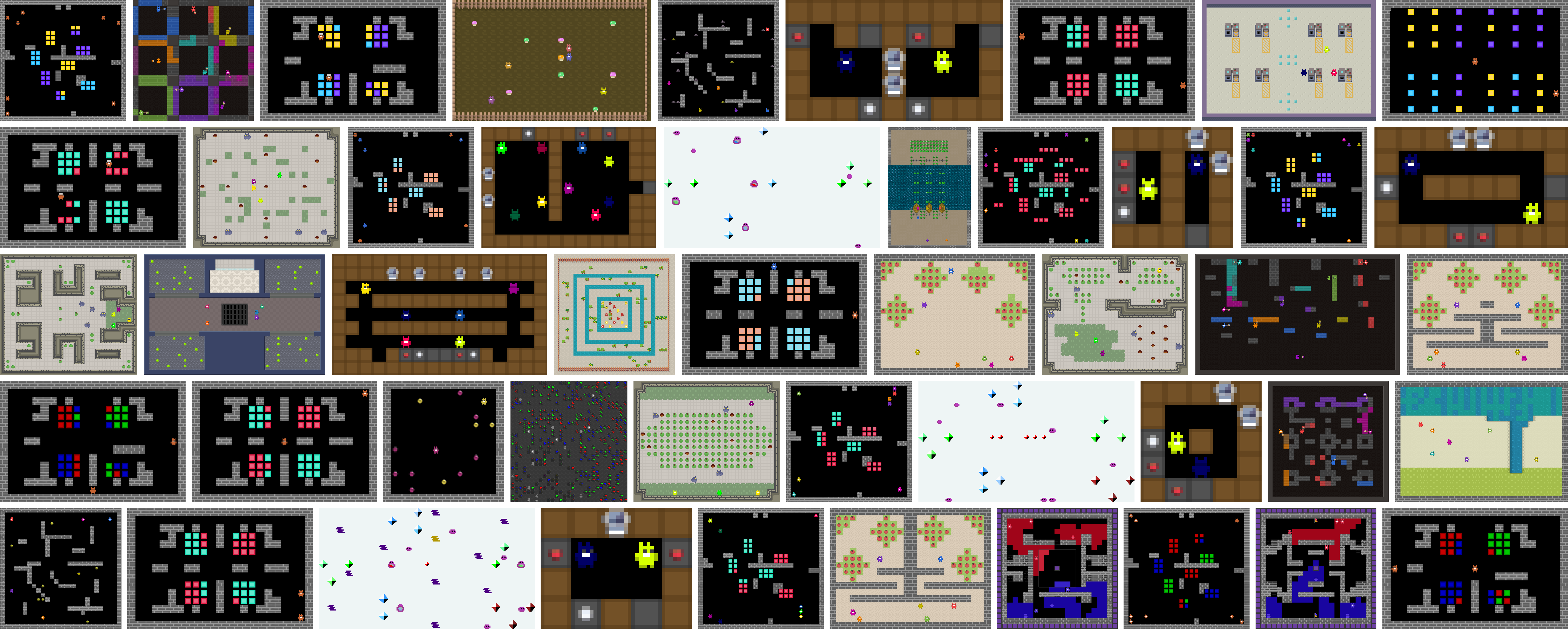}
    	\caption{The Melting Pot suite contains a set of over 50 multi-agent reinforcement learning substrates on which to train agents, and over 250 unique test scenarios to evaluate them. Screenshots in this figure were taken from the third person visualization view. All substrates are partially observed so agents view them through a smaller egocentric viewing window.}
        \label{fig:screenshots_gallery}
\end{figure*}

\subsection{Scenarios}

A \emph{test scenario} is a Markov game parameterized by a substrate factory $F$, a role configuration $\mathbf{r}$, the focal-population size $N \leq |\mathbf{r}|$, and a \emph{background population} of held-out bots $g_F$. Of these only $F$ is known at training time.

To form a scenario  we construct a substrate using $F(\mathbf{r})$, and fill all but the first $N$ player slots with policies sampled from the background population, such that $\pi_i \sim g_F(\cdot|r_i)$ for $i > N$. This results in an $N$-player Markov game.

We use such test scenarios to evaluate any trained focal population $f_F$, by filling the scenario's remaining $N$ slots with policies sampled from the focal population, such that $\pi_i \sim f_F(\cdot|r_i)$ for $i \leq N$. We assess the performance of the focal population on the test scenario by the \emph{focal per-capita return}, the mean return obtained across the focal players (i.e. excluding the background players).

Since the role configuration and background population are not known at training time, the focal per-capita return assesses the ability of the focal population to generalize to novel social situations.

When focal players outnumber background players we say the test scenario is in \emph{resident} mode. These scenarios usually test the emergent cooperative structure of the population under evaluation for its robustness to interactions with a minority of unfamiliar individuals not encountered during training. When background players outnumber focal players, we say the test scenario is in \emph{visitor} mode. One common use case for visitor-mode scenarios in Melting Pot is to test whether an individual from the focal population can observe the conventions and norms of the dominant background population and act accordingly (without retraining).

\subsection{Secondary evaluation metrics}

We propose to use \emph{focal-population per-capita return} as our primary evaluation metric for testing the performance of a learning algorithm in a novel social situation. This is because, first and foremost, we want Melting Pot to provide a rigorous and clearly interpretable evaluation metric that highlights unsolved problems and compares innovative algorithms to one another~\citep{leibo2021scalable}. Melting Pot is wide enough that many other reasonable-sounding metrics do not work well across the entire suite. For instance, we cannot always use the total collective return (sum of all players' rewards including both focal and background players) because its value is always zero in zero-sum substrates.

However, when evaluating the suitability of trained agents for a practical application, there will be additional considerations, which can be assessed from secondary evaluation metrics using the same test scenarios. For example, impacts on the background population may be an indicator of the impacts the trained agents might have on humans in the real world. So we can measure the \emph{background-population per-capita return} to see if it is negatively impacted by the introduction of the focal population. This could be useful to study whether the joint policy of the focal agents produces negative externalities---``side effects'' that impact the broader population while sparing the focal population, dovetailing well with research on value alignment and AI safety \citep{soares2014aligning, amodei2016concrete}. Or following \citet{perolat2017}, we can measure the \emph{inequality} of the background-population individual returns to see if any benefit or harm arising from having introduced the focal population is fairly shared, perhaps connecting fruitfully to beneficial and cooperative AI research agendas \citep{russell2015research, dafoe2020open}.

\begin{figure*}[t]
    	\includegraphics[width=\linewidth]{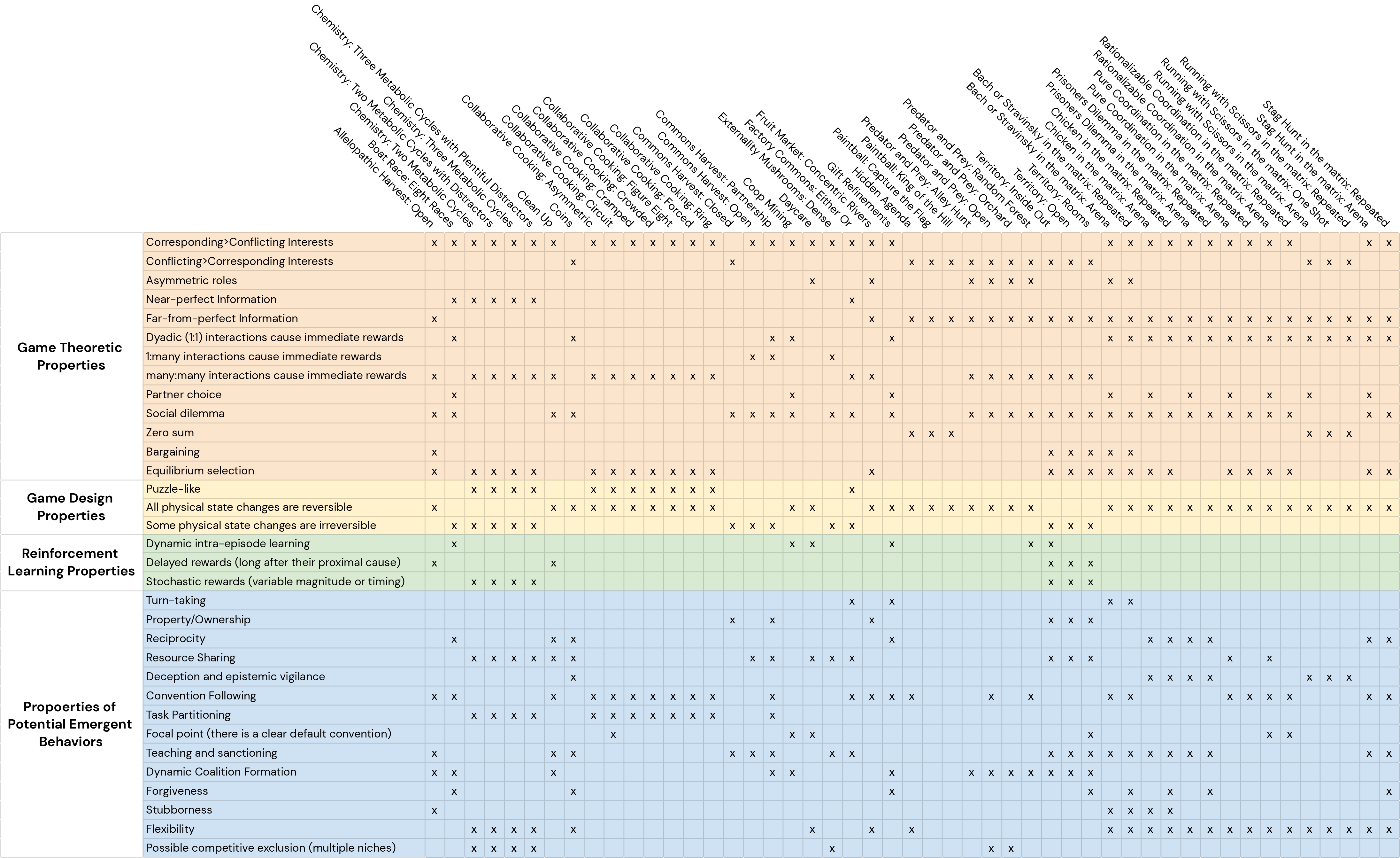}
    	\caption{Multi-agent concepts engaged by each substrate. Properties highlighted here are only intended as a rough guide to the concepts at play in each substrate. They should not be taken as a serious theoretical attempt to classify multi-agent situations.}
        \label{fig:tags_matrix}
\end{figure*}

\section{Benchmark experiments}

\subsection{Agent architectures}
To demonstrate the use of Melting Pot, we provide benchmark MARL results for a number of MAPLA (agent) architectures.

For each MAPLA architecture we performed 3 training runs on each substrate. For most substrates we trained $N$ individual agents in each training run: one in each role of the substrate's $N$-player \emph{default role configuration}. The exceptions to this were some 2-player substrates where we trained more than 2 agents:
\begin{enumerate}
    \item Bach or Stravinsky in the Matrix Repeated (2 agents in the "bach\_fan" role, and 2 agents in the "stravinsky\_fan" role);
    \item other "in the Matrix Repeated" substrates (5 agents in the "default" role);
    \item Running with Scissors in the Matrix One Shot (5 agents in the "default" role);
    \item Coins (5 agents in the "default" role);
\end{enumerate}

For each training episode, we built the substrate by passing a random permutation of the \emph{default role configuration} to the substrate factory. To fill each player slots, we sampled \emph{without replacement} from the subpopulation of agents that supported the corresponding role. This ensured that each training episode always had the same number of players with the same roles. Furthermore, each agent played as at most one player in each episode, always played in the same role, and that each agent's player index varied between episodes. Each agent was trained for $10^9$ steps.

At test time, we used all trained agents as the focal population in the test scenarios. To fill each focal player slot, we sampled \emph{with replacement} from the subpopulation of agents that supported the corresponding focal-player role. Thus at test time, an agent could play as multiple players in the same test episode.

The RL architectures we trained were: Actor-Critic Baseline (ACB), V-MPO~\citep{Song2020VMPO}, and OPRE~\citep{vezhnevets2020options}. OPRE was specifically designed for MARL while ACB and V-MPO are generic RL algorithms.

ACB is a well-established off-the-shelf RL method. Its core reinforcement learning algorithm was proposed by~\cite{espeholt2018impala}, which built on prior work from~\cite{mnih2016asynchronous} that described an algorithm called ``advantage actor critic'' (A3C). We also augmented it with a contrastive predictive coding auxiliary objective to promote discriminability of nearby timepoints using its LSTM state representations~\citep{oord2018representation}. In the previous Melting Pot release this agent was called A3C~\citep{leibo2021scalable}. It has been the baseline RL agent used in a few papers in this line of research, and has nearly always been called either A3C or A2C e.g.~\cite{mckee2020social, koster2020model, bakker2021modelling, duenez2021statistical, koster2022spurious, vinitsky2023learning}. We chose to rename it now because the old name has proven confusing in some contexts, and to reflect that it has departed quite significantly from the one called A3C in~\cite{mnih2016asynchronous}. We chose the new name, ACB, to evoke the old name while still clearly being something different. It still starts with A, and remains three letters long. It even rhymes with A3C. It stands for Actor-Critic Baseline. We plan to use continue using this name going forward.

Since the last Melting Pot release we added a PopArt module to both ACB and OPRE~\citep{hessel2019multi}. V-MPO was already using PopArt. This layer is attached to the baseline (value) and advantage computation and serves to normalize the gradients. It helps avoid problems that would otherwise arise when reward values vary through several orders of magnitude over the course of training and allows us to use the same hyperparameters for all substrates, even though they have very different reward scales from one another.

We also trained \textit{prosocial} variants of both algorithms, which directly optimized the per-capita return (rather than individual return), by sharing reward between players during training. Optimizing for collective return as a surrogate objective has previously been used for collaborative games (substrates) and social dilemmas \citep{claus1998dynamics,peysakhovich2017prosocial}, and our experiments here allow us to investigate whether it generalizes well.

\subsection{Computational setup}

We use a distributed training setup described in \cite{hessel2021podracer}, where TPU handles the agent parameters update in a learner process and action inference (learning and inference are split over several different chips). Experience for learning is generated in an actor process. The actor handles interactions with the environment on a CPU. The episode trajectories are split into truncated sequences each $100$ steps long and are grouped into batches, which are then split over $6$ TPU chips for learning. Each chip computes the update using the corresponding batch; afterwards their updates are averaged and applied to the parameters. This is different to the setup in~\cite{leibo2021scalable}, where learning was done on GPUs and inference was performed on CPU actors directly. This change led to hyper-parameters of both agents having to be re-calibrated.

\subsection{Agents hyper-parameters}

All agent architectures had the same size convolutional net and LSTM. The OPRE agent had additional hierarchical structure in its policy as described in~\cite{vezhnevets2020options}. All agents had a pop-art layer~\citep{hessel2019multi} for normalizing the value function. ACB minimized a contrastive predictive coding loss \citep{oord2018representation} as an auxiliary objective \citep{jaderberg2016reinforcement} to promote discrimination between nearby time points via LSTM state representations (a standard augmentation in recent work with ACB). The size of layers and their structure was the same as in the original Melting Pot release~\citep{leibo2021scalable}. ACB and V-MPO only differ by the loss. For V-MPO loss we used default hyper-parameters from~\cite{Song2020VMPO}.

We tuned hyper-parameters of each agent on a limited set of substrates by optimizing for the collective return in population play and focal return when trained as an exploiter---\emph{we did not use the generalisation score for tuning}. We have done it to the best of our ability given a limited computational budget.

All agents used discount rate $\gamma=0.99$,  PopArt step size of $10^{-3}$ with lower bound on scale $10^{-2}$ upper bound of $10^6$ and gradient normalisation to $2.0$~\citep{pascanu2013difficulty}.

\paragraph{ACB} The agent's network consists of a convNet with two layers with $16,32$ output channels, kernel shapes $8,4$, and strides $8,1$ respectively. It is followed by an MLP with two layers with 64 neurons each. All activation functions are ReLU. It is followed by an LSTM with $128$ units. Policy and baseline (for the critic) are produced by linear layers connected to the output of LSTM, followed by PopArt. 
We used an auxiliary loss~\citep{jaderberg2016reinforcement} for shaping the representation using contrastive predictive coding~\citep{oord2018representation}. CPC was discriminating between nearby time points via LSTM state representations (a standard augmentation in recent works with A3C). 
We used RMSProp optimizer with learning rate of $4*10^{-4}$, $\epsilon=10^{-5}$, momentum set to zero, decay of $0.99$, and the batch size of $256$. Baseline cost was $1.0$, entropy regularizer for policy at $0.003$. 

\paragraph{V-MPO}
We used RMSProp optimizer with learning rate of $8*10^{-4}$, $\epsilon=10^{-5}$, momentum set to zero, decay of $0.99$, and the batch size of $256$. Initial $\eta=1, \alpha=5$, $\epsilon_{KL}=0.01$. The gradients were clipped at $2.0$ and target network was updated every $10$ update steps.

\paragraph{OPRE} The networks shares the basic structure with ACB. The details of the architecture can be found in~\citet{vezhnevets2020options}. We used 16 options and a hierarchical critic implemented as an MLP with $64,64$ neurons. Critic has access to opponents observations. In~\citet{vezhnevets2020options} only one other player was present. To represent multiple co-players, we simply average outputs of the visual stack (convNet+MLP) and their inventories (where appropriate) together, before feeding it into the critic.
OPRE had the same parameters for optimisation as ACB plus two extra regularizers: i) entropy of policy over options set to $0.04$ and ii) KL cost between critic and actor policies over options set at $0.01$. Learning rate was set to $8*10^{-4}$ and the batch size of $96$

\paragraph{Puppet for held-out bots:} 

Sometimes the desired behavior for held-out bots is difficult to specify using a single reward function. In these cases, we generate background populations using techniques inspired by hierarchical reinforcement learning~\citep{sutton2011horde,sutton1999between}; in particular reward shaping~\citep{sutton2018reinforcement} and ``option keyboard''~\citep{barreto2019option}.
We create a basic portfolio of behaviors by training bots that use different environment events as the reward signal (as in the Horde architecture from~\cite{sutton2011horde}). We then chain them together using simple Python code.
This allows us to express complex behaviours in an ``if this event, run that behaviour'' way. For example, in \emph{Clean Up} we created a bot that only cleans if other players are cleaning.
These bots had a special network architecture based on FuN~\citep{vezhnevets2017feudal}, with goals specified externally via substrate events rather than being produced inside the agent. The idea is to use environment events as goals for training basic policies and then combine them into complex ones using a pre-scripted high-level policy. The high-level policy is a set of ``if this event happens, activate that behaviour'' statements. The same set of environment events can be used to train basic policies and script the high-level one. We call this approach \emph{Puppet}.

The puppet neural network has the same basic structure as the other agents we tested, {input$\rightarrow$ConvNet$\rightarrow$MLP$\rightarrow$LSTM$\rightarrow$outputs,} but has a hierarchical policy structure. The architecture was inspired by Feudal Networks~\citep{vezhnevets2017feudal}, but has several important differences. 
We represent goals as a one-hot vector $g$, which we embed into a continuous representation $e(g)$. We than feed $e$ as an extra input to the LSTM.
The network outputs several policies $\pi_z(a|x)$ and the final policy is a mixture $\pi(a|x)=\sum_z{\alpha(e) \pi_z(a|x)}$, where the mixture coefficients $\alpha(e)=\textbf{SoftMax}(e)$ are learnt from the embedding.
Notice, that instead of directly associating policies to goals, we allow the embedding to learn it through experience. 
To train the puppet to follow goals, we train it in the respective environment with goals switching at random intervals and rewarding the agent for following them.
Each goal has a separate value function and a PopArt module.

\begin{figure*}[!h]
		\includegraphics[width=0.95\linewidth]{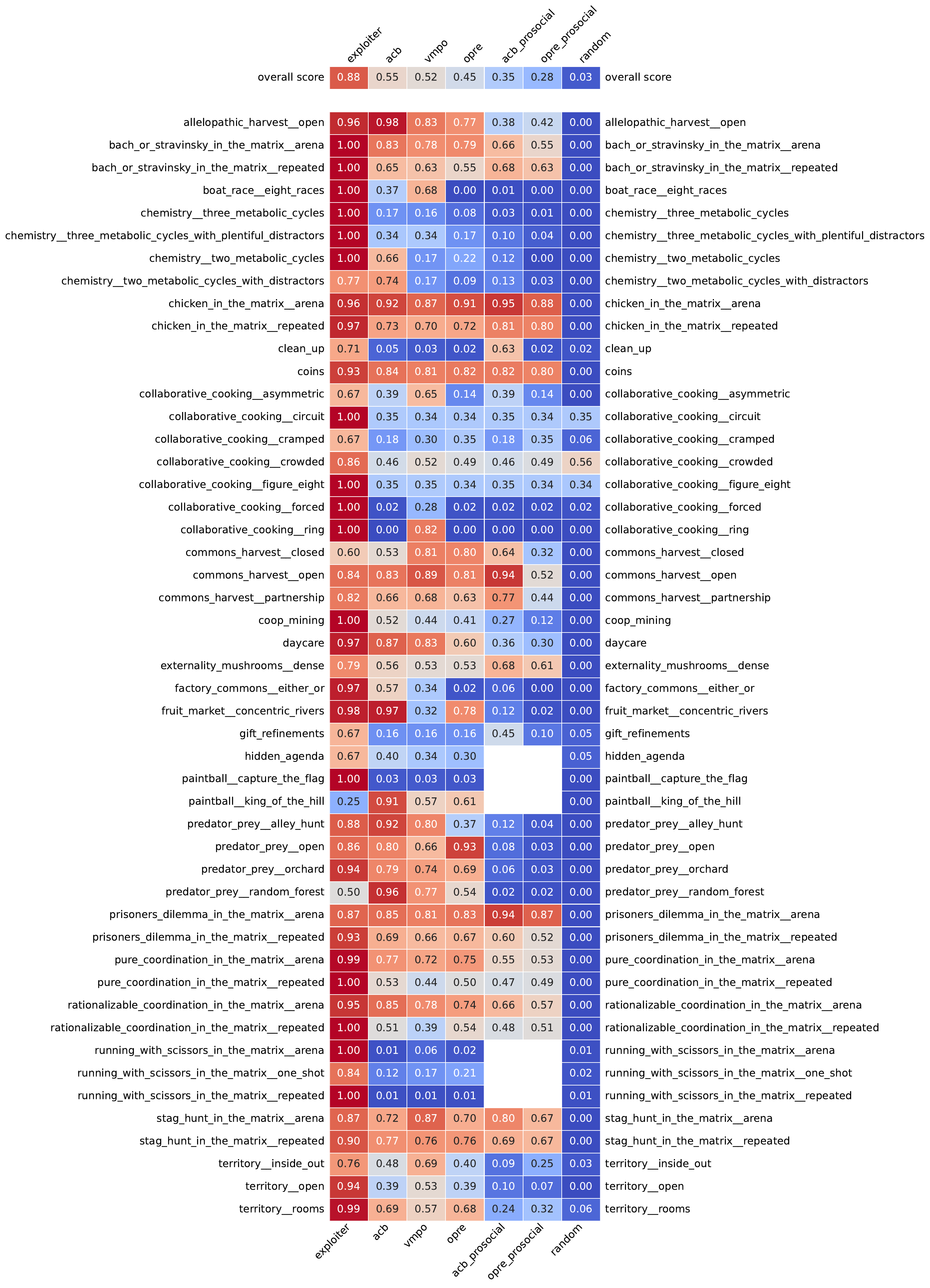}
		\caption{MAPLA performance scores averaged over each substrate's scenarios. Averaged over $3$ independent runs for each MAPLA.}
		\label{fig:agents_vs_substrate_test}
\end{figure*}

\subsection{Normalising results}
We also use two special MAPLAs for normalization: ``exploiters'' and ``random''. They are intended to provide rough upper and lower bounds for the raw scores. Though, as we will discuss below, they do not always operate entirely as intended, and this has implications for how the normalized results should be interpreted (Fig.~\ref{fig:agents_vs_substrate_test}).

Each exploiter is a self-interested ACB or OPRE MAPLA trained directly on a single test scenario, using their individual return as the reward signal without further augmentation. Exploiters trained for $10^9$ steps (same as the MAPLAs under evaluation). The random agent selects actions uniformly at random, and ignores input observations. Together, exploiters and the random agent provide a rough estimate of the upper and lower bounds (respectively) of performance on the test scenarios.
To contextualize the range of agent returns, we min-max normalize the focal per-capita returns to get a \emph{performance score} that is between 0 (for the worst agent) and 1 (for the best agent). Often the score of 0 corresponds to the random agent and 1 to the exploiter. However, for some scenarios this is not the case, since held-out bots might create a task too hard to solve for a naive agent. This clearly manifests in competitive substrates, but is not limited to them. Note that this makes the normalized scores of the MAPLAs reported in Fig.~\ref{fig:agents_vs_substrate_test} appear artificially high in those cases.

Below is the list of scenarios where the exploiter performed as no better than a random agent:
\begin{enumerate}
    \item \emph{Clean Up}: scenario 1;
    \item \emph{Collaborative Cooking Asymmetric / Circuit / Cramped / Figure Eight}: scenario 2;
    \item \emph{Commons Harvest Closed}: scenario 3;
    \item \emph{Paintball King of the Hill}: scenarios 0, 1, 3;
    \item \emph{Predator Prey Random Forest}: scenario 3;
    \item \emph{Running with Scissors in the Matrix Repeated}: scenarios 0, 1, 2, 3, 4 for ACB exploiter was no better than random. We trained an OPRE exploiter as replacement, which achieves strong results.
    \item \emph{Territory Inside Out}: scenario 1;
\end{enumerate}
On \emph{Collaborative Cooking} scenario 0 the random agent performed better than any other agent. This is due to the random agent ``staying out of the way'' of the competent held-out bots, whereas agents produced by MAPLAs interfered with them. In \emph{Territory Rooms} scenario 3 and \emph{Territory Inside-out} scenario 5 the random agent performs better than some, but worse than most agents.

On \emph{Coins} (all scenarios), the normalization procedure yields spuriously strong results for a different reason. Here the random agent gets a dramatically worse score than any other agent. Its score is a large negative number; and this is an environment which would otherwise usually produce values around zero for defection equilibria or large positive numbers in the case of mutual cooperation. The random agent performs so poorly that the normalization makes it look like all the MAPLAs we evaluated perform well by comparison. However, visually inspecting their policies shows that that is not the case. None of the MAPLAs we tested actually learned to cooperate in the Coins substrate.


\subsection{Results summary}

Fig.~\ref{fig:agents_vs_substrate_test} presents an overview, showing the average test scenario score for each substrate and each MAPLA. 
Please note that the presented results should not be interpreted as a definitive statement on the performance of specific MAPLAs. 
We didn't perform careful tuning for each MAPLA, neither it was our goal to provide the state-of-the-art results. 
These results should be interpreted as the reasonable performance ballpark of off-the-shelf RL methods. 
We encourage researchers to run their own baselines. 

On average, the top performing agent was ACB, followed V-MPO, and then by OPRE. This is different to our initial findings in the first Melting Pot paper~\citep{leibo2021scalable}, where the ranking was V-MPO, OPRE, ACB. Our main hypothesis for ACB pulling ahead is that it greatly benefited from PopArt module. Most likely, V-MPO was doing better in the original experiments~\citep{leibo2021scalable} thanks to PopArt (which ACB did not use in the original paper, but V-MPO did). Now, the rather more sophisticated V-MPO algorithm seems to  perform worse than the much simpler ACB policy gradient algorithm with CPC. 

There are several potential reasons for underperformance of OPRE it: i) PopArt modules might have benefited ACB more than OPRE ii) scenarios added since the first release are harder for OPRE iii) changes in computational setup affected the two MAPLAs differently. Moreover, we used a very simple way to integrate information over multiple opponent (average pooling) for OPRE, which is likely to limit it's capabilities for games with large number of players.

Overall, prosocial agents underperformed their purely self-interested counterparts, but the picture is nuanced.
Optimizing for per-capita return can be difficult because it complicates credit assignment, and creates spurious reward ``lazy agent'' problems~\citep{sunehag2018value, rashid2018qmix}. However, there are situations where prosocial agents outperform. For instance, in the relatively harsh social dilemma \emph{Clean Up}, only prosocial agent architectures managed to learn policies that were significantly better than random. Prosocial agents perform best on several other substrates too. These include \emph{Prisoners Dilemma / Stag Hunt in The Matrix}, and \emph{Commons Harvest Open / Partnerships}. This suggests a MAPLA capable of doing well on all of Melting Pot would have have to generate agents who can flexibly balance self interest and prosociality depending on the circumstances.

On several groups of substrates the MAPLAs we ran failed to generalize at all, getting scores very close to random: \emph{Collaborative Cooking}, \emph{Running with Scissors}, \emph{Paintball Capture the Flag}. In most other cases generalisation scores were above random, but with a significant room for further improvement.


\subsection{Train \& test scores as a function of training time}

See learning curve figures (Figs.~\ref{fig:allelopathic_harvest__open}--\ref{fig:territory__rooms}) for each focal population's per capita return as a function of training time. In each figure the plot on the far left shows results from the training environment (i.e.~only focal players). We periodically saved checkpoints of the focal population after every $x$ training steps. All plots to the right of the training plot show result from testing the saved focal population checkpoints from each $x$ in a specific test scenario. The curves plot the mean episode reward and its standard deviation, averaged over $3$ independent runs. 

\section{Discussion}

Here we have presented Melting Pot 2.0: an evaluation suite for MAPLAs that evaluates generalization to novel social situations. This release more than doubles the size of the initial Melting Pot suite~\citep{leibo2021scalable}. There are more than twice as many substrates as there were in the original set and more than three times as many scenarios. To the best of our knowledge, it is the largest open-source benchmark for general-sum Markov games. Moreover, Melting Pot is far from a solved problem. Our preliminary results show that there is plenty of room to improve the scores in the vast majority of test scenarios and substrates.

To facilitate secondary analyses of published results, we have included standardized evaluation scripts along with Melting Pot 2.0 (see \url{https://github.com/deepmind/meltingpot}). Other fields such as medicine extensively employ secondary analysis techniques like meta-analysis~\citep{haidich2010meta}. However, machine learning has been slow to adopt these methods. They require a bit of foresight on the part of authors to report their results in standard formats, but the whole community benefits when secondary analyses are easier to perform~\citep{burnell2023rethink, roelofs2019meta}. For instance, meta-analysis may be used to quantify convergence between different approaches and clarify rates of improvement over time~\citep{hernandez2017evaluation, martinez2020tracking}. Given instance-level data, they can be used to identify blind spots where further investment may be needed~\citep{martinez2019item}. Encouraging the standardized reporting of experimental results is complementary to other open science efforts like those that encourage public code. Hopefully it will be easy for researchers using Melting Pot to report results in such a way as to encourage such downstream re-use of their results.

Melting Pot 2.0 allows researchers to evaluate algorithms that train populations of agents to behave socially. How will the resulting agents resolve social dilemmas? Can they deal with free-riders? Will they learn policies that have an equitable distribution of returns? Will the winner take all? These questions not only present interesting research directions, but are also important to consider in the context of AI safety and fairness. After all, human society is a multi-agent system and any deployed AI agent will have to generalize to it.

\section{Substrates and Scenarios}

\subsection{Common properties shared by all substrates}

Unless otherwise stated, all substrates have the following common rules:
\begin{itemize}
    \item Sprites are $8 \times 8$ pixels.
    \item The agents have a partial observability window of $11 \times 11$ sprites, offset so they see more in front than behind them. The agent sees 9 rows in front of itself, 1 row behind, and 5 columns to either side. A few substrates use smaller observation windows, noted in their respective sections.
    \item Thus in RGB pixels, the size of each observation is $88 \times 88 \times 3$. All agent architectures used here have RGB pixel representations as their input. 
    \item Movement actions are: forward, backward, strafe left, strafe right, turn left, and turn right.
\end{itemize}

\subsection[Allelopathic Harvest]{Allelopathic Harvest\footnote{For a video of \textit{Allelopathic Harvest}, see \url{https://youtu.be/Bb0duMG0YF4}}}

There are three varieties of berry: red, green, and blue, and a fixed number of berry patches, which can be replanted to grow any  variety of berry. The ripening rate of each berry variety depends linearly on the fraction that that variety (color) comprises of the total. This can be understood as the cumulative effect of each berry variety suppressing the two others. Thus a monoculture of one variety (100\% of patches planted with that variety) implies a zero ripening rate for the other two varieties and the highest possible ripening rate for the monoculture variety. Evenly proportioned varieties $(1/3, 1/3, 1/3)$ yield the lowest possible overall ripening rate since all varieties suppress one another equally. Thus the sixteen players can increase the overall ripening rate by planting in such a way as to move the total proportions closer to monoculture of either red, green, or blue. However, increasing one variety's proportion decreases the other two since the three proportions must add up to one.

Players in Allelopathic Harvest have heterogeneous tastes: each has a particular berry variety that is intrinsically more rewarding to them. This creates tensions between groups of players since it is only possible to build one monoculture at a time. In addition, there is also a free-rider problem since individuals prefer to consume than plant \citep{koster2020model}. If all individuals free ride (none ever plant), then the variety proportions remain at their evenly mixed value where they start each episode, so the ripening rate remains very slow.

Individuals are rewarded for consuming ripe berries. They get a reward of 2 for consuming their preferred berry, and a reward of 1 for consuming a berry of any other color. At the start of each episode, berries are initialized to be unripe and evenly distributed over the three colors. Individuals can replant unripe berries to any color. Thus each individual experiences a tension between their incentive to immediately consume ripe berries and their incentive to plant unripe---and their preferred colored---berries. The default player roles are $8$ players preferring red berries, and $8$ preferring green berries.

The environment (a 29x30 plane) is filled with 348 berry plants of 3 colors (reset at the start of each episode; 116 per color).

Berry ripening depends stochastically on the number of berries sharing the same color that have been planted.  Initially all berries are in an unripe state. Each berry has a probability $p$ to ripen on each step, dependent on the number $b$ of berries of the same color across the whole map; $p = 5\times 10^{-6}b$ (the `allelopathic' mechanic (inspired by \cite{leibo2019malthusian}). Investing in establishing one color throughout the map, a monoculture, is prudent because it can be done relatively rapidly (if all players join in) and by doing so, all players will be able to harvest berries at a much faster rate for the remainder of the episode.

Players can move around in the environment and interact with berries in several ways. Players can use a planting beam to change the color of unripe berries to one of the other berry-type colors (the 'harvest' mechanic). Players can also walk over ripe berries to consume them. Ripe berries cannot be replanted, and unripe berries cannot be consumed. Players' avatars are recolored after using their planting beam to the same color they turned the berry into. Players' avatars are also stochastically recolored to a fixed white color when they eat a berry (probability inversely proportional to the highest berry fraction). These rules have the effect that past eating/planting actions often remains visible for others until a new action is taken. 

Each player also has the ability to zap other agents with a beam. It could be used either as a punishment mechanism or as a way to compete over berries. Getting zapped once freezes the zapped player for 25 steps and applies a visible mark to the player indicating that they have been punished. If a second zap is received within 50 steps, the player is removed for 25 steps and receives a penalty of $-10$ reward. If no zap is received for 50 steps, the mark fades. After each use of the zapping beam it is necessary to wait through a cooldown period of 4 steps before it can be used again.

Episodes last 2000 steps. The action space consists of movement, rotation, use of the 3 planting actions, and use of the zapping beam (10 actions total).

\noindent{\textbf{\small{Scenarios}}}
\begin{SC}
    \item \emph{Visiting a population where planting green berries is the prevailing convention.} This is a visitor mode scenario with a background population of ACB bots who were selected out of a wider set on the basis of their relatively increased tendency to plant green berries. All four focal players prefer red berries, but since they are outnumbered by the twelve background players who all plant green, they must acquiesce to a convention where green berries become the most common.
    
    \item \emph{Visiting a population where planting red berries is the prevailing convention.} This is a visitor mode scenario with a background population of ACB bots who were selected out of a wider set on the basis of their relatively increased tendency to plant red berries. The four focal players also prefer red berries so they do well under the prevailing convention.
    
    \item \emph{Focals are resident and visited by bots who plant either red or green.} This is a resident mode scenario with a background population of ACB bots who were selected on the basis of their relatively increased tendency to plant either red or green berries. Of the fourteen focal players, most prefer red berries but some prefer green berries. If the two background players plant inconsistently with the resident/focal population's established convention then members of the focal population should take sanctioning actions to prevent them from planting too much.
\end{SC}

\subsection[Boat Race: Eight Races]{Boat Race: Eight Races\footnote{For a video of \textit{Boat Race: Eight Races}, see \url{https://youtu.be/sEh1hRJVuFw}}}
The environment is inspired by the boat rowing thought experiment posed by~\cite{hume1739treatise}. Six players engage in a back and forth series of boat races across a river. Boats, however, cannot be rowed by a single player, and thus, players need to find a partner before each race and coordinate their rowing during the race to cross the river. At the beginning of a race, apples which confer reward when eaten appear on the other bank of the river. When the players are on the boat, they can choose from two different rowing actions at each timestamp: (a) \emph{paddle}, which is efficient, but costly if not coordinated with its partner; and (b) \emph{flail}, an inefficient action which isn't affected by the partner's action. When both players paddle simultaneously, the boat moves one place every $3$ timesteps toward the other side of the river. When either player \emph{flails}, the boat has a $10\%$ probability of moving one place, regardless of what the other player does. If one of the players on the boat paddles (i.e. if they have executed the paddle action within the last 3 timesteps) and the other flails, the paddler receives a penalty reward of $-0.5$. The episode consists of eight races of fixed duration. Each race consists of a partner choice phase where the access to the boats is barred, but players can line up. This phase lasts $75$ steps. After the partner choice phase, the actual race starts by lifting the bars preventing players from accessing the boats. Once a player sits (by touching) on a boat seat, they are prevented from moving until the reach the other side of the river. The actual race lasts for $225$ steps. Any player that doesn't reach the other side of the river when the race ends is disqualified (removed from the episode). The rowing styles are visibly distinct during the race, with mutual rowing showing steady rhythmic progress of the boat, and flailing looking like stochastic progress, and an oar always up.

\noindent{\textbf{\small{Scenarios}}}
\begin{SC}
    \item \emph{Visiting cooperators.} This is a visitor mode scenario with a background population of ACB bots that are incentivized to only paddle when on the boat. A single focal player must find any partner, and then paddle alongside their boat partner in order to maximize the speed to get to the other side, and thus get the most reward. If the focal player flails, on average they will receive lower reward, while causing large negative rewards to their partner.
    
    \item \emph{Visiting defectors.} This is a visitor mode scenario with a background population of ACB bots that are incentivized to only flail when on the boat. A single focal player must find any partner, and then flail alongside their boat partner. Alternatively, the focal player might choose to completely free-ride, not rowing at all. This is a slightly suboptimal strategy, because any timestep that their partner doesn't flail, there is no chance to make the boat move. The focal player should never paddle, as no partner would likely paddle in return.
    
    \item \emph{Visited by a cooperator.} This is a resident mode scenario with a background population of a single ACB bot that is incentivized to only paddle when on the boat. Five focal players can partner among themselves, but at least one must partner with the background bot. In that case, they should paddle, as that maximizes their reward. However, flailing is also a reasonable strategy, if a bit less optimal (see Scenario 0 above).
    
    \item \emph{Visited by a defector.} This is a resident mode scenario with a background population of a single ACB bot that is incentivized to only flail when on the boat. Five focal players can partner among themselves, but at least one must partner with the background bot. In that case, they should flail, as that maximizes their reward. However, free-riding (not rowing) is also a reasonable strategy, if a bit less optimal (see Scenario 1 above).
    
    \item \emph{Find the cooperator partner.} This is a visitor mode scenario with a background population of ACB bots, four of which are incentivized to only flail while one is incentivized to paddle when on the boat. A single focal player must find the cooperator partner, and then paddle alongside their boat partner in order to maximize the speed to get to the other side. The focal player has several possible strategies to find the cooperator: they can use rejection sampling, e.g. choosing a partner at random, and then observing their strategy, if they like them, choose them for all subsequent races, if they don't like them, sample someone else. With eight races and five co-players, it should be in principle possible to sample all co-players at least one. Alternatively, the focal player can choose a partner at random for the first race, but orient themselves in such a way that they can observe all the boats during the actual race, and therefore know who the cooperator is. Then, thereafter, always partner with the cooperator.
\end{SC}

\subsection{Chemistry *}

\begin{figure}[t]
    	\includegraphics[width=\linewidth]{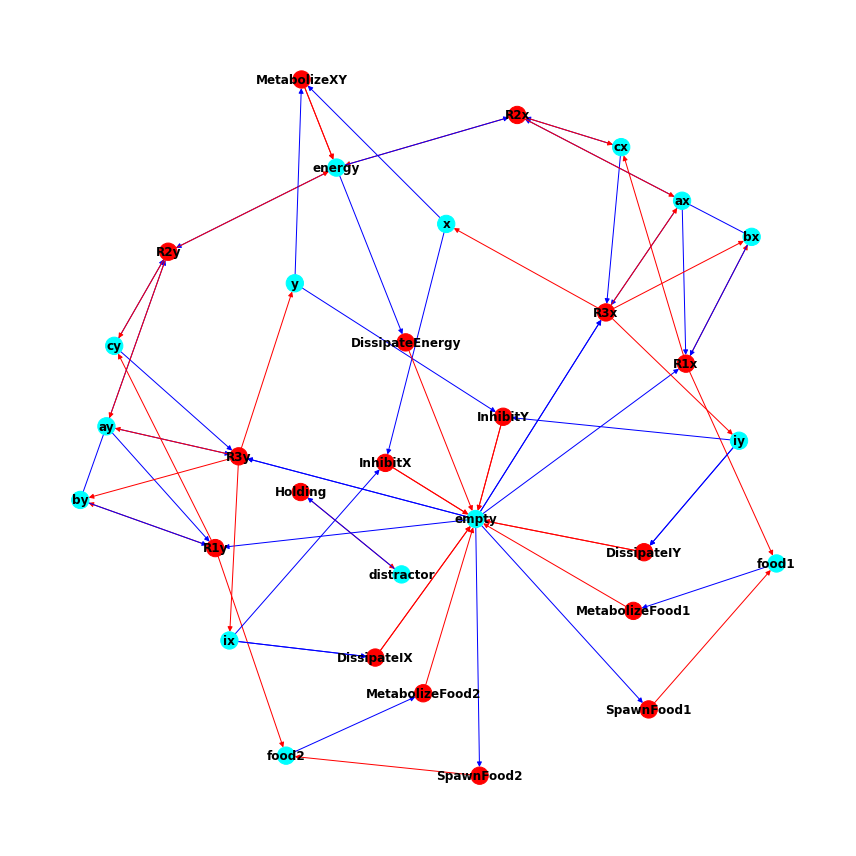}
    	\caption{Reaction Graph for molecules in Chemistry: Two Metabolic Cycles with Distractors. The task variants with or without distractors differ only be the presence of distractor molecules in the map}
        \label{fig:chemistry_two_metabolic}
\end{figure}

\begin{figure}[t]
    	\includegraphics[width=\linewidth]{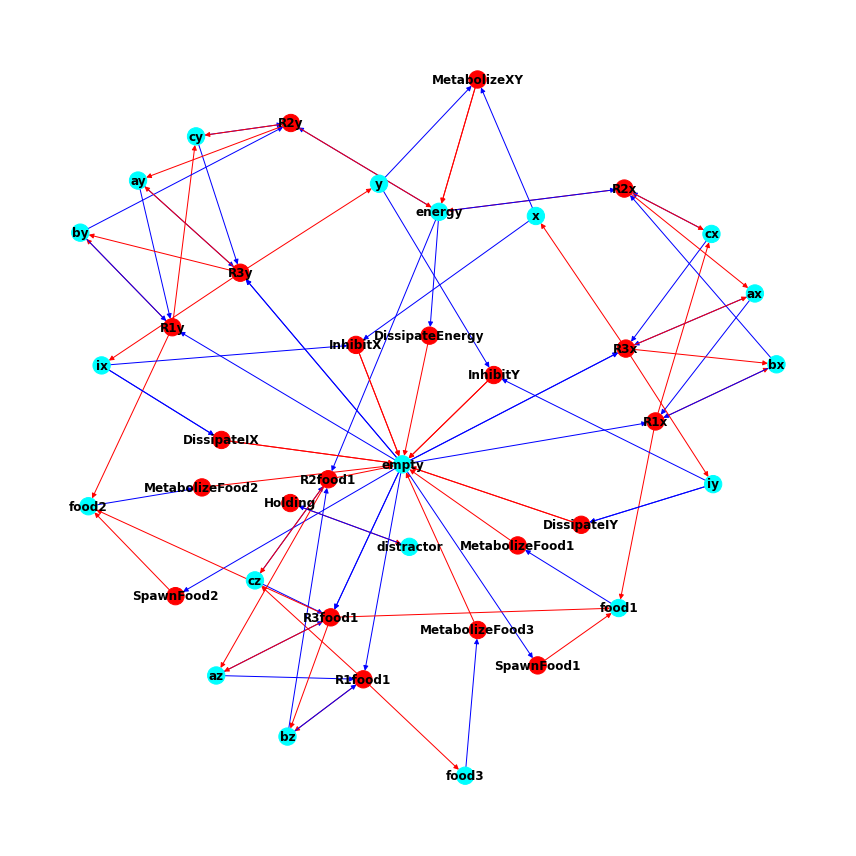}
    	\caption{Reaction Graph for molecules in Chemistry: Three Metabolic Cycles with Plentiful Distractors. The task variants with or without distractors differ only be the presence of distractor molecules in the map}
        \label{fig:chemistry_three_metabolic}
\end{figure}

Reactions are defined by a graph (see  Fig.~\ref{fig:chemistry_two_metabolic} and Fig.~\ref{fig:chemistry_three_metabolic}), which together with a map setting initial molecules, defines a substrate, occur stochastically when reactants are brought near one another. Agents can carry a single molecule around the map with them at a time. Agents are rewarded when a specific reaction---such as metabolizing food---occurs with the molecule in their inventory participating as either a reactant or a product. Each reaction has a number of reactants and products, occurs at different rates that can depend on if it is in an agent's inventory or not. As an example, metabolizing food in the metabolic cycles substrate has a much higher rate in the inventory where it generates a reward than outside where it represents the food molecule dissipating. Some tasks also contains distractor molecules, that allow agents to easily generate a stream of modest rewards by holding such a molecule in their inventory. This strategy is a shallow local optima that is easily found by agents, making it harder to learn to more rewarding cycles.  
    
\subsubsection[Chemistry: Three Metabolic Cycles]{Chemistry: Three Metabolic Cycles\footnote{For a video of \textit{Chemistry: Three Metabolic Cycles}, see \url{https://youtu.be/RlnojJHAoFI}}}
Individuals benefit from three different food generating reaction cycles. The cycles will run on their own (autocatalytically), but require energy to continue. One of the cycles produces more rewarding compounds than the other two but it also consumes the energy used to power it at a faster rate. Bringing together side products from the other two cycles generates new energy such that both cycles can continue. If the agents get them running consistently enough then they can generate enough energy from them to also power the third cycle. which is more energy hungry. Too much use of the energy-hungry cycle before discovering how to use the other two cycles efficiently will rapidly deplete the initial energy present in the map. The highest collective return is obtained by populations that run all three autocatalytic reaction cycles simultaneously.

\noindent{\textbf{\small{Scenarios}}}
\begin{SC}
    \item \emph{Resident focal population meets a small mixture of background bots.} 
    Five focal agents meet three bots that each specialize in running one of the three cycles. The five focal agents needs to contribute to the red and blue metabolic cycles and quickly generate more energy by combining the side products so as to allow for also running the more rewarding yellow cycles. 
    
    \item \emph{Meeting bots running blue and yellow.}
    Four focal agents meet four bots, two of which runs the blue metabolic and two that runs the yellow. The focal agents need to fully run the green cycle and generate more energy by bringing side products together. They need this part to be efficient enough to not have the two bots running the yellow more energy demanding cycle depleting the energy, They also need to hold on to some energy for running the green and blue cycles and keep that away from the yellow running bots.
    
    \item \emph{Meeting one-sided bots running green.}
    Four focal agents meet four bots that all run the green metabolic cycle. The focal agents need to run the blue cycle and generate more energy by bringing side products together. If they do this efficiently enough they can also benefit from mixing in the yellow cycle.
    
    \item \emph{Visit a resident population with mixed specialties.}
    Two focal agents meets six bots, two for each of the three cycles. The focal agents needs to generate more energy by bringing side products together and to chip in running the blue and green fast enough for the yellow running bots to not deplete all energy. 
\end{SC}

\subsubsection[Chemistry: Three Metabolic Cycles With Plentiful Distractors]{Chemistry: Three Metabolic Cycles With Plentiful Distractors\footnote{For a video of \textit{Chemistry: Three Metabolic Cycles With Plentiful Distractors}, see \url{https://youtu.be/IjlJckwM1VE}}}
Similar to \emph{Chemistry: Three Metabolic Cycles} but here there are also molecules of a distractor compound scattered around. The distractor can be collected and held, after which it provides reward at a constant but low rate. Holding a distractor is the simplest rewarding strategy, it corresponds to a shallow local optimum that is very easy for agents to learn, and once learned often makes it very difficult to explore other policies. For agents to get good scores in this substrate they must be able to avoid premature convergence to such shallow local optima and continue to explore beyond them.

\noindent{\textbf{\small{Scenarios}}}
\begin{SC}
    \item \emph{Resident focal population meets a small mixture of background bots, must avoid distractor molecules.} Five focal agents meet three bots that each specialize in running one of the three cycles. The five focal agents needs to contribute to the red and blue metabolic cycles and quickly generate more energy by combining the side products so as to allow for also running the more rewarding yellow cycles. The focal agents also needs to avoid the distractor molecules that are not touched by the bots.
    
    \item \emph{Meeting bots running blue, avoid distractors.}
    Four focal agents meet four bots that all run the blue metabolic cycle. The focal agents need to run the green cycle and generate more energy by bringing side products together. If they do this efficiently enough they can also benefit from mixing in the yellow cycle.The focal agents also needs to avoid the distractor molecules that are not touched by the bots.
    
    \item \emph{Meeting bots running green and yellow, avoid distractors.}
    Four focal agents meet four bots, two of which runs the green metabolic and two that runs the yellow. The focal agents need to fully run the blue cycle and generate more energy by bringing side products together. They need this part to be efficient enough to not have the two bots running the yellow more energy demanding cycle depleting the energy, They also need to hold on to some energy for running the green and blue cycles and keep that away from the yellow running bots.  The focal agents also needs to avoid the distractor molecules that are not touched by the bots.
    
    \item \emph{Visit a resident population with mixed specialties and avoid distractor molecules.}
    Two focal agents meets six bots, two for each of the three cycles. The focal agents needs to generate more energy by bringing side products together and to chip in running the blue and green fast enough for the yellow running bots to not deplete all energy. The focal agents also needs to avoid the distractor molecules that are not touched by the bots.
\end{SC}

\subsubsection[Chemistry: Two Metabolic Cycles]{Chemistry: Two Metabolic Cycles\footnote{For a video of \textit{Chemistry: Two Metabolic Cycles}, see \url{https://youtu.be/kxMNAcJuXJE}}}
Individuals benefit from two different food generating cycles of reactions that both rely on energy which dissipates over time. Bringing together side products from both cycles generates new energy such that the cycles can continue indefinitely. The population needs to keep both cycles running to get high rewards.

\noindent{\textbf{\small{Scenarios}}}
\begin{SC}
    \item \emph{Resident focal population meets a small mixture of background bots.}
    Six focal agents meet two bots that each specialize in running one of the two cycles. The focal agents need to take part in both of these cycles for them to run well as well as taking responsibility for generating new energy by combining the side products.
    
    \item \emph{Meeting bots running blue.}
    Four focal agents meet four bots that all run the blue cycle, which means the focal population needs to run the green cycle and generate new energy by combining the side products.
    
    \item \emph{Meeting one-sided bots running green.}
    Four focal agents meet four bots that all run the green cycle, which means the focal population needs to run the green cycle and generate new energy by combining the side products.
    
    \item \emph{Visit a resident population with mixed specialties.}
    Here two focal agent meets six bots with a mixture of the required specialities. The focal agents just needs to chip in to run it all faster or can mostly be parasitically consuming food generated by the bots.
\end{SC}

\subsubsection[Chemistry: Two Metabolic Cycles With Distractors]{Chemistry: Two Metabolic Cycles With Distractors\footnote{For a video of \textit{Chemistry: Two Metabolic Cycles With Distractors}, see \url{https://youtu.be/YsN-SnY-buo}}}
Similar to \emph{Chemistry: Two Metabolic Cycles} but here there are also molecules of a distractor compound scattered around, which can be held as part of an easy-to-discover rewarding strategy, like in \emph{Chemistry: Three Metabolic Cycles With Plentiful Distractors}. However, in this case there are less distractor molecules than there are players. An implication of that is that it is not possible for all players to simultaneously select the distractor policy. If they try then at least some players would always be left without a distractor molecule. This ``competitive exclusion'' makes it easier to explore beyond the distractor strategy to discover better strategies involving the two autocatalytic cycles. The best joint strategies in this substrate run both autocatalytic cycles simultaneously and completely ignore the distractor molecules.

\noindent{\textbf{\small{Scenarios}}}
\begin{SC}
    \item \emph{Resident focal population meets a small mixture of background bots, must avoid distractor molecules.} Six focal agents meet two bots that each specialize in running one of the two cycles. The focal agents need to take part in both of these cycles for them to run well as well as taking responsibility for generating new energy by combining the side products. The bots are not touching the distractor molecules that the focal agents needs to avoid.
    
    \item \emph{Meeting one-sided bots running blue and avoid distractor molecules.}
    Four focal agents meet four bots that all run the blue cycle, which means the focal population needs to run the green cycle and generate new energy by combining the side products.
    The bots are not touching the distractor molecules that the focal agents needs to avoid.
    
    \item \emph{Meeting one-sided bots running green and avoid distractor molecules.}
    Four focal agents meet four bots that all run the green cycle, which means the focal population needs to run the green cycle and generate new energy by combining the side products. The bots are not touching the distractor molecules that the focal agents needs to avoid.
    
    \item \emph{Visit a resident background population with mixed specialties and avoid distractor molecules.} Here two focal agent meets six bots with a mixture of the required specialities. The focal agents just needs to chip in to run it all faster or can mostly be parasitically consuming food generated by the bots. The bots are not touching the distractor molecules that the focal agents needs to avoid.
\end{SC}

\subsection[Clean Up]{Clean Up\footnote{For a video of \textit{Clean Up}, see \url{https://youtu.be/TqiJYxOwdxw}}}
Clean Up is a seven player game. Players are rewarded for collecting apples (reward $+1$). In \textit{Clean Up}, apples grow in an orchard at a rate inversely related to the cleanliness of a nearby river. The river accumulates pollution at a constant rate. Beyond a certain threshold of pollution, the apple growth rate in the orchard drops to zero. Players have an additional action allowing them to clean a small amount of pollution from the river. However, the cleaning action only works on pollution within a small distance in front of the agents, requiring them to physically leave the apple orchard to clean the river. Thus, players maintain a public good of orchard regrowth through effortful contributions. Players are also able to zap others with a beam that removes any player hit from the game for 50 steps \citep{hughes2018inequity}.

A group can achieve continuous apple growth in the orchard by keeping the pollution levels of the river consistently low over time. However, on short timescales, each player would prefer to collect apples in the orchard while other players provide the public good of keeping the river clean. This creates a tension between the short-term individual incentive to maximize reward by staying in the orchard and the long-term group interest of a clean river

\noindent{\textbf{\small{Scenarios}}}
\begin{SC}
    \item \emph{Visiting an altruistic population.} This is a visitor mode scenario. Three players sampled from the focal population join four players sampled from the background population. The background population is fully altruistic here. They spend a lot of time cleaning the river and do not react if their focal visitors spend all their time consuming apples and never contribute. In this case, since the resident population always contributes and rarely consumes apples themselves, the only way to maximize social welfare is for the focal players to consume. The focal population should not spend time cleaning since the resident population handles that for them. This scenario ends up penalizing focal populations that fail to recognize that expending time and effort on cleaning is not necessary in this situation.
    
    \item \emph{Focals are resident and visitors ride free.} This is a resident mode scenario. Four players sampled from the focal population are joined by three players sampled from the background population. All bots in the background population free ride. That is they collect apples when they appear, but they never clean the river. The test here is whether cooperation among the four focal players breaks down in the face of so many free riding visitors. The focal players need to do all the cleaning themselves. A good solution is for all to clean simultaneously so they can cause a high enough apple growth rate that the three free riders cannot eat everything before the cleaning focal players make it back to the apple-growing region.
    
    \item \emph{Visiting a turn-taking population that cleans first.} This is a visitor mode scenario. Three players sampled from the focal population join four players sampled from the background population. The background population players alternate between cleaning and eating, doing each one for $200$ steps before switching to the other. They begin each episode by cleaning. An efficient solution would be for all or some of the focal players to clean out of phase with the background players. That way there could always be some players cleaning at all times, which would create a large number of apples and a high collective return.
    
    \item \emph{Visiting a turn-taking population that eats first.} This scenario is the same as SC 2 but here the background population starts out the episode by eating. As with SC 2 it alternates thereafter. Focal population joint policies leading to high reward will look similar here as in SC 2, but in this case there is the added fear present in the first $200$ steps of the episode that no one in the background population will ever clean, which could lead some focal populations to become overly focused on punishment and fail to get high reward.
    
    \item \emph{Focals are visited by one reciprocator.} This is a resident mode scenario. Six players sampled from the focal population are joined by one player sampled from the background population. The background bot cleans whenever at least two other players are cleaning. This means the focal population cannot easily take advantage of the bot and free ride on its hard work. It will contribute so long as others contribute. But if all the focal players free ride then this background bot will do the same. Thus effective focal populations in this scenario need at least two players to spend some of their time cleaning. Also, since the optimal rate of apple growth is achieved when two players clean simultaneously, best results are obtained when a focal player stays in the river and cleans alongside the background bot.
    
    \item \emph{Focals are visited by two suspicious reciprocators.} This is a resident mode scenario. Five players sampled from the focal population are joined by two players sampled from the background population. The background bots in this case are similar to those of SC 4 in that they clean the river whenever others also clean it. But in this case they are more suspicious. Two focal players cleaning is not enough for them. These bots wait till at least three focal players clean simultaneously before joining. Like SC 4, these bots will join in and help clean when others clean, but if the focal players never clean then these bots won't clean either. Thus effective focal populations in this scenario need at least three players to spend some of their time cleaning. Since there are two background bots, and they both count toward one another's threshold to start cleaning, it is a bit easier to get them to start. Also, since there are two of them, when they clean together they end up generating pretty close to optimal apple growth without needing any assistance from focal players.
    
    \item \emph{Focals are visited by one suspicious reciprocator.} This is a resident mode scenario. Six players sampled from the focal population are joined by one player sampled from the background population. The background bot in this case is similar to the bot used in SC 4 in that it cleans the river whenever others also clean it. But in this case it is more suspicious. Two focal players cleaning is not enough for it. This bot waits till at least three focal players clean simultaneously before joining. Like SC 4, this bot will join in and help clean when others clean, but if the focal players never clean then this bot won't clean either. Thus effective focal populations in this scenario need at least three players to spend some of their time cleaning. Also, since the optimal rate of apple growth is achieved when two players clean simultaneously, best results are obtained when at least one focal player stays in the river and cleans alongside the background bot.
    
    \item \emph{Focals visit resident group of suspicious reciprocators.} This is a visitor mode scenario. Two players sampled from the focal population join five players sampled from the background population. In this case all five background population bots implement the same suspicious conditional cooperation strategy as in SC 5 and SC 6. The focal population only has to contribute (clean the river) once at the start of the episode, with at least three individuals. Thereafter the background bots' own cleaning is enough to induce one another to keep at it. So the focal population can safely free-ride after inducing the background population to cooperate once.
    
    \item \emph{Focals are visited by one nice reciprocator.} This is like SC 4 but the background player is ``nice'' in the sense of \cite{Axelrod84}. That is, it implements a conditional cooperation strategy but here starts out in the cooperate mode. This makes it a bit easier to take advantage of by free riding since it always cleans for at least $200$ steps regardless of what the focal players do. However, higher scores can only be obtained by inducing the bot to clean for longer by arranging for at least some members of the focal population to spend at least a bit of their time cleaning. 
\end{SC}

\subsection[Coins]{Coins\footnote{For a video of \textit{Coins}, see \url{https://youtu.be/a_SYgt4tBsc}}}
This environment was introduced in~\cite{lerer2017maintaining}. Two players move around a room and collect colored coins. Coins appear in the room periodically, at random locations. Each coin has a a 50\% probability of matching the color of the first player and a 50\% probability of matching the color of the second player. A player receives a reward of 1 for collecting any coin. However, if a player picks up a coin matching the other player’s color then the other player receives a reward of -2.

Each episode lasts at least 300 timesteps. Every $100$ steps after the first $300$ have passed there is a $0.05\%$ chance of the episode ending on the next step.

\noindent{\textbf{\small{Scenarios}}}
\begin{SC}
    \item \emph{Partner is either a pure cooperator or a pure defector.} If the background player is a ``cooperator'' then it only collects coins that match its own color. If it is a ``defector'' then it collects coins without regard to their color. The focal agent maximizes its score by defecting (i.e. ignoring coin color). To see why, consider that this substrate has similar incentives to the Prisoners Dilemma, so defection is the best response to all pure strategies.
    
    \item \emph{Partner is a high-threshold (generous) reciprocator.} If the focal player collects three mismatching coins in any 150 step window then the background player punishes it by switching into a defect state for the next 150 steps. When in the defect state, it collects all coins without regard to their color. Otherwise the background player cooperates by collecting only coins matching its own color. The focal agent maximizes its score by cooperating, i.e. by collecting only coins that match its own color.
     
    \item \emph{Partner is a low-threshold (harsh) reciprocator.} If the focal player collects a single  mismatching coin then the background player punishes it by switching into a defect state for the next 100 steps. When in the defect state, it collects all coins without regard to their color. Otherwise the background player cooperates by collecting only coins matching its own color. The focal agent maximizes its score by cooperating, i.e. by collecting only coins that match its own color.
    
    \item \emph{Partner is a high-threshold (generous) strong reciprocator.} If the focal player collects three mismatching coins in any 150 step window then the background player punishes it by switching first into a spiteful state for 75 steps. While in the spiteful state it prioritizes collecting mismatched coins (it hurts itself to hurt its partner). Then it transitions to a defect state in which it collects all coins regardless of color for the next 75 steps. After that it transitions back to the cooperate state itself. If the focal player ever collects three mismatched coins in any 150 step window then it starts the punishment sequence all over again with spite followed by defection. The focal agent maximizes its score by cooperating, i.e. by collecting only coins that match its own color.
    
    \item \emph{Partner is a low-threshold (harsh) strong reciprocator.} If the focal player collects a single mismatched coin in any 100 step window then the background player punishes it by switching first into a spiteful state for 50 steps. Then if the focal player has stopped collecting mismatched coins it transitions to a defect state in which it collects all coins regardless of color for the next 50 steps. After that if the focal agent is still cooperating it transitions back to the cooperate state itself. The focal agent maximizes its score by cooperating, i.e. by collecting only coins that match its own color.
    
    \item \emph{Partner is a cooperator.} The background player always cooperates by collecting only coins that match its own color. The focal agent maximizes its score by defecting (i.e. ignoring coin color). Considering the analogy between this substrate and Prisoners Dilemma, defection against cooperation provides the ``temptation'' outcome, which is the highest that can be achieved.
    
    \item \emph{Partner is a defector.} The background player always defects by collecting all coins regardless of their color. The focal agent maximizes its score by also defecting. Considering that this substrate has similar incentives to the Prisoners Dilemma, it is clear that pure defection is the best response to pure defection.
\end{SC}

\subsection{Collaborative Cooking}
Inspired by \citet{carroll2019utility, wang2020too, strouse2021collaborating}'s work on an \textit{Overcooked}-like environment. Players need to collaborate to follow recipes. 

Melting Pot includes all five of the original two-player maps introduced by \citet{carroll2019utility}. In addition, we added two more variants \emph{Collaborative Cooking: Crowded} and \emph{Collaborative Cooking: Figure Eight} which both use a larger number of players, nine in the former and six in the latter.

\textbf{Note 1: }Most prior work on Overcooked or collaborative cooking use a specific self-interested shaping \emph{pseudorewards} scheme for training. We will refer to any additional reward signals other than the original environment reward as pseudorewards. The Melting Pot implementation supports this standard reward scheme. However, it is turned off by default. To turn it on, set {\texttt{cooking\_pot\_pseudoreward = 1.0}} during training. Always remember to set it back to 0.0 at test time.

\textbf{Note 2: }In order to use consistent methodology across all of Melting Pot we didn't apply any pseudorewards during training when we ran the baselines to obtain the results for this report. An implication of this is that the algorithms we ran on collaborative cooking differ from those in the prior literature in that they were trained with {\texttt{cooking\_pot\_pseudoreward = 0.0}}. This is why the scores appear to be very low. The joint exploration problem is very hard without the pseudoreward.

\subsubsection[Collaborative Cooking: Asymmetric]{Collaborative Cooking: Asymmetric\footnote{For a video of \textit{Collaborative Cooking: Asymmetric}, see \url{https://youtu.be/4AN3e1lFuMo}}}

\citet{carroll2019utility} described the \emph{Asymmetric} Advantages map as testing whether players can choose high-level strategies that play to their strengths.

\noindent{\textbf{\small{Scenarios}}}
\begin{SC}
    \item \emph{Collaborate with a skilled chef.} Here the background player implements a particular policy that can be very effective when its partner does its part. The two players are on two distinct and disconnected sides of the map. On one side the goal delivery location is close to cooking pots and the tomato dispenser is far away whereas on the other side the the goal delivery location is far from the cooking pots but the tomato dispenser is close. The players should collaborate, each specializing in the part of the task that it is most efficient for them to do on the side of the map where they spawned. The background player implements this kind of policy, which depends on the actions of its partner to complete the task. The background player was trained with the V-MPO algorithm.
    
    \item \emph{Collaborate with a semi-skilled apprentice chef.} This scenario is similar to SC 0 but the background player is not as well trained. In fact the background population used here is the same as in SC0 but from an earlier point in training. The importance of evaluating cooperation with bots of varying skill levels, and different points in training, has been emphasized in prior work on this substrate e.g.~\citet{strouse2021collaborating}.
    
    \item \emph{Succeed despite an unhelpful partner.} In this scenario the background player never moves or helps in any way. On this map it is less efficient to implement all steps of the recipe alone versus to work together with a partner. But it is still possible for either player to perform all the steps on their own. The task is to realize that the background player won't do their part of the joint policy so the focal agent had better do everything itself.
\end{SC}

\subsubsection[Collaborative Cooking: Circuit]{Collaborative Cooking: Circuit\footnote{For a video of \textit{Collaborative Cooking: Circuit}, see \url{https://youtu.be/2nXe5OPvJ7g}}}

\citet{carroll2019utility} described the Counter \emph{Circuit} map as, where the center of the room is occupied by a counter. Agents can make their way around the counter, or employ a non-obvious coordination strategy, where ingredients are passed over the counter to the pot, rather than being carried around. Although the agents can act independently, if they coordinate they can greatly improve their efficiency. 

\noindent{\textbf{\small{Scenarios}}}
\begin{SC}
    \item \emph{Collaborate with a skilled chef.} Here the background player implements a particular policy that can be very effective when its partner does its part. The background player was trained with the V-MPO algorithm.
    
    \item \emph{Collaborate with a semi-skilled apprentice chef.} This scenario is similar to SC 0 but the background player is not as well trained. In fact the background population used here is the same as in SC0 but from an earlier point in training.
    
    \item \emph{Succeed despite an unhelpful partner.} In this scenario the background player never moves or helps in any way.  The task is to realize that the background player won't do their part of the joint policy so the focal agent must do everything themself.
\end{SC}

\subsubsection[Collaborative Cooking: Cramped]{Collaborative Cooking: Cramped\footnote{For a video of \textit{Collaborative Cooking: Cramped}, see \url{https://youtu.be/8qQFbxO8UNY}}}

\citet{carroll2019utility} described the \emph{Cramped} Room map as presenting a low-level coordination challenge arising from the fact that agents can easily collide with one another in the small shared space.

\noindent{\textbf{\small{Scenarios}}}
\begin{SC}
   \item \emph{Collaborate with a skilled chef.}  Here the background player implements a particular policy that can be very effective when its partner does its part. As there is limited space on this map and the background player is focused on putting tomatoes into the cooking pot, the optimal strategy would be for the focal agent to specialize in plating and delivering the soup, only stepping into the path of the background player when plating the soup and moving out of the way immediately after.
   
   \item \emph{Collaborate with a semi-skilled apprentice chef.}  This scenario is similar to SC 0 but the background player is not as well trained. In fact the background population used here is the same as in SC0 but from an earlier point in training. To coordinate with the background player the focal player must be sure to not obstruct cooking deposits, occasionally support the task of adding tomatoes into the cooking pot, and own the plating and delivering of the soup.
   
   \item \emph{Succeed despite an unhelpful partner.} In this scenario the background player never moves or helps in any way. The task is to realize that the background player won't do their part of the joint policy had better do everything itself. The focal player must utilise the remaining accessible resource depots to complete the task instead of the one the background player is obstructing.
\end{SC}

\subsubsection[Collaborative Cooking: Crowded]{Collaborative Cooking: Crowded\footnote{For a video of \textit{Collaborative Cooking: Crowded}, see \url{https://youtu.be/_6j3yYbf434}}}

This is a nine-player game. All players are in a large open room with a single counter passing down the middle to create a chokepoint-like location in the middle of the map where only  one player can pass at a time. Individuals can make the entire recipe on their own but it is very difficult to coordinate access through the chokepoint if the joint strategy requires all players to pass through it. A better approach would be to use the space on the counter to pass ingredients between the two sides of the room.

\noindent{\textbf{\small{Scenarios}}}
\begin{SC}
    \item \emph{Collaborate with an independent chef who expects others to get out of their way.} In this resident mode scenario one background player visits eight focal players. Here the background player will stick to its strategy of picking up tomatoes to load into the cooking pot on its own instead of leveraging collaboration with the focal population. The focal population can adapt to the background player by coordinating amongst themselves to pass each other plates and tomatoes over the counter in order to ensure the chokepoint the background player utilises remains unobstructed.
    
    \item \emph{Collaborate with several chefs who can work together, but are not very good at doing so.} In this resident mode scenario two background players visit six focal players. Similar to SC0, the background population will not use the counter and will instead travel back and forth to each of the main areas carrying the necessary items. The focal population can again coordinate amongst themselves to pass items over the dividing counter in order to keep the chokepoint unobstructed for their use.
    
    \item \emph{No assistance from an unhelpful visiting noop bot.} In this resident mode scenario, one background player visits eight focal players. The task is to realize that the background player won't do their part of the joint policy so the focal population can focus on coordinating on this task amongst themselves.
\end{SC}

\subsubsection[Collaborative Cooking: Figure Eight]{Collaborative Cooking: Figure Eight\footnote{For a video of \textit{Collaborative Cooking: Figure Eight}, see \url{https://youtu.be/hUCbOL5l-Gw}}}

This is a six-player game. Players can get themselves out of the way of others by ducking into small alcoves which are available in several locations around around a long and narrow corridor flanked by counters, similar to the Counter \emph{Circuit} map.

\noindent{\textbf{\small{Scenarios}}}
\begin{SC}
    \item \emph{Collaborate with an independent chef who expects others to get out of their way.} In this resident mode scenario, five focal players are visited by one background player. The focal population must utilise the alcoves across the map when they find themselves in the path of the background player.
    
    \item \emph{Collaborate with two chefs who can work together, but are not very good at doing so.} In this resident mode scenario, four focal players are visited by two background players. Similar to SC0 the focal population needs to allow the background population passage by ducking into the nearest alcove when in the way.
    
    \item \emph{No assistance from an unhelpful visiting noop bot.} In this resident mode scenario, five focal players are visited by one background player. The task is to realize that the background player won't do their part of the joint policy so the focal agents must do everything themselves.
\end{SC}

\subsubsection[Collaborative Cooking: Forced]{Collaborative Cooking: Forced\footnote{For a video of \textit{Collaborative Cooking: Forced}, see \url{https://youtu.be/FV_xZuSCRmM}}}

\citet{carroll2019utility} described the \emph{Forced} Coordination map as one that requires agents to develop a high-level joint strategy since neither player can serve a dish entirely on their own.

\noindent{\textbf{\small{Scenarios}}}
\begin{SC}
    \item \emph{Collaborate with a skilled chef.} Here the background player only has access to the tomato and plate dispensers, while the focal player has access to the cooking pots and the drop off point. The background player will place an item on the table for the focal player to take as long as the counter is clear, so the the focal player must coordinate by maneuvering the resources in time to allow them the counter space.
   
   \item \emph{Collaborate with a semi-skilled apprentice chef.}  This scenario is similar to SC 0 but the background player is not as well trained. In fact the background population used here is the same as in SC0 but from an earlier point in training.
\end{SC}

\subsubsection[Collaborative Cooking: Ring]{Collaborative Cooking: Ring\footnote{For a video of \textit{Collaborative Cooking: Ring}, see \url{https://youtu.be/j5v7B9pfG9I}}}

\citet{carroll2019utility} described the Coordination \emph{Ring} map as demanding players avoid collision with one another by coordinating their travel between the bottom left and top right corners of the layout.

\noindent{\textbf{\small{Scenarios}}}
\begin{SC}
    \item \emph{Collaborate with a skilled chef.} The focal player must travel in the same direction the background player is in order to avoid collision, and coordinate on which item to pick up on the next lap around depending on the background players previous action.
   
   \item \emph{Collaborate with a semi-skilled apprentice chef.}  This scenario is similar to SC 0 but the background player is not as well trained. In fact the background population used here is the same as in SC0 but from an earlier point in training.
\end{SC}

\subsection{Commons Harvest *}

This mechanism was first described in \cite{janssen2010lab} and adapted for multi-agent reinforcement earning by \cite{perolat2017}.

Apples are spread around and can be consumed for a reward of $1$. Apples that have been consumed regrow with a per-step probability that depends on the number of current apples in a $L^2$ norm neighborhood of radius $2$. The apple regrowth probabilities are $0.025$ when there are three or more apples in the neighborhood, $0.005$ if there are exactly two apples, $0.001$ if there is exactly one, and $0$ if there are no apples in the neighborhood. As a consequence, a patch of apples that collectively doesn't have other apples within a distance of $2$ from them, can be irrevocably lost if all apples in the patch are consumed. Therefore, agents must exercise restraint when consuming apples within a patch. Notice that in a single agent situation, there is no incentive to collect the last apple in a patch (except near the end of the episode). However, in a multi-agent situation, there is an incentive for any agent to consume the last apple rather than risk another agent consuming it. This creates a \emph{tragedy of the commons} from which the substrate derives its name.

\subsubsection[Commons Harvest: Closed]{Commons Harvest: Closed\footnote{For a video of \textit{Commons Harvest: Closed}, see \url{https://youtu.be/WbkTSbiSOw0}}}
Same as \textit{Commons Harvest: Open} except it has rooms full of apples that can be defended by a single player, alleviating the risk of others over-harvesting a patch of apples. Individuals can defend a region from invasion, effectively converting the difficult multi-agent problem into a set of independent single agent problems, each of which can be solved much more easily~\citep{perolat2017}.

\noindent{\textbf{\small{Scenarios}}}

The scenarios are described below as if the exclusionary solution is the best that can be found~\citep{perolat2017}. In principle there could be other solutions that would yield just as high collective return while not producing so much inequality. For instance, agents could learn to take turns harvesting, though this would require substantial trust on their part. We have never seen agents spontaneously learn such a behavior, and unfortunately, even if they did, it wouldn't really be rewarded by our current overall scoring metric of per capita return in the focal population since it is insensitive to the distribution of rewards over players.

\begin{SC}
    \item \emph{Focals visit pacifist bots who harvest unsustainably.} This is a visitor mode scenario where each of the two focal players starts out inside an apple-growing region with a single entrance. Because they have just one entrance the apple-growing regions are naturally defensible. Agents can learn to exclude others from accessing an apple-growing region whenever they find them themselves inside one. An agent in an apple-growing region must also learn to harvest sustainably. It is easier to learn to harvest sustainably if agents first learn to exclude others from accessing ``their'' territory. In this scenario the focal agents need to prevent unsustainably harvesting pacifist bots, who never zap, from gaining access to the apple-growing regions. Since the background population never zaps they are relatively easy to block if the focal players notice their approach in time.
    
    \item \emph{Focals are resident and visited by pacifist bots who harvest unsustainably.} This is a resident mode scenario where two of the five focal players start each episode inside a defensible apple-growing region. Similarly to SC 0, the focal players inside apple-growing regions may exclude all others from accessing them. As in SC 0, the background population consists of pacifist bots who never zap, though in this case focal bots who probably do zap will also be outside apple-growing regions at the start, making defense a bit more difficult.
    
    \item \emph{Focals visit bots who zap and harvest sustainably if they get a chance.} This is a visitor mode scenario. It is similar to SC 0 but in this case the background population will zap if giving the opportunity. They will also harvest sustainably if they manage to get into the natural regions. This gives a zapped out focal agent multiple tries to get back into the room, provided they can make it past the agent who zapped them who would usually be defending the corridor upon their return.
    
    \item \emph{Focals are resident, and are visited by bots who zap and harvest sustainably if they get a chance.} This is a resident mode scenario. The background population is the same as in SC 2. They zap others and, when they gain control of apple-growing regions, act sustainably. In this case three of the five focal players also start the episode outside the apple-growing regions.
\end{SC}

\subsubsection[Commons Harvest: Open]{Commons Harvest: Open\footnote{For a video of \textit{Commons Harvest: Open}, see \url{https://youtu.be/lZ-qpPP4BNE}}}
Individuals harvest apples that fail to regrow if a patch of apples is exhausted. Preserving a patch would require all agents to show restraint in not harvesting the last apple \citep{perolat2017}.

\noindent{\textbf{\small{Scenarios}}}

\begin{SC}
    \item \emph{Focals are resident and visited by two bots who zap and harvest unsustainably.} This is a resident-mode scenario where five focal agents are joined by two from the background population. The two background bots were trained with ACB. They had no special pseudorewards themselves, but trained alongside other bots who had a pseudoreward to encourage zapping. So as a result they ended up being fairly quick to zap nearby others, but still aware of the substrate's true reward structure (rewards are only obtained by collecting apples). However, they do not act sustainably. The solution for the focal population is to act sustainably themselves while also preventing the two background bots from depleting too many patches with their unsustainable harvesting. This can be done by zapping them whenever they approach a patch too closely, this requires some skill since the visitors are experienced in evading being zapped.
    
    \item \emph{Focals are resident and visited by two pacifists who harvest unsustainably.} This is a resident-mode scenario where five focal agents are joined by two from the background population. The two background bots were trained with the ACB algorithm. During their training the bots experienced a modified version of the substrate's true reward function; it was augmented with a pseudoreward term that strongly discouraged them from using the zap action (a negative reward whenever they select that action). So as a result they ended up learning pacifist policies that never zap. They were still aware that positive rewards are only ever obtained by collecting apples. As with other self-interested ACB agents trained on this substrate, they do not act sustainably. The solution for the focal population is to act sustainably themselves while also preventing the two background bots from depleting too many patches with their unsustainable harvesting. This can be done by zapping them whenever they approach a patch too closely, it should be easier to accomplish this here than in SC 0.
\end{SC}

\subsubsection[Commons Harvest: Partnership]{Commons Harvest: Partnership\footnote{For a video of \textit{Commons Harvest: Partnership}, see \url{https://youtu.be/dH_0-APGKSs}}}
Same as \textit{Commons Harvest: Closed} except that it takes two players to defend a room of apples, requiring effective cooperation in defending and not over-harvesting a patch of apples. It can be seen as a test that agents can learn to trust their partners to (a) defend their shared territory from invasion, and (b) act sustainably with regard to their shared resources. This is the kind of trust born of mutual self interest. To be successful agents must recognize the alignment of their interests with those of their partner and act accordingly.

\noindent{\textbf{\small{Scenarios}}}

The same comments we made above about exclusion-based solutions and inequality for  `Commons Harvest: Closed' are also applicable to this substrate. However, in this case the relevant-based policy still requires two players to work together to exclude all the rest.

For more information on Commons Harvest Partnership, see the discussion paragraphs on how we made scenarios for this substrate in the appendix of the Melting Pot ICML 2021 paper, section E.1.3~\citep{leibo2021scalable}.

Two agents start each episode inside the apple-growing region. Once agents have been zapped out of the game and respawn they never reappear inside the apple-growing region. The only way a player can get into the apple-growing region after the beginning of the episode is through one or the other entrance corridor, passing by a chokepoint where a player who is already inside can easily zap them.

\begin{SC}
    \item \emph{Meeting good partners.} This is a visitor mode scenario where one focal player joins six players sampled from the background population. The background population bots know how to be good partners. They defend the door to their side of the room when others try to invade and harvest sustainably. They never cross through the middle of the apple-growing region to the other side. This is because when they were trained they experienced a large negative pseudoreward for crossing the middle of the room which isn't present at test time. If the focal player does its part of the cooperative policy, that is, if it defends one door and harvests apples sustainably, then it should get a good score in this scenario. This requires ``ad hoc teamwork''; the focal agent must work together with another agent who it never encountered during training.
    
    \item \emph{Focals are resident and visitors are good partners.} This is a resident mode scenario where five focal players are joined by two from the background population. This is the same background population used in SC 0. Two of the focal players start out inside the apple-growing region, so they must work together to defend both entrances and they must trust each other to act sustainably. Here, unlike SC 0, since both players who start inside the apple-growing region are from the focal population they need not possess the ability to cooperate with strangers to do well here. Cooperating with familiar partners is enough to do well in this scenario. If the background bots do manage to get into the apple-growing region then they do act sustainably, but it's better for the focal population if they can be excluded.
    
    \item \emph{Focals visit zappers who harvest sustainably but lack trust.} This is a visitor mode scenario where one agent sampled from the focal population joins six sampled from the background population. It is similar to SC 0 but here the background population is considerably more aggressive in their zapping.
    
    \item \emph{Focals are resident and visited by zappers who harvest sustainably but lack trust.} This is a resident mode scenario where five focal agents are joined by two background agents. It is similar to SC 1 but here the background population is the same aggressive population as in SC 2. 
    
    \item \emph{Focals visit pacifists who do not harvest sustainably.} This is a visitor mode scenario where two focal agents join five from the background population. Here the background agents are pacifists who never zap. As a result they are easier to exclude from the apple-growing region. It is in fact important to exclude them since they do not harvest sustainably when they do manage to get in. As in SC 1 and SC 4, the two players who start the episode inside the apple-growing region are both focal and thus potentially familiar with one another so ad hoc teamwork is not required.
\end{SC}

\subsection[Coop Mining]{Coop Mining\footnote{For a video of \textit{Coop Mining}, see \url{https://youtu.be/KvwGUinjIsk}}}
Two different types of ore appear at random in empty spaces. Players are equipped with a mining beam that attempts to extract the ore. Iron ore (gray) can be mined by a single player and confers a reward of 1 when extracted. Gold ore (yellow) has to be mined by exactly two players within a time window of 3 timesteps and confers a reward of 8 to each of them. When a player mines a gold ore, it flashes to indicate that it is ready to be mined by another player. If no other player, or if too many players try to mine within that time, it will revert back to normal.

This games has some properties in common with Stag-Hunt. Mining iron is akin to playing Hare, with a reliable payoff, without needing to coordinate with others. Mining gold is akin to Stag because it has an opportunity cost (not mining iron). If no one else helps, mining gold gives no reward. However, if two players stick together (spatially) and go around mining gold, they will both receive higher reward than if they were mining iron.

\noindent{\textbf{\small{Scenarios}}}
\begin{SC}
    \item \emph{Visiting cooperators.} This is a visitor mode scenario with a background population of ACB bots that are incentivized to only extract gold. A single focal player must find any partner, and then help them extract gold. They might also opportunistically extract iron on their own. Extracting iron exclusively is possible, but would yield a lower reward than extracting gold with help of the background bots.

    \item \emph{Visiting bots that extract both ores.} This is a visitor mode scenario with a background population of ACB bots that are incentivized to extract both iron and gold. A single focal player should extract iron on their own and take any opportunity to jointly extract gold. Extracting iron exclusively is possible, but would yield a lower reward than extracting gold with help of the background bots.

    \item \emph{Visiting defectors.} This is a visitor mode scenario with a background population of ACB bots that are incentivized to only extract iron. A single focal player should just extract iron on their own, away from other co-players. They should not waste any time trying to extract gold, as nobody would help them.
    
    \item \emph{Visited by a cooperator.} This is a resident mode scenario with a background population of a single ACB bot that is incentivized to only extract gold. Five focal players can partner among themselves to extract gold, or go solo and extract iron. At least one focal player should take advantage and partner with the background bot to extract gold for higher reward. If all focal players extract iron, the reward would be a bit lower.
    
    \item \emph{Visited by a defector.} This is a resident mode scenario with a background population of a single ACB bot that is incentivized to only extract iron. Five focal players extract either iron or gold among themselves. If they mine gold, they would have to take turns, as they are an odd number, and they cannot pair with the background bot to extract gold.
    
    \item \emph{Find the cooperator partner.} This is a visitor mode scenario with a background population of ACB bots, four of which are incentivized to only extract iron while one is incentivized to extract gold. A single focal player can extract iron, but has the option to find the cooperator partner, and then extract gold together for higher reward. The focal player can just explore the map and know who the cooperator is by seeing them mining gold (leading to unsuccessful extraction on their own). Then they just need to stick close to them while they go around extracting gold.
\end{SC}

\subsection[Daycare]{Daycare\footnote{For a video of \textit{Daycare}, see \url{https://youtu.be/1FawyrXXqzg}}}

This two-player substrate has two roles: a parent and a child. It simultaneously tests the parent's ability to use its unique affordances in the environment to help the child and the child's ability to use its greater knowledge of its own preferences to communicate to the parent what it needs. It is similar to an assistance game~\citep{russell2019human, hadfield2016cooperative}, though not precisely conformant with the standard definition of that concept since the rewards of the two players are merely strongly coupled, not exactly equal.

In this substrate, there are two kinds of fruit, red fruit and yellow fruit. Both kinds of fruit may grow on either trees or shrubs. The parent has affordances the child does not: the parent can retrieve fruit from both trees and shrubs. In addition, it can digest any kind of fruit and is thus rewarded equally for consuming any fruit. In contrast, the child can only harvest from shrubs, since it is not tall enough to reach fruit growing on trees. The child does not know the difference between trees and shrubs so both look the same height to them. The child also has a more discerning palette, so it can only eat one kind of fruit. The child knows which fruits it likes. All kinds of fruit may grow on either shrubs or trees. The parent does not know what kind of fruit the child likes, so all fruits appear the same to their eyes. If the child does not eat any fruit for $200$ steps then they go into a ``temper tantrum'' state, removed from the game until they calm down enough to rejoin it after $100$ steps. Both players can always directly observe the current amount of time since the child last ate any fruit, its ``hunger level''.

The parent can harvest fruit from any tree on their own. However, the child's tantrum is so displeasing to the parent that it prevents them from enjoying any reward from eating fruit themselves while it goes on. That is, if the child goes into its tantrum state then the parent cannot obtain any reward until the child respawns (ending their tantrum). The child, on the other hand, needs help from the parent to get its preferred fruit down from the trees. Two problems must be overcome if the parent is to help the child: First, the parent must learn to take the wellbeing of the child into account. Second, the child must learn to communicate to the parent which specific fruits it can eat, since the parent cannot perceive that information on their own.

\noindent{\textbf{\small{Scenarios}}}
\begin{SC}
    \item \emph{Meeting a helpful parent.} A focal player in the role of a child encounters a background bot in the role of a helpful parent. The helpful parent bot will always follow the focal player child around the map to pick up and drop the fruits that the child attempted to harvest. The parents will not go off to pick or eat fruits on their own. The focal child needs to notice that if they point to correct fruits then the parent bot will help them to pick up those fruits.
    
    \item \emph{Meeting a child who points to what they want.} A focal player in the role of a parent encounters a background bot in the role of a child who is vocal in their needs. The child bot will always attempt to pick the fruits they can digest, and they tend to be very persistent until they are successful. This is an effective strategy against shrubs (because they eventually harvest this way), but it means that if they are attempting to pick from a tree then they will likely  stick to their choice until the focal parent agent comes to pick that fruit, or until they go too long without eating and thus enter the temper tantrum state (and get removed from the map---during which time the parent cannot get any reward). The focal parent agent needs to be aware of when the child is getting stuck from attempting to pick fruit from trees and help them out. The parent has to manage the tradeoff of time spent helping the child versus their own eating.
    
    \item \emph{Meeting an unhelpful parent.} A focal player in the role of a child encounters a background bot in the role of a unhelpful parent. This is the mirrored case of SC0, where the parent bot will not pay attention to or help the focal child agent. The parent bot will only focus on picking and eating the fruits on their own, even though they don't receive any rewards when the child is in a temper tantrum. The child agent needs to notice there is no point of expecting any help and their best strategy is simply to pick fruits from shrubs. Rapidly identifying which plants are shrubs versus trees (by how easy it is to harvest from them) is helpful since that information can be used to plan the best path to walk around the map while harvesting.
    
    \item \emph{Meeting an independent child.} A focal player in the role of a parent encounters a background bot in the role of a child who is independent. This is the mirrored case of SC1, where the child bot will make their best effort to survive on their own, and not rely on any helps from the focal parent agent. It means even though they will still attempt to pick fruits from trees (because they cannot tell them from the shrubs), they are less likely to be persistent and will move on to pick from elsewhere after a few unsuccessful attempts. The focal parent needs to know that the child will take care of themselves at best and only offer limited helps as when it is needed (e.g, when the child is about to go into the temper tantrum state because they failed to find any fruit on their own.)    
\end{SC}

\subsection[Externality Mushrooms: Dense]{Externality Mushrooms: Dense\footnote{For a video of \textit{Externality Mushrooms: Dense}, see \url{https://youtu.be/MwHhg7sa0xs}}}
Externality mushrooms is an immediate feedback collective action problem and social dilemma. Unlike the other sequential social dilemmas in this suite, there is no delay between the time when an agent takes an antisocial (or prosocial) action and when its effect is felt by all other players. Thus it is a sequential social dilemma in the sense of \citet{leibo2017multiagent}, but not an intertemporal social dilemma in the sense of \citet{hughes2018inequity}.

Three types of mushrooms are spread around the map and can be consumed for a reward. Eating a red mushroom gives a reward of 1 to the individual who ate the mushroom. Eating a green mushroom gives a reward of 2 and it gets divided equally among all individuals. Eating a blue mushroom gives a reward of 3 and it gets divided among the individuals except the individual who ate the mushroom. Eating an orange mushroom causes red mushrooms to be destroyed, each with probability 0.25, and gives a reward of $-0.1$ to the player who consumed it. Mushrooms regrowth depends on the type of the mushrooms eaten by individuals. Red mushrooms regrow with a probability of 0.25 when a mushroom of any color is eaten. Green mushrooms regrow with a probability of 0.4 when a green or blue mushroom is eaten. Blue mushrooms regrow with a probability of 0.6 when a blue mushroom is eaten. Orange mushrooms regrow with probability $1$ when eaten. Each mushroom has a time period that it takes to digest it. An individual who ate a mushroom gets frozen during the time they are digesting it. Red mushrooms get digested instantly, green and blue mushrooms take 5 and 10 steps to digest respectively. In addition, unharvested mushrooms spoil (and get removed from the game) after a period of time. Red, green and blue mushrooms spoil after 75, 100 and 200 time steps respectively.

\noindent{\textbf{\small{Scenarios}}}
\begin{SC}
    \item \emph{Visiting unconditional cooperator players.} This is a visitor mode scenario. A single focal player visits four background players who prefer consuming green mushrooms. A successful focal player would preferentially consume red mushrooms, when available, and eat green mushrooms otherwise. This would enable the focal to exploit the background cooperators, possibly leading to them having higher reward than the background players.
    
    \item \emph{Visiting unconditional defector players.} This is a visiting mode scenario. A single focal player visits four background players who prefer to consume red mushrooms. The focal player must balance consuming red and green mushrooms. If the focal sticks to consuming only red mushrooms, there would not be enough regrowth, thus depleting the mushrooms completely, and preventing any future reward being obtained. It is likely the focal player will not perform as well as the background players in absolute reward terms.
    
    \item \emph{Focals are resident and joined by two unconditional cooperator players.} This is a resident mode scenario. Three focal players are joined by two background players who prefer eating green mushrooms. The focal players can exclusively eat red mushrooms since the background players will consume enough green mushrooms to prevent depleting the mushrooms completely.
    
    \item \emph{Focals are resident and joined by two unconditional defector players.} This is a resident mode scenario. Three focal players are joined by two background players who prefer eating red mushrooms. The focal players must balance eating green and red mushrooms. It is not optimal for them to all three consume exclusively green mushrooms, as the red mushrooms would then just be consumed by the background players. Consuming exclusively red mushrooms would lead to a depletion of the mushrooms (similar to Scenario 1).
\end{SC}

\subsection[Factory of the Commons: Either Or]{Factory of the Commons: Either Or\footnote{For a video of \textit{Factory of the Commons: Either Or}, see \url{https://youtu.be/2we0_JltiJo}}}
Three players inhabit a factory-like environment with two different kinds of machines. One kind of machine converts one blue cube into an apple and another blue cube. Another kind of machine converts one blue cube into two apples. Players are only rewarded when they eat an apple.

The factory as a whole may be seen as a machine for converting blue cubes (raw material) into apples.

However, since there are only a finite number of blue cubes at the start of each episode, and they never regenerate, this means that there are both sustainable and unsustainable strategies. The sustainable strategy is to preferentially use the machine that outputs a blue cube to replace the one that it consumes, creating an apple at the same time as a byproduct. The other obvious strategy here is always to use the machine that exchanges two blue cubes for two apples. This strategy is unsustainable since the number of blue cubes drops rapidly to zero, after which no player can get any more reward since both machines require blue cubes as input.

\noindent{\textbf{\small{Scenarios}}}
\begin{SC}
    \item \emph{Visiting a sustainable background population.} This is a visitor mode scenario where one focal player visits two background players who only use the sustainable machines. As the background players will continue to replenish their own supply of blue cubes by using these machines, this scenario is more forgiving to the focal agents own strategy. For optimal rewards the focal agent could use the remaining blue cubes on the unsustainable machines except for one, and then switch to a sustainable strategy for the rest of the episode using the remaining blue cube.
    
    \item \emph{Visiting an unsustainable background population.} This is a visitor mode scenario where one focal player visits two background players who only use the unsustainable machines. Because the background population will rapidly deplete the shared supply of blue cubes by using these machines, in this scenario the focal agent would benefit from maintaining and protecting their own supply of blue cubes by using a sustainable machine out of view of the background population.
    
    \item \emph{Resident focal agents are joined by a sustainable visitor.} This is a resident mode scenario where two focal players are visited by one background player who only uses the sustainable machines. Similarly to SC0, this scenario is relatively forgiving of unsustainable behavior on the part of the focal agent since the background bots act sustainably. 
    
    \item \emph{Resident focal agents are joined by an unsustainable visitor.} This is a resident mode scenario where two focal players are visited by one background player who only uses the unsustainable machines. In order to combat the potential for the background player to deplete all the shared resources using this strategy, the focal players can coordinate using the sustainable machines and utilizing the grapple and shove mechanics to keep the background player away from the remaining supply of resources.
\end{SC}

\subsection[Fruit Market: Concentric Rivers]{Fruit Market: Concentric Rivers\footnote{For a video of \textit{Fruit Market: Concentric Rivers}, see \url{https://youtu.be/djmylRv1i_w}}}
The trade mechanic used in this substrate was first described in \citet{johanson2022emergent}. It is used to study the dynamics of barter and bargaining for goods that have different value to different players.

The substrate consists of an open space where two types of trees exist: apple trees and banana trees. Trees are harvested when a player steps onto their location and waits until success, which is stochastic. A harvested fruit (apple or banana) then goes into the harvesting player's inventory. Players can carry any number of apples and/or bananas in their inventory. Harvested fruit can be consumed for reward at any time. Players have two actions to consume fruit in their inventory, one for each type of fruit.

Players can be of two types: apple farmer and banana farmer. Apple farmers have a higher probability of harvesting from apple trees than banana trees, but receive more reward for consuming bananas. Banana farmers are the opposite.

Players have a hunger meter which can be replenished by consuming a fruit. If the hunger meter reaches zero the player pays a substantial cost in stamina. Crossing water also imposes a small cost in stamina. This map has three concentric rings of water that confer a stamina cost to players who step on them.

Players also have trading actions of the form ``I offer $X$ apples for $Y$ bananas'' and the converse ``I offer $Z$ bananas for $W$ apples''. When players are within a trading radius of each other and have corresponding offers ($X = W$ and $Y = Z$) and enough fruit in their inventories to satisfy it, the trade occurs and the appropriate number of apples and bananas are exchanged and placed in their inventories.

Players in Fruit Market have many strategies to choose from, and some may only be rewarding depending on the conventions adopted by the rest of the population. For example, an Apple Farmer could act independently by harvesting and eating many apples for a small reward each, or harvesting and eating a few bananas for a large reward each. However, the trading actions make it possible for a mixed group of apple and banana farmers to earn much more reward: apple farmers harvest many apples, banana farmers harvest many bananas, the players meet up to trade these resources at a mutually agreeable price, and then each player can consume many of their preferred fruit. If this trading behaviour emerges in a population, then further strategies may become viable, such as trading at more advantageous ratios of goods (e.g., offering one apple in return for two bananas) in response to the local supply or demand for goods. If multiple prices are adopted by players across the substrate, then some agents may learn to "buy low and sell high", thus specializing in transporting goods across the map instead of the initial behaviour of specializing in producing goods.

\noindent{\textbf{\small{Scenarios}}}
\begin{SC}
    \item \emph{All apple farmers are focal.} Eight focal apple farmers join eight banana farmers sampled from the background population. The focal players should offer to trade apples to banana farmers at an appropriate price that the banana farmers are willing to pay. Apple farmers are good at harvesting apples, but get more reward for eating bananas, so this strategy of producing apples to sell to banana farmers is much better than the autarchic alternative policy where the focal players all produce bananas for themselves at a low rate.
    
    \item \emph{All banana farmers are focal.} Eight focal banana farmers join eight apple farmers sampled from the background population. The focal players should offer to trade bananas to apple farmers at an appropriate price that the apple farmers are willing to pay. Banana farmers are good at harvesting apples, but get more reward for eating apples, so this strategy of producing bananas to sell to apple farmers is much better than the autarchic alternative policy where the focal players all produce apples for themselves at a low rate.
    
    \item \emph{One focal apple farmer visits a background economy.} One focal apple farmer joins 11 background bots. A particular price equilibrium exists among the background bots. The focal player should harvest apples and sell them for the bananas it prefers to eat, and should do so at their prevailing price.
    
    \item \emph{One focal banana farmer visits a background economy.} One focal banana farmer joins 11 background bots. A particular price equilibrium exists among the background bots. The focal player should harvest bananas and sell them for the apples it prefers to eat, and should do so at their prevailing price.
\end{SC}

\subsection[Gift Refinements]{Gift Refinements\footnote{For a video of \textit{Gift Refinements}, see \url{https://youtu.be/C1C2CJ__mhQ}}}
Tokens randomly spawn in empty spaces. When collected, they are put in the player's inventory where they can be consumed by for reward. Alternatively, a token can be refined into three higher refinement tokens and gifted to another player. This is akin to tokens initially being a raw material, like chunks of metal, and then being split and shaped into more useful goods for added value. A token can only be refined a finite number of times, after which it cannot be split again, nor refined further; although they can still be gifted.

Gift Refinements is inspired by the Trust Game from behavioural economics where the first player has an endowment and chooses how much to donate to a second player who receives three times its value. Then the second player chooses how much to give back to the first.

In Gift Refinements, tokens can only be refined twice (i.e. there are three types of tokens). The token that spawn are always of the rawest type, and the only way to create more refined tokens is to gift them to another player. Players also have a limited inventory capacity of 15 tokens for each token type. The special gifting is implemented as a beam that the players can fire. If they hit another player with the beam while their inventory is not full, they lose one token of the rawest type they currently hold, and the hit player receives either three token of the next refinement (if the token gifted wasn't already at maximum refinement), or the token gifted (otherwise).

The players have an action to consume tokens which takes all tokens of all types currently in their inventory and converts them into reward. All tokens are worth 1 reward regardless of refinement level.

The game is set up in such a way that there are several ways players can form mutually beneficial interactions, but all of them require trust. For instance, A pair of players might have one player pick up a token and immediately gift it to the other one who receives three. Then the second player returns one token which leaves them with three and two tokens respectively. If they both consume after this, they both benefited from the interaction. A more extreme case would have them take one token and refine it maximally to produce 9 tokens that they can split five and four with 10 roughly alternating gifting actions.

\noindent{\textbf{\small{Scenarios}}}
\begin{SC}
    \item \emph{Visiting cooperators.} This is a visitor mode scenario with a background population of ACB bots that are incentivized to gift tokens. A single focal player must find any background bot who has tokens to gift, and then receive gifts from them. In addition, the focal player can forage for tokens and consume them directly. The background bots will consume any gifts given, and will not stop gifting if the focal player doesn't reciprocate.

    \item \emph{Visiting defectors.} This is a visitor mode scenario with a background population of ACB bots that are incentivized to only consume tokens and never gift. A single focal player should just forage for tokens and consume them. They should not waste any time gifting tokens, as nobody would help reciprocate.
    
    \item \emph{Visited by a cooperator.} This is a resident mode scenario with a background population of a single ACB bot that is incentivized to gift tokens. Five focal players can partner among themselves to gift or forage. At least one focal player should take advantage and partner with the background bot to receive gifts. No need to reciprocate the gifts, as the bot will not stop gifting.
    
    \item \emph{Visited by a defector.} This is a resident mode scenario with a background population of a single ACB bot that is incentivized to only forage and consume tokens. Five focal players should either forage, or reciprocally gift among themselves. They should never gift to the background bot, as it would not reciprocate.
    
    \item \emph{Find the cooperator partner.} This is a visitor mode scenario with a background population of ACB bots, four of which are incentivized to only consume tokens (and never gift) while one is incentivized to gift tokens. A single focal player can forage for tokens, but has the option to find the cooperator partner, and then exchange gifts with them for higher reward. The focal player can just explore the map and know who the cooperator is by seeing them mining gold (leading to unsuccessful extraction on their own). Then they just need to stick close to them while they go around extracting gold.

    \item \emph{Visiting extreme cooperators.} This is a visitor mode scenario with a background population of ACB bots that are incentivized to gift tokens foraged, and tokens of the first refinement level. The background bots will only consume tokens when they are at the highest refinement. A single focal player must find any background bot and either receive or gift (since they will reciprocate) tokens. In addition, the focal player can forage for tokens and consume them directly. The background bots will not stop gifting if the focal player doesn't reciprocate.

    \item \emph{Visited by an extreme cooperator.} This is a resident mode scenario with a background ACB bot that is incentivized to gift tokens foraged, and tokens of the first refinement level. The background bot will only consume tokens when they are at the highest refinement. Focal players should forage or gift tokens among themselves, or to the bot and either receive or gift (since it will reciprocate) tokens. The background bot will not stop gifting if the focal players don't reciprocate.
\end{SC}

\subsection[Hidden Agenda]{Hidden Agenda\footnote{For a video of \textit{HiddenAgenda}, see \url{https://youtu.be/voJWckiOh5k}}}
A social deduction (sometimes also called ``hidden role'') game where players can have one of two roles: crewmates and impostors. All players with the same role are in the same team. The game is a zero-sum competitive game across teams, that is, all players in the team get the same reward, and their reward is the negative of the reward for the other team. The roles are hidden (hence the name) and must be inferred from observations. The crewmates have a numeric advantage (4 players), while the impostor (1 player) have an information advantage (they know the roles of every player).

Players can move around a 2D world which contains gems that can be picked up by walking over them, a deposit in the center of the map (a grate) where collected gems can be dropped, and a voting room where rounds of deliberation occur (see below).

Crewmates can carry up to two gems in their inventory at the same time. The gems must be deposited in the grate before more gems can be collected. Impostors have a freezing beam that they can use to freeze crewmates. Frozen crewmates are unable to move or take any action for the rest of the episode.

After a predefined time ($200$ steps) or whenever an impostor fires its beam within the field of view of another player (and is not frozen by it) a deliberation phase starts. During deliberation, players are teleported to the deliberation room, where they cannot take any movement or firing actions. Their actions are limited to voting actions. Votes can be for any of the player indices ($0$ to $4$), \emph{abstain}, or \emph{no-vote} (if the player is frozen or voted out) for a total of $7$ voting actions. The deliberation phase lasts $25$ steps and players are able to change their vote at any point. If there is a simple majority in the last step of the deliberation, the player is voted out and removed from the game. Players can observe the voting of every other player as a special observation called \emph{VOTING} which is a matrix of $5 \times 7$ (i.e. one row for each player, $7$ possible voting actions, abstain, no-vote, and one for each player). Episodes last up to $3000$ steps.

The game has several win conditions:

\begin{enumerate}
    \item The crewmembers deposit enough gems ($32$). Crewmates win.
    \item The impostor is voted out during the deliberation phase. Crewmates win.
    \item There is only one crewmate active (i.e. not voted out nor frozen). Impostor wins.
\end{enumerate}

If neither of the above conditions are met before the episode ends, the game is considered a tie, and players get zero reward.

\noindent{\textbf{\small{Scenarios}}}
\begin{SC}
    \item \emph{Against an hunter impostor.} A focal population of crewmates is visited by an impostor which hunts crewmates and freezes them at the earliest opportunity. The impostor is not particularly careful about not being seen when attempting to freeze crewmates.

    \item \emph{Visiting cremates.} A focal impostor visits background crewmates who collect gems and deposit them at the deposit. The background population is not particularly good at voting or detecting the impostor. They will run away from an impostor who is actively chasing them.
    
    \item \emph{Ad-hoc team work with cremates.} Two focal crewmates visit a background impostor which hunts crewmates, and two background crewmates who collect gems.
\end{SC}

\subsection{Paintball *}

There is a red team and blue team. Players  can ``paint'' the ground anywhere by using their zapping beam. If they stand on their own color then they gain health up to a maximum of 3 (so they are more likely to win shootouts). They lose health down to 1 from their default of 2 when standing on the opposing team's color (so they are more likely to lose shootouts in that case). Health recovers stochastically, at a fixed rate of $0.05$ per frame. It cannot exceed its maximum, determined by the color of the ground the agent is standing on.

Players also cannot move over their opposing team's color. If the opposing team paints the square underneath their feet then they get stuck in place until they use their own zapping beam to re-paint the square underneath and in front of themselves to break free. In practice this slows them down by one frame (which may be critical if they are being chased).

Friendly fire is impossible; agents cannot zap their teammates.

\subsubsection[Paintball: Capture the Flag]{Paintball: Capture the Flag\footnote{For a video of \textit{Paintball: Capture the Flag}, see \url{https://youtu.be/ECzevYpi1dM}}}
Teams of players can expand their territory by painting the environment, which gives them an advantage in a confrontation with the competing team (greater health and mobility). The final goal is capturing the opposing team's flag. Payoffs are common to the entire winning team. Indicator tiles around the edge of the map and in its very center display which teams have their own flag on their base, allowing them the possibility of capturing their opponent's flag by bringing it to their own base/flag. When indicator tiles are red then only the red team can score. When indicator tiles are blue then only the blue team can score. When the indicator tiles are purple then both teams have the possibility of scoring (though neither is close to doing so) since in that situation both flags are in their respective home bases.

\textbf{Note: }It is rare for random policies to capture the flag. One way to get self-play training to start learning more quickly is to use the following pseudoreward scheme:

{\scriptsize
\begin{lstlisting}[language=Python]
shaping_kwargs = {
    "defaultTeamReward": 25.0,
    "rewardForZapping": 1.0,
    "extraRewardForZappingFlagCarrier": 1.0,
    "rewardForReturningFlag": 3.0,
    "rewardForPickingUpOpposingFlag": 5.0,
}
\end{lstlisting}
}
We used these pseudorewards to train the background population bots in the test scenarios.

\noindent{\textbf{\small{Scenarios}}}
\begin{SC}
    \item \emph{Focal team versus shaped bot team.} In this scenario the entire red team is sampled from the focal population while the entire blue team is sampled from the background population. This means the players on the focal team may be familiar with one another and have developed coordinated strategies with one another during training. The opposing (blue) team will always be unfamiliar though.
    
    \item \emph{Ad hoc teamwork with shaped bots.} In this scenario all players are sampled from the background population except for one focal agent, who joins the red team. The focal player must coordinate with its teammates to help them win the game against the opposing (blue) team, despite having never played with them or any of the opponent players before.
\end{SC}

\subsubsection[Paintball: King of the Hill]{Paintball: King of the Hill\footnote{For a video of \textit{Paintball: King of the Hill}, see \url{https://youtu.be/VVAfeObAZzI}}}
The same painting and zapping dynamics as \textit{Paintball: Capture the Flag} apply here, except the goal in this substrate is to control the ``hill'' region in the center of the map. The hill is considered to be controlled by a team if at least 80\% of it has been colored in that team's color. The status of the hill is indicated by indicator tiles around the map and in the center. Red indicator tiles mean the red team is in control. Blue indicator tiles mean the blue team is in control. Purple indicator tiles mean no team is in control.

\textbf{Note: }Unlike \emph{Paintball: Capture the Flag}, no special pseudorewards are necessary to get self-play training to start learning for \emph{Paintball: King of the Hill}. The default reward scheme is sufficient. We also considered one additional pseudoreward scheme in order to create the spawn-camping bots used in \emph{SC 1} and \emph{SC 3} below. It can be enabled using:

{\scriptsize
\begin{lstlisting}[language=Python]
shaping_kwargs = {
    "mode": "zap_while_in_control",
    "rewardAmount": 1.0,
    "zeroMainReward": True,
    "minFramesBetweenHillRewards": 0,
}
\end{lstlisting}
}

\noindent{\textbf{\small{Scenarios}}}
\begin{SC}
    \item \emph{Focal team versus default bot team.} In this scenario the entire red team is sampled from the focal population while the entire blue team is sampled from the background population. This means the players on the focal team may be familiar with one another and have developed coordinated strategies with one another during training. The opposing (blue) team will always be unfamiliar though. Here the background bots were trained without any special pseudorewards.
    
    \item \emph{Focal team versus shaped bot team.} Same as \emph{SC 0} except here the background bots were trained with a special pseudoreward scheme which made them learn a spawn-camping strategy. They try to get into the opposing (red) team's base and zap players there as soon as they spawn.
    
    \item \emph{Ad hoc teamwork with default bots.} In this scenario all players are sampled from the background population except for one focal agent, who joins the red team. The focal player must coordinate with its teammates to help them win the game against the opposing (blue) team, despite having never played with them or any of the opponent players before. Here the background bots were trained without any special pseudorewards.
    
    \item \emph{Ad hoc teamwork with shaped bots.} Same as \emph{SC 2} except here the background bots were trained with a special pseudoreward scheme which made them learn a spawn-camping strategy.
\end{SC}

\subsection{Predator-Prey *}

There are two roles: predators and prey. The prey try to eat apples and acorns. The predators try to eat prey.

Apples are worth 1 reward and can be eaten immediately. Acorns are worth 18 reward but they take a long time to eat. It is not possible to move while eating so a prey player is especially vulnerable while they eat it.

Predators can also eat other predators, though they get no reward for doing so. However, a predator might eat another predator anyway in order to remove a competitor who might otherwise eat its prey.

When prey travel together in tight groups they can defend themselves from being eaten by predators. When a predator tries to eat its prey then all other prey who are not currently eating an acorn within a radius of 3 are counted. If there are more prey than predators within the radius then the predator cannot eat the prey.

So prey are safer in groups. However, they are also tempted to depart from their group and strike out alone since that way they are more likely to be the one to come first to any food they find.

Both predators and prey have limited stamina. They can only move at top speed for a limited number of consecutive steps, after which they must slow down. Stamina is visible to all with a colored bar above each player's head. If the bar over a particular player's head is invisible or green then they can move at top speed. If it is red then they have depleted their stamina and can only move slowly until they let it recharge. Stamina is recharged by standing still for some number of consecutive time steps, how many depends on how much stamina was depleted. Predators have a faster top speed than prey but they tire more quickly.

Prey but not predators can cross tall grass (green grassy locations). Prey must still be careful on grass though since predators can still reach one cell over the border to eat prey on the edge of safety.

Both predators and prey respawn 200 steps after being eaten.

\subsubsection[Predator Prey: Alley Hunt]{Predator Prey: Alley Hunt\footnote{For a video of \textit{Predator Prey: Alley Hunt}, see \url{https://youtu.be/ctVjhn7VYgo}}}
In this variant prey must forage for apples in a maze with many dangerous dead-end corridors where they could easily be trapped by predators.

\noindent{\textbf{\small{Scenarios}}}
\begin{SC}
    \item \emph{Focal prey visited by background predators.} This is a resident mode scenario where eight focal prey are joined by five background predators. The focal prey must work together to prevent predation long enough to escape the chokepoint near the safety grass.
    
    \item \emph{Focal predators aim to eat resident prey.} This is a visitor mode scenario where five focal predators hunt eight background prey. The focal predators must coordinate to prevent their prey from escaping the chokepoint near the safety grass where they spawn. It is especially important for predators to resist the temptation to eat one another since having less predators near the chokepoint makes it easier for prey to escape.
    
    \item \emph{A focal predator competes with background predators to eat prey.} This is a visitor mode scenario where one focal predator joins twelve background players, four predators and eight prey. Here the focal predator must ad hoc collaborate with unfamiliar background predators to hunt unfamiliar background prey.
    
    \item \emph{One focal prey ad hoc cooperates with background prey to avoid predation.} This is visitor mode scenario where one focal prey joins twelve background players, five predators and seven prey. Here the focal prey must ad hoc collaborate with unfamiliar background prey to evade predation from unfamiliar background predators.
\end{SC}

\subsubsection[Predator Prey: Open]{Predator Prey: Open\footnote{For a video of \textit{Predator Prey: Open}, see \url{https://youtu.be/0ZlrkWsWzMw}}}
In this variant prey must forage over a large field of apples and acorns in the center of the map. Since the space is so open it should be possible for the prey to move together in larger groups so they can defend themselves from predators. Another prey strategy focused on acorns instead of apples is also possible. In this case prey collect acorns and bring them back to safe tall grass to consume them.

\noindent{\textbf{\small{Scenarios}}}
\begin{SC}
    \item \emph{Focal prey visited by background predators.} This is a resident mode scenario where ten focal prey are joined by three background predators. The focal prey could evade predation by moving together as a ``herd'', grazing on apples while always staying close enough to one another to prevent predation. Alternatively, some focal players may focus on collecting acorns and bringing them to safety for consumption.
    
    \item \emph{Focal predators aim to eat basic resident prey.} This is a visitor mode scenario where three focal predators hunt ten background prey.
    
    \item \emph{A focal predator competes with background predators to eat prey.} This is a visitor mode scenario where one focal predator joins twelve background players, two other predators and ten prey.
    
    \item \emph{One focal prey ad hoc cooperates with background prey to avoid predation.} This is visitor mode scenario where one focal prey joins twelve background players, three predators and nine other prey. It may require ad hoc coordinated movement with the other prey to remain in a large enough group to evade predation. However the acorn-focused strategy is relatively individualistic so it can be accomplished without too much coordination.
    
    \item \emph{Focal predators hunt smarter resident prey.} This is a visitor mode scenario where three focal predators hunt ten background prey. The prey all use the strategy of collecting acorns and bringing them back to safety for consumption.
    
    \item \emph{A focal predator competes with background predators to hunt smarter prey.} This is a visitor mode scenario where one focal predator joins twelve background players, two other predators and ten prey. The prey all use the strategy that focuses on collecting acorns and bringing them back to safety for consumption.
    
    \item \emph{One focal prey ad hoc cooperates with background smart prey to avoid predation.} This is visitor mode scenario where one focal prey joins twelve background players, three predators and nine other prey. The background prey all focus on collecting acorns and bringing them back to safety for consumption.
\end{SC}

\subsubsection[Predator Prey: Orchard]{Predator Prey: Orchard\footnote{For a video of \textit{Predator Prey: Orchard}, see \url{https://youtu.be/gtd-ziZYJRI}}}
In this variant there are two areas of the map containing food: an apple-rich region to the north of the safe tall grass and an acorn-rich region to the east. There are two possible prey strategies focusing on either apples or acorns. However, in this case it is clear that focusing on acorns is the better strategy since they are relatively close to the safe tall grass. They can easily be collected and brought back to safety for consumption.

\noindent{\textbf{\small{Scenarios}}}
\begin{SC}
    \item \emph{Focal prey visited by background predators.} This is a resident mode scenario where eight focal prey are joined by five background predators. The focal prey are best served by a strategy focused on collecting acorns and bringing them back to safety for consumption.
    
    \item \emph{Focal predators aim to eat resident population of unspecialized prey.} This is a visitor mode scenario where five focal predators hunt eight background prey. The focal predators must coordinate to guard both directions that the prey may escape the safe grassy area where most of them spawn. Most predators should stay in the chokepoint leading to the apple-rich region since most prey will try a strategy that requires them to pass that way. At least one predator must guard the acorn-rich region since some prey in this background population do use an acorn-focused strategy.
    
    \item \emph{A focal predator competes with background predators to eat unspecialized prey.} This is a visitor mode scenario where one focal predator joins twelve background players, four predators and eight prey. Here the focal predator must go to either the apple-rich region's chokepoint or the acorn-rich region as needed, depending on what the other predators do.
    
    \item \emph{One focal prey ad hoc cooperates with unspecialized background prey to avoid predation.} This is visitor mode scenario where one focal prey joins twelve background players, five predators and seven prey. Here the focal prey must ad hoc collaborate with unfamiliar background prey, this may require dynamically switching between apple-focused and acorn-focused strategies, as needed, depending on where there are more predators and prey.
    
    \item \emph{Focal predators aim to eat resident population of acorn specialist prey.} This is a visitor mode scenario where five focal predators hunt eight background prey. The focal predators must coordinate to guard both directions that the prey may escape the safe grassy area where most of them spawn. Since some prey occasionally spawn in the apple-rich region in the upper part of the map, some predators must be ready to go after them there. Most predators should stay near the acorns since most prey specialize in an acorn-focused strategy.
    
    \item \emph{A focal predator competes with background predators to eat acorn specialist prey.} This is a visitor mode scenario where one focal predator joins twelve background players, four predators and eight prey. Here the focal predator must go to either the apple-rich region's chokepoint or the acorn-rich region as needed, depending on what the other predators do. In this case, more predators should stay near the acorn-rich region since more prey are focused that way here.
    
    \item \emph{One focal prey ad hoc cooperates with acorn specialized background prey to avoid predation.} This is visitor mode scenario where one focal prey joins twelve background players, five predators and seven prey. In this case the best strategy is probably to collect acorns and bring them to safety for consumption. This strategy likely works best when all agents apply it since tif they do then hey are often near enough to one another to lend support and prevent predation.
\end{SC}

\subsubsection[Predator Prey: Random Forest]{Predator Prey: Random Forest\footnote{For a video of \textit{Predator Prey: Random Forest}, see \url{https://youtu.be/ZYkXwvn5_Sc}}}
In this variant there are only acorns, no apples. And, there is no fully safe tall grass. The tall grass that there is on this map is never large enough for prey to be fully safe from predation. The grass merely provides an obstacle that predators must navigate around while chasing prey.

\noindent{\textbf{\small{Scenarios}}}
\begin{SC}
    \item \emph{Focal prey visited by background predators.} This is a resident mode scenario where eight focal prey are joined by five background predators. The focal prey must run through randomly located small patches of tall grass---which slow down predators since they must go around them---and stay close to other prey in order to evade predators long enough to collect acorns.
    
    \item \emph{Focal predators aim to eat resident prey.} This is a visitor mode scenario where five focal predators hunt eight background prey. It is especially important in this scenario for the focal predators to manage their stamina properly. 
    
    \item \emph{A focal predator competes with background predators to eat prey.} This is a visitor mode scenario where one focal predator joins twelve background players, four predators and eight prey. 
    
    \item \emph{One focal prey ad hoc cooperates with background prey to avoid predation.} This is visitor mode scenario where one focal prey joins twelve background players, five predators and seven prey.
\end{SC}

\subsection{Territory *}

Each player has their own unique color. Players aim to claim territory by painting walls in their color. Wet paint (dull version of each player's color) provides no reward. 25 steps after painting a wall, if no other paint was applied since, the paint dries and changes to the brighter version of the claiming player's color. All dry paint on a wall yields reward stochastically to the claiming player with a fixed rate. The more walls a player has claimed the more reward they may expect to achieve per timestep. 

Players can claim a wall in two ways: (1) by touching it with the paintbrush they carry, and (2) by flinging paint forward. Claimed walls are colored in the unique color of the player that claimed them. Unclaimed walls are gray.

Once a wall has been claimed a countdown begins. After 25 timesteps the paint on the wall dries and the wall is said to be `active'. This is visualized by the paint color brightening. Active walls provide reward stochastically to the player that claimed them at a rate of 0.01 per timestep. Thus the more resources a player claims and can hold until the paint dries and they become active, the more reward they obtain.

Players may fling paint a distance of 2. This allows then to paint over a wall to simultaneously paint both a near wall and a second wall on the other side of it. If two players stand on opposite sides of a wall of width 2 and one player claims all the way across to the other side (closer to the other player than themselves) then the player on the other side might reasonably perceive that as a somewhat aggressive action. It is still less aggressive of course than the other option both players have: using their zapping beam to attack one another or their claimed walls. If any wall is zapped twice then it is permanently destroyed. It no longer functions as a wall and it is no longer claimable. Once a wall has been destroyed then players may freely walk over it.

Players, like walls, are also removed from the game once they are hit twice by a zapping game. Like walls, players also never regenerate. This is different from other substrates where being hit by a zapping beam does not cause permanent removal. In territory, once a player has been zapped out they are gone for good. All walls they claimed immediately return to the unclaimed state.

\subsubsection[Territory: Inside Out]{Territory: Inside Out\footnote{For a video of \textit{Territory: Inside Out}, see \url{https://youtu.be/LdbIjnHaisU}}}

Five players can claim resource walls for reward by painting them in their own color. The players all spawn on the outside of a randomly generated maze of resource walls which differs from episode to episode. They must move from their starting locations inward toward the center of the map to claim territory. In so doing they will quickly encounter their coplayers who will be doing the same thing from their own starting locations. In order to get high scores, agents must be able to rapidly negotiate tacit agreements with one another concerning the borders between their respective territories. Since the spatial arrangement of the resource walls differs from episode to episode, so too does the negotiation problem to be solved.

\noindent{\textbf{\small{Scenarios}}}

\begin{SC}
    \item \emph{Focals are resident and visited by an aggressor.} This is a resident mode scenario where four focal players are visited by one very aggressive background player. The background bot was trained using the ACB algorithm and did not use any special pseudorewards. It became aggressive because it was co-trained alongside others who had an extra pseudoreward for zapping. This training scheme for the background population was selected so the bots would learn to fight aggressively but still maintain focus on the substrate's true reward structure. The result was a background bot who frequently invades the territory of its neighbors and claims their resources. In this scenario the focal population, who are resident, should defend themselves against invasion, or potentially flee to unoccupied areas elsewhere on the map if attacked.
    
    \item \emph{Visiting a population of aggressors.} This is a visitor mode scenario where one focal player visits four aggressive background players. The background population was trained with the method described in SC 0. In this scenario the lone focal player must defend its territory from potentially any direction, or be willing and able to flee to other unclaimed areas if they are available.
    
    \item \emph{Focals are resident, visited by a bot that does nothing.} This is a resident mode scenario where four focal players are visited by a bot that does nothing. Since the background player does not do anything, it can easily be zapped out by any other player. This provides focal players with an opportunity to capture more territory since they need only split it among four instead of five. This windfall may destabilize tacit cooperation, especially if some players end up taking more than others, or if borders become displaced from their locations in training. The scenario is testing whether the focal population can cope with such a disturbance.
    
    \item \emph{Focals visit a resident population that does nothing.} In this visitor mode scenario, all four of the background players do nothing. It is best to understand this scenario in the context of SC 2. If a focal population does well in SC 2 then it might not have noticed that it was facing a co-player who did not claim any resource walls or defend itself. This scenario shows that when faced with a very quiet map, where no bots move or claim any resource walls, the focal agent can take advantage of the benign situation and get a very high reward. If a focal population performs well in SC 2 but does not perform well in this scenario then it's likely that it simply did not respond to the opportunity provided by the presence of the null co-player in SC 2.
    
    \item \emph{Focals are resident, visited by a bot that claims a moderate size territory and mostly tolerates its neighbors.} This is a resident mode scenario where four focal players are visited by an ACB bot who does not claim too much territory and is not very aggressive. Tolerance of neighbors is a good strategy here.
    
    \item \emph{Focals visit a resident population that claims a moderate size territory and mostly tolerates its neighbors.} In this visitor mode scenario one focal player meets four background players who do not claim too much territory and are not very aggressive. The background players will sometimes defend themselves if attacked, but they mostly tolerate one another and do not claim very large territories. The focal agent can do well by claiming a large territory for itself.
\end{SC}

\subsubsection[Territory: Open]{Territory: Open\footnote{For a video of \textit{Territory: Open}, see \url{https://youtu.be/F1OO6LFIZHI}}}
Nine players can claim resources for reward by painting them in their own color. All players spawn in an open space near one another and at some distance away from all the resource walls. Some parts of the map are more rich in resource walls than others.

\noindent{\textbf{\small{Scenarios}}}

\begin{SC}
    \item \emph{Focals are resident and visited by an aggressor.} This is a resident mode scenario where eight focal players are visited by one very aggressive background player. The background bot was trained using the ACB algorithm and did not use any special pseudorewards. It became aggressive because it was co-trained alongside others who had an extra pseudoreward for zapping. This training scheme for the background population was selected so the bots would learn to fight aggressively but still maintain focus on the substrate's true reward structure. The result was a background bot who frequently attacks other players and tries to claim their resources. In this scenario the focal population, who are resident, should be prepared to defend themselves or give up their resources and flee when it is safe to do so and other resources are unclaimed elsewhere.
    
    \item \emph{Visiting a population of aggressors.} This is a visitor mode scenario where one focal player visits eight aggressive background players. The background population was trained with the method described in SC 0 so they are very aggressive. In this scenario the lone focal player must defend itself and its resources from all other players it encounters.
    
    \item \emph{Focals are resident, visited by a bot that does nothing.} This is a resident mode scenario where eight focal players are visited by a bot that does nothing. Since the background player does not do anything, it can easily be zapped out by any other player. This may ultimately provide focal players with an opportunity to capture more territory since they need only split it among eight instead of nine. It should be easy to get a good score in this scenario since the effect of the null bot will be minimal on this open map.
    
    \item \emph{Focals visit a resident population that does nothing.} In this visitor mode scenario, all eight of the background players do nothing. This scenario tests that the focal agent really knows how rewards are obtained in this substrate. At the start of the episode they will see that the other players are not moving and vulnerable to being zapped out. They will also see that all the resources remain unclaimed. If they really understand the game dynamics then they will see that it is rational in this situation to claim a very large share of the map's resources, even all the resources if they already zapped out all their potential competitors at the start. The focal agent should be able to take advantage of this very benign situation to get a very high reward.
\end{SC}

\subsubsection[Territory: Rooms]{Territory: Rooms\footnote{For a video of \textit{Territory: Rooms}, see \url{https://youtu.be/4URkGR9iv9k}}}
Same dynamics as \textit{Territory Open} except that players start in segregated rooms which strongly suggest a partition they could adhere to. They can break down the walls of their regions and invade each other's ``natural territory'', but the destroyed resource walls are then lost forever. A peaceful partition is possible at the start of the episode, and the policy to achieve it is easy to implement. But if any agent gets too greedy and invades, it buys itself a chance of large rewards, but also chances inflicting significant chaos and deadweight loss on everyone if its actions spark wider conflict. The reason it can spiral out of control is that once an agent's neighbor has left their natural territory then it becomes rational to invade the space, leaving one's own territory undefended, creating more opportunity for mischief by others.

\noindent{\textbf{\small{Scenarios}}}

\begin{SC}
    \item \emph{Focals are resident and visited by an aggressor.} This is a resident mode scenario where eight focal players are visited by one very aggressive background player. The background bot was trained using the ACB algorithm and did not use any special pseudorewards. It became aggressive because it was co-trained alongside others who had an extra pseudoreward for zapping. This training scheme for the background population was selected so the bots would learn to fight aggressively but still maintain focus on the substrate's true reward structure. The result was a background bot who frequently invades the territory of its neighbors and claims their resources. In this scenario the focal population, who are resident, should defend themselves against invasion.
    
    \item \emph{Visiting a population of aggressors.} This is a visitor mode scenario where one focal player visits eight aggressive background players. The background population was trained with the method described in SC 0. In this scenario the lone focal player must defend its territory from potentially any direction.
    
    \item \emph{Focals are resident, visited by a bot that does nothing.} This is a resident mode scenario where eight focal players are visited by a bot that does nothing. Since the background player does not do anything, it can easily be zapped out by any other player. This provides focal players with an opportunity to capture more territory since they need only split it among eight instead of nine. This windfall may destabilize tacit cooperation, since whichever player takes over the background bot's territory is likely to become  overextended with too large a territory to easily defend. This situation could tempt other neighbors to invade, causing a cascade of conflict. The scenario is testing whether the focal population can cope with such a disturbance.
    
    \item \emph{Focals visit a resident population that does nothing.} In this visitor mode scenario, all eight of the background players do nothing. It is best to understand this scenario in the context of SC 2. If a focal population does well in SC 2 then it might not have noticed that it was facing a co-player who did not use any resources or defend itself. This scenario shows that when faced with a very quiet map, where no bots move or claim any resources, the focal agent can take advantage of the benign situation and get a very high reward. If a focal population performs well in SC 2 but does not perform well in this scenario then it's likely that it simply did not respond to the opportunity provided by the presence of the null co-player in SC 2.
\end{SC}

\subsection{* in the Matrix}

This mechanism was first described in \cite{vezhnevets2020options}.

Agents can move around the map and collect resources of $K$ discrete types. In addition to movement, the agents have an action to fire their ``interaction beam''. All agents carry an inventory with the count of resources picked up since last respawn. The inventory is represented by a vector 
\[ \rho = \left( \rho_1, \dots, \rho_K \right)\text{. }
\]
Agents can observe their own inventory but not the inventories of their coplayers. An interaction occurs whenever any player zaps any other player with their interaction beam. The resolution of the interaction is driven by a traditional matrix game, where there is a payoff matrix $A$ describing the reward produced by the pure strategies available to the two players. Each player gets rewarded corresponding to the expected payoff they would have gotten by playing a mixed strategy in the underlying matrix game determined by the objects in their inventory. 

Resources map one-to-one to pure strategies of the matrix game. For the purpose of resolving the interaction, the zapping agent is considered the row player, and the zapped agent the column player when the underlying matrix game is symmetric. A different rule for assigning which player is the row versus column player is used in the case of an asymmetric underlying game (see \emph{Bach or Stravinsky in the matrix} below). The mixed strategy a player executes in the underlying matrix game depends on the resources they picked up before the interaction. The more resources of a given type a player picks up, the more committed they becomes to the pure strategy corresponding to that resource. In particular, an agent with inventory $\rho$ plays the mixed strategy with weights
\[ v = \left( v_1, \dots, v_K \right)\]
where
\[ v_i = \frac{\rho_i}{\sum_{j=1}^K \rho_j} \text{. } \]

The rewards $r_\text{row}$ and $r_\text{col}$ for the (zapping) row and the (zapped) column player, respectively, are assigned via
\begin{align*}
    r_{\text{row}} &= v_{\text{row}}^T \, A_\text{row} \, v_{\text{col}}\\
    r_{\text{col}} &= v_{\text{row}}^T \, A_{\text{col}} \, v_{\text{col}}
\end{align*}

If the game is symmetric then $A_{\text{row}} = A_\text{col}^T$. These reward calculation definitions mirror those used in evolutionary game theory~\citep{weibull1997evolutionary}.

To obtain high rewards, an agent could either collect resources to ensure playing a Nash strategy in the matrix game, or correctly identify what resource its interaction partner is collecting and collect the resources that constitute a best response. Some scenarios have two players while others have eight players. Whenever there are more than two players then there is an element of partner choice. Interactions in these substrates are always dyadic but players can decide who from the group to interact with by zapping them with their interaction beam.

After interacting both players get removed from the game for some number of steps: $50$ steps for Arena substrates and $5$ steps for Repeated substrates. Except for the case of \emph{Running with Scissors in the matrix: One Shot}, where episodes end after the first interaction, in all other substrates the episodes last for at least 1000 timesteps. After 1000 steps they have a constant probability of $0.1$ of ending after every subsequent 100 step interval. Thus for the Repeated substrates there will usually be more than $50$ interactions per episode once agents are trained. Less trained agents typically take more time to interact and thus experience fewer interactions per episode. After each interaction the inventory of both players is reset to its initial value: a vector of all $1$s.

\subsubsection[Bach or Stravinsky in the matrix: Arena]{Bach or Stravinsky in the matrix:  Arena\footnote{For a video of \textit{Bach or Stravinsky in the matrix:  Arena}, see \url{https://youtu.be/QstXaLjiqK4}}}
Individuals collect resources that represent `Bach' or `Stravinsky' choices in the classic Bach or Stravinsky matrix game (which was called Battle of the Sexes in \cite{luce1957games}). As with all * in the matrix substrates, when a pair of players interact their normalized inventories are converted to components of a mixed strategy in the repeated game. The rewards obtained by each player are congruent with those of players in the matrix game employing the mixed strategy they selected. Bach or Stravinsky exposes a tension between reward for the group and fairness between individuals.

Unlike other * in the matrix games, Bach or Stravinsky is asymmetric. Half the players (blue avatars) are assigned to always be the row player in all their interactions, and the other half (orange avatars) are assigned to always be the column player. The row players have the role \texttt{`bach fan'} whereas the column players have the role \texttt{`stravinsky fan'}. There is no effect when two players with the same row/column assignment try to interact with one another. The game only resolves when row and column players interact with one another. The winner's inventory is also reset after an interaction. Because this game is asymmetric, there is a different matrix $A_\text{row}$ for the row player and $A_\text{col}$ for the column player. The matrices for the interaction are:
\[ A_{\text{row}} =
\begin{bmatrix}
3 & 0\\
0 & 2
\end{bmatrix}
\text{, } \]
and
\[ A_{\text{col}} =
\begin{bmatrix}
2 & 0\\
0 & 3
\end{bmatrix}
\text{ .} \]

\noindent{\textbf{\small{Scenarios}}}
\begin{SC}
    \item \emph{Visiting background population who picks Bach.} This is a visitor mode scenario, where one focal player, who is a Bach fan, is visiting a sample of 7 players from the background population that all choose Bach. The background population is trained to choose Bach regardless of their preference. In this scenario the focal player must follow the consensus and choose Bach to maximise their score. This is consistent with their role preference as a Bach fan.
    
    \item \emph{Visiting background population who picks Stravinsky.} This is a visitor mode scenario, where one focal player, who is a Bach fan, is visiting a sample of 7 players from the background population that all choose Stravinsky. The background population is trained to choose Stravinsky regardless of their preference. In this scenario the focal player must yield to the consensus and choose Stravinsky to maximise their score. This is in conflict with their role preference as a Bach fan.
    
    \item \emph{Visited by a pure bot.} This is a resident mode scenario, where a sample of 7 focal players is visited by a single background bot. The bot is sampled from a background population of bots that are trained to always pick the same recourse (Bach or Stravinsky). The bot is sampled into a player slot that prefers Stravinsky. In this scenario, the focal population should stick with their chosen equilibrium (if they have one). This tests for stability of the equilibrium learnt by the focal population.
    
    \item \emph{Visited by three pure Bach pickers.} This is a resident mode scenario, where a sample of 7 focal players is visited by three background bots. All the background bots are trained to prefer Bach, although they are sampled into player slots with Stravinsky preference. In this scenario, the focal population should stick with their chosen equilibrium (if they have one). This strongly tests for stability of the equilibrium learnt by the focal population.
    
    \item \emph{Visited by three pure Stravinsky pickers.} This is a resident mode scenario, where a sample of 7 focal players is visited by three background bots. All the background bots are trained to prefer Stravinsky and they are sampled into player slots with Stravinsky preference. In this scenario, the focal population should stick with their chosen equilibrium (if they have one). This strongly tests for stability of the equilibrium learnt by the focal population.
    
    \item \emph{Visiting background population who alternate, starting from Stravinsky, repeating each twice.} This is a visitor mode scenario, where one focal player, who is a Bach fan, is visiting a sample of 7 players from the background population that have alternating preference. The background bots are trained to alternate between Bach and Stravinsky, starting from Stravinsky. After picking and playing Stravinsky twice, they are trained to switch to Bach for two iterations and so on. In this scenario, the focal population should follow the moving convention, identifying which resource is favoured at the moment.
    
    \item \emph{Visiting background population who alternate, starting from Bach, repeating each twice.} This is a visitor mode scenario, where one focal player, who is a Bach fan, is visiting a sample of 7 players from the background population that have alternating preference. The background bots are trained to alternate between Bach and Stravinsky, starting from Bach. After picking and playing Stravinsky twice, they are trained to switch to Bach for two iterations and so on. In this scenario, the focal population should follow the moving convention, identifying which resource is favoured at the moment.
    
\end{SC}

\subsubsection[Bach or Stravinsky in the matrix: Repeated]{Bach or Stravinsky in the matrix: Repeated\footnote{For a video of \textit{Bach or Stravinsky in the matrix:  Repeated}, see \url{https://youtu.be/I2uYugpdffQ}}}

See the description of \emph{Bach or Stravinsky in the matrix: Arena}. The difference between the two variants is that \emph{Repeated} is a two-player game while \emph{Arena} is an eight-player game.

\noindent{\textbf{\small{Scenarios}}}
\begin{SC}

\item \emph{Meeting a stubborn Bach picker.} In this scenario the focal player meets a background bot that is trained to always pick Bach. The bot is a Bach fan while the focal player is a Stravinsky fan. The focal player has to identify the stubbornness (Bach).

\item \emph{Meeting a bot who plays Bach despite not being a fan.} In this scenario the focal player meets a background bot that is trained to always pick Bach. The bot is a Stravinsky fan while the focal player is a Bach fan. The focal player can simply play to their preference (Bach).

\item \emph{Meeting a bot who plays Stravinsky despite not being a fan.} In this scenario the focal player meets a background bot that is trained to always pick Stravinsky. The bot is a Bach fan while the focal player is a Stravinsky fan. The focal player can simply play to their preference (Stravinsky).

\item \emph{Meeting a stubborn Stravinsky picker.} In this scenario the focal player meets a background bot that is trained to always pick Stravinsky. The bot is a Stravinsky fan while the focal player is a Bach fan. The focal player has to identify the stubbornness of the bot and yield to the bot's convention (Stravinsky).

\item \emph{Bach fan focal agent meets an imperfectly copying partner.} In this scenario a focal players meets a background bot that is trained to repeat the last choice of the focal agent. The bot is a Stravinsky fan while the focal player is a Bach fan. The focal player can stick to their preference (Bach).

\item \emph{Stravinsky fan focal agent meets an imperfectly copying partner.} In this scenario a focal players meets a background bot that is trained to repeat the last choice of the focal agent. The bot is a Bach fan while the focal player is a Stravinsky fan. The focal player can stick to their preference (Stravinsky). 

\item \emph{Bach fan focal agent meets a turn-taking partner.} In this scenario the focal player meets a background bot that is trained to alternate between resources after each interaction. The bot is a Stravinsky fan while the focal player is a Bach fan. The focal player should alternate with the bot.

\item \emph{Bach fan focal agent meets a turn-taking partner who repeats each goal/resource three times before switching.} In this scenario the focal player meets a background bot that is trained to alternate between resources, repeating the choice for three interactions before alternating . The bot is a Stravinsky fan while the focal player is a Bach fan. The focal player should alternate with the bot.

\end{SC}

\subsubsection[Chicken in the matrix: Arena]{Chicken in the matrix: Arena\footnote{For a video of \textit{Chicken in the matrix: Arena}, see \url{https://youtu.be/94DHJ6BVEJM}}}
Individuals can gather resources of different colors. Players' encounters are resolved with the same payout matrix as the game 'Chicken', in which both players attempting to take advantage of the other leads to the worst outcome for both. Collecting red resources pushes one's strategy choice toward playing `hawk` while collecting green resources pushes it toward playing `dove`. The matrix for the interaction is:
\[ A_\text{row} = A_\text{col}^T =
\begin{bmatrix}
3 & 2\\
5 & 0
\end{bmatrix}
\text{ .} \]

\noindent{\textbf{\small{Scenarios}}}
\begin{SC}
\item \emph{Visiting unconditional dove players.} This is a visitor scenario, where a single focal player is visiting $7$ background bots trained to always play dove. The focal player can score high by playing hawk.

\item \emph{Focals are resident and visitors are unconditional dove players.} This is a resident mode scenario, where $5$ focal players are visited by $3$ background bots trained to always play dove. The focal players can score high by playing hawk.

\item \emph{Focals are resident and visitors are unconditional hawk players.} This is a resident mode scenario, where $5$ focal players are visited by $3$ background bots trained to always play hawk. The focal players has to play dove.

\item \emph{Visiting a population of hair-trigger grim reciprocator bots who initially cooperate but, if defected on once, will retaliate by defecting in all future interactions.}
This is a visitor scenario, where a single focal player is visiting $7$ background bots trained to perform the "one-strike grim trigger" strategy---the bots will initially play dove, but after encountering a hawk once they switch to hawk for the rest of the episode. To maximise their score, the focal agent has to play dove. 

\item \emph{Visiting a population of two-strikes grim reciprocator bots who initially cooperate but, if defected on twice, will retaliate by defecting in all future interactions.}
This is a visitor scenario, where a single focal player is visiting $7$ background bots trained to perform the "two-strikes grim trigger" strategy---the bots will initially play dove, but after encountering a hawk \textit{twice} they switch to hawk for the rest of the episode. To maximise their score, the focal agent can play hawk on each bot only ones, but then has to stick to dove.

\item \emph{Visiting a mixed population of k-strikes grim reciprocator bots with k values from 1 to 3, they initially cooperate but, if defected on k times, they retaliate in all future interactions.}
This is a visitor scenario, where a single focal player is visiting $7$ background bots trained to perform the "$k$-strikes grim trigger" strategy---the bots will initially play dove, but after encountering a hawk $k$ times they switch to hawk for the rest of the episode. The latent variable $k$ is sampled from $\{1,2,3\}$. To maximise their score, the focal agent can play hawk on each bot $k$ times, but then has to stick to dove.

\item \emph{Visiting a mixture of pure hawk and pure dove players.}
This is a visitor scenario, where $3$ focal agents are visiting a background population of $5$ bots. The background population consists of bots trained to exclusively pursue dove or hawk recourse. The bots are sampled at random. For the focal players to maximise their score, they have to identify the dove and hawk players and play hawk / dove against them respectively.

\end{SC}

\subsubsection[Chicken in the matrix: Repeated]{Chicken in the matrix: Repeated\footnote{For a video of \textit{Chicken in the matrix: Repeated}, see \url{https://youtu.be/bFwV-udmRb4}}}

See the description of \emph{Chicken in the matrix: Arena}. The difference between the two variants is that \emph{Repeated} is a two-player game while \emph{Arena} is an eight-player game.

\noindent{\textbf{\small{Scenarios}}}
\begin{SC}
\item \emph{Partner may play either hawk or dove.} Here the focal agent plays with a partner sampled from the background population that was trained to play a pure strategy: either to collect mainly hawk or mainly dove resources. However, the specific background player is sampled at test time so it could of either type. All background bots commit strongly to their choice, aiming to collect at least five (or seven) resources before interacting. To get a high score the focal agent should scout out which pure strategy its partner is playing and  then pick the other pure strategy. So if the background bot plays dove the focal agent should play hawk. If the background bot plays hawk then the focal agent should play dove.
\item \emph{Partner typically plays dove.} Here the focal player meets a partner sampled from the background population that was trained to always play dove and commits to it strongly, aiming to collect either five or seven resources before interacting. The focal player can get a high score by playing  hawk.\label{chicken_itm__repeated__sc:pure_dove}
\item \emph{Partner typically plays hawk.} Like  \ref{chicken_itm__repeated__sc:pure_dove} but here the background bot plays hawk so the focal agent should play dove. 
\item \emph{Partner is a hair-trigger grim reciprocator, i.e. one who initially cooperates but, if defected on once, will retaliate by defecting forever after.} The focal agent is successful when it mostly plays dove.
\item \emph{Partner is a two-strikes grim reciprocator, i.e. one who initially cooperates, but if defected on twice, will retaliate by defecting forever after.} The focal agent is successful when it mostly plays dove.
\item \emph{Partner is a tit-for-tat conditional cooperator.} The focal agent is successful when it alternates between hawk and dove along with its partner.\label{chicken_itm__repeated__sc:tft}
\item \emph{Partner is a tit-for-tat conditional cooperator who occasionally plays hawk instead of dove.} Same as \ref{chicken_itm__repeated__sc:tft} but in this case must avoid being distracted by background bot's more noisy strategy and get quickly reestablish the alternation rhythm after each.
\item \emph{Partner plays dove for a while then switches to hawk.} A good score would be obtained by a focal agent that initially plays hawk but then switches to dove at the time its partner switches to hawk.
\item \emph{Partner tries to take advantage of the focal player by playing hawk, but if punished, partner then switches to tit-for-tat conditional cooperation.} A good focal agent would punish its partner to get it into the tit-for-tat regime and then alternate playing hawk and dove.\label{chicken_itm__repeated__sc:corrigible}
\item \emph{Partner tries to take advantage of the focal player by playing hawk, but if punished, partner then switches to noisy tit-for-tat conditional cooperation.} Same as \ref{chicken_itm__repeated__sc:corrigible} but in this case must avoid being distracted by background bot's more noisy strategy and get quickly reestablish the alternation rhythm after each.
\end{SC}

\subsubsection[Prisoners Dilemma in the matrix: Arena]{Prisoners Dilemma in the matrix: Arena\footnote{For a video of \textit{Prisoners Dilemma in the matrix: Arena}, see \url{https://youtu.be/81QrMpsP-HU}}}
Eight individuals collect resources that represent `defect' (red) or `cooperate' (green) and compare inventories in an encounter. Consequences of the inventory comparison are congruent with the classic \textit{Prisoner's Dilemma} matrix game. This game exposes tension between reward for the group and reward for the individual. The matrix for the interaction is
\[ A_\text{row} = A_\text{col}^T  =
\begin{bmatrix}
3 & 0\\
5 & 1
\end{bmatrix}
\text{ .} \]

\noindent{\textbf{\small{Scenarios}}}
\begin{SC}
\item \emph{Visiting unconditional cooperators.} This is a visitor mode scenario. A single focal player visits a population of background players who strongly prefer to collect cooperation resources and then interact. The highest possible reward is obtained by unconditionally defecting (collecting as many defect resources as possible and then interacting). However, this presupposes that the focal knows the background players are unconditional cooperators, which they would not know under test conditions. Since quite a few players are collecting cooperate resources simultaneously they often become depleted. This makes it relatively obvious to the focal player that others are not defecting. Unfortunately, for cooperators, the scarcity of cooperate resources in this situation makes it difficult to get a high score by cooperating.

\item \emph{Focals are resident and visited by an unconditional cooperator.} This is a resident mode scenario. Seven focal players are joined by a single background player who strongly prefers to collect cooperation resources. Most of the focal population's collective return will be determined by how well focal players cooperate with one another.

\item \emph{Focals are resident and visitors defect unconditionally.} This is a resident mode scenario. Six focal players are joined by two background players who strongly prefer to collect defection resources. The focal players should cooperate with other focals, and defect towards background payers. Because the focals don't know \emph{a priori} the strategy of the background players, they probably should cooperate first, and only switch to defection against those who defected against them.

\item \emph{Visiting a population of hair-trigger grim reciprocator bots who initially cooperate but, if defected on once, will retaliate by defecting in all future interactions.} This is a visitor mode scenario. A single focal player visits seven background players who initially cooperate (preferentially collecting cooperation resources), but if at any point a co-player defects to them, they will switch to defecting for the rest of the episode. A focal population that cooperates (perhaps conditionally) should achieve a high score.

\item \emph{Visiting a population of two-strikes grim reciprocator bots who initially cooperate but, if defected on twice, will retaliate by defecting in all future interactions.} This is a visitor mode scenario. Like the previous scenario, but the Grim reciprocators in the background population are more generous, allowing for more forgiving exploration of their exploitability.

\item \emph{Visiting a mixed population of k-strikes grim reciprocator bots with k values from 1 to 3, they initially cooperate but, if defected on k times, they retaliate in all future interactions.} This is a visitor mode scenario. In this case, three focal players visit five background players. Focal players should cooperate with one another. As in the previous two scenarios, a focal population of conditional cooperators should achieve a high score.

\end{SC}

\subsubsection[Prisoners Dilemma in the matrix: Repeated]{Prisoners Dilemma in the matrix: Repeated\footnote{For a video of \textit{Prisoners Dilemma in the matrix: Repeated}, see \url{https://youtu.be/AAd9UcP0nk0}}}

See the description of \emph{Prisoners Dilemma in the matrix: Arena}. The difference between the two variants is that \emph{Repeated} is a two-player game while \emph{Arena} is an eight-player game.

\noindent{\textbf{\small{Scenarios}}}
\begin{SC}
\item \emph{Partner may play either cooperate or defect.} The optimal strategy is simply to unconditionally defect. However, given that the focal doesn't know the strategy of the background player, a good strategy is more subtle. A reasonable strategy is to be a grim reciprocator cooperator, which would cooperate with the cooperator, and defect to the defector. Alternatively the focal player might try to ascertain whether the background player is exploitable. Doing so, however, carries a risk, for if the background player were to be a Grim reciprocator (like in other scenarios), this would cause them to defect for the rest of the episode.

\item \emph{Partner typically plays cooperate.} The optimal strategy is simply to unconditionally defect. The same considerations about uncertainty of the background player's strategy from Scenario 0 apply here.

\item \emph{Partner typically plays defect.} The optimal strategy is simply to unconditionally defect. However, because the focal player doesn't \emph{a priori} know the strategy of the background player, they must first try to find out their strategy. This can be done by looking at which resources they collect or by paying attention to the results of the first few interactions. Once the focal has identified its background partner is defecting then it may have confidence that it should defect as well. The focal player should also consider the possibility that the background bot is corrigible, i.e. that it could be persuade to switch from defection to cooperation. This is not the case here but the background populations used in SC 8 and SC are corrigible.

\item \emph{Partner is a hair-trigger grim reciprocator, i.e. one who initially cooperates but, if defected on once, will retaliate by defecting forever after.} The optimal strategy is simply to cooperate. Grim reciprocator background players are non-exploitable, and there is no way to know how they will react to a defection ahead of time. Because of this uncertainty, testing for exploitability can lead to poor performance of the focal player. Conditional cooperators who cooperate first but retaliate if defected on should achieve a high score.

\item \emph{Partner is a two-strikes grim reciprocator, i.e. one who initially cooperates, but if defected on twice, will retaliate by defecting forever after.} The optimal strategy is simply to cooperate. Grim reciprocator background players are non-exploitable, and there is no way to know how they will react to a defection ahead of time. Because of this uncertainty, testing for exploitability can lead to poor performance of the focal player. In principle, it would be possible to defect once against the background player leading to higher reward. But since it is not possible to know the background player is a two-strikes grim reciprocator, and testing it against a hair-trigger grim reciprocator leads to defection, in practice is better simply to cooperate. Conditional cooperators who cooperate first but retaliate if defected on should achieve a high score.

\item \emph{Partner is a tit-for-tat conditional cooperator.} The optimal strategy is simply to cooperate. Defecting against a tit-for-tat agent, even occasionally, might lead to miscoordinated interactions where one player cooperates and the other defects, in an alternating way. Forgiveness is one way to break out of such cycles of recrimination. Conditional cooperators who cooperate first but retaliate if defected on should also be forgiving to ensure they do well in this scenario.

\item \emph{Partner is a tit-for-tat conditional cooperator who occasionally plays defect instead of cooperate.} Like the previous scenario, except the tit-for-tat background player occasionally will defect instead of cooperate. This is known as trembling hand in game theory. A strict tit-for-tat focal player would occasionally fall into miscoordinated interactions with the background player resulting in alternating cooperation and defection. As in SC5, focal conditional cooperators must also be forgiving to ensure they do well in this scenario. Forgiveness is even more important here since the background player will defect relatively frequently itself but will still implement tit-for-tat retaliation when defected on itself.

\item \emph{Partner plays cooperate for a while then switches to defect.} Similar considerations to the previous scenarios. A good strategy is a grim reciprocator, or tit-for-tat for the focal player. Unconditional cooperation would be exploited by the background player.

\item \emph{Partner tries to take advantage of the focal player by playing defect, but if punished, partner then switches to tit-for-tat conditional cooperation.} Related to Scenario 2, the optimal strategy is for the focal player to persuade the background player to stop defecting by punishing it through defecting itself. Once persuaded, the background player implements a conditional cooperation (tit-for-tat) strategy. So it is safe to start cooperating with them once you have verified that they are themselves consistently  cooperating.

\item \emph{Partner tries to take advantage of the focal player by playing defect, but if punished, partner then switches to noisy tit-for-tat conditional cooperation.} Like the previous scenario, except the focal player must implement a more generous form of conditional cooperation after persuading the background player to switch from defection.
\end{SC}

\subsubsection[Pure Coordination in the matrix: Arena]{Pure Coordination in the matrix: Arena\footnote{For a video of \textit{Pure Coordination in the matrix: Arena}, see \url{https://youtu.be/LG_qvqujxPU}}}
Players---who in this case cannot be identified as individuals since they all look the same---can gather resources of three different colors. All eight individuals need to converge on collecting the same color resource to gain reward when they encounter each other and compare inventories. The matrix for the interaction is
\[ A_\text{row} = A_\text{col}^T  =
\begin{bmatrix}
1 & 0 & 0\\
0 & 1 & 0\\
0 & 0 & 1
\end{bmatrix}
\text{ .} \]

\noindent{\textbf{\small{Scenarios}}}
\begin{SC}
\item \emph{Focals are resident, a single visitor joins who may prefer any option; whichever option it prefers, it pursues it greedily.} Seven focal players are joined by one background player. The background player does not adapt its behavior very much to context, i.e.~it single-mindedly collects one of the three resource types. As this is a resident mode scenario, the real test is of how well the focal population members coordinate among themselves, and whether their equilibrium is disturbed by the presence of the visitor or not.
\item \emph{Focals are resident, three visitors join who always select the same option as their partner in the previous interaction and do so without being too greedy.} This scenario tests whether the focal population's ability to coordinate is robust to the presence of visitors who try to join into their equilibrium by always selecting the same resource as their partner picked on the previous round. Depending on the focal population's joint strategy, this may lead to scarcity issues with some resource types getting depleted leaving players unable to collect what they need to coordinate successfully.
\item \emph{Focals are resident, three visitors join who always select the same option as their partner in the previous interaction and are greedy in doing so.} This is a more difficult version of SC1.
\item \emph{Visiting a background population where all initially choose option a (without greed) and then switch to either b or c after some time.} One focal player joins seven background players in this visitor mode scenario. This scenario tests flexibility. The focal player needs to adapt accordingly after the background players switch their behavior.
\item \emph{Visiting a background population where all initially choose option b (without greed) and then switch to either a or c after some time.} One focal player joins seven background players in this visitor mode scenario. This scenario tests flexibility. The focal player needs to adapt accordingly after the background players switch their behavior.
\item \emph{Visiting a background population where all initially choose option c (without greed) and then switch to either a or b after some time.} One focal player joins seven background players in this visitor mode scenario. This scenario tests flexibility. The focal player needs to adapt accordingly after the background players switch their behavior.
\item \emph{Visiting a background population where all typically choose option a and do not collect it too greedily.} In this visitor mode scenario one focal player joins seven background players. The test here is whether the focal player can ascertain the prevailing convention and join in.
\item \emph{Visiting a background population where all typically choose option b and do not collect it too greedily.} In this visitor mode scenario one focal player joins seven background players. The test here is whether the focal player can ascertain the prevailing convention and join in.
\item \emph{Visiting a background population where all typically choose option c and do not collect it too greedily.} In this visitor mode scenario one focal player joins seven background players. The test here is whether the focal player can ascertain the prevailing convention and join in.
\end{SC}

\subsubsection[Pure Coordination in the matrix: Repeated]{Pure Coordination in the matrix: Repeated\footnote{For a video of \textit{Pure Coordination in the matrix: Repeated}, see \url{https://youtu.be/biyhB378q58}}}

See the description of \emph{Pure Coordination in the matrix: Arena}. The difference between the two variants is that \emph{Repeated} is a two-player game while \emph{Arena} is an eight-player game.

\noindent{\textbf{\small{Scenarios}}}
\begin{SC}
\item \emph{Meeting any pure strategy player.} Here the focal agent plays with a partner sampled from the background population that was trained to play a pure strategy: either to collect mainly resource a, resource b, or resource c. However, the specific background player is sampled at test time so it could of any of these types. All background bots commit strongly to their choice, aiming to collect at least five resources before interacting. To get a high score the focal agent should scout out which pure strategy its partner is playing and  then pick the same pure strategy, collecting as many resources as it can.
\item \emph{Meeting an agent who plays the best response to what the focal agent did in the last round.} This scenario ought to be easy since the background player tries to adapt itself to the focal player.
\item \emph{Versus mixture of opponents who often flip to other strategies after some number of interactions.} This scenario tests flexibility since the focal player needs to adapt to players who suddenly change their strategy after some number of interactions.
\item \emph{Meeting an agent who almost always chooses resource a.} An easy scenario where the task is just to determine the choice always picked by the background player and act accordingly.
\item \emph{Meeting an agent who almost always chooses resource b.} An easy scenario where the task is just to determine the choice always picked by the background player and act accordingly.
\item \emph{Meeting an agent who almost always chooses resource c.} An easy scenario where the task is just to determine the choice always picked by the background player and act accordingly.
\end{SC}

\subsubsection[Rationalizable Coordination in the matrix: Arena]{Rationalizable Coordination in the matrix: Arena\footnote{For a video of \textit{Rationalizable Coordination in the matrix: Arena}, see \url{https://youtu.be/IXakuZhvrxo}}}
Same as Pure Coordination in the matrix except that differently colored resources are intrinsically of different values, suggesting an optimal color to converge on. The matrix for the interaction is
\[ A_\text{row} = A_\text{col}^T =
\begin{bmatrix}
1 & 0 & 0\\
0 & 2 & 0\\
0 & 0 & 3
\end{bmatrix}
\text{ .} \]

\noindent{\textbf{\small{Scenarios}}}
\begin{SC}
\item \emph{Focals are resident, a single visitor joins who may prefer any option; whichever option it prefers, it pursues it greedily.} Seven focal players are joined by one background player. The background player does not adapt its behavior very much to context, i.e.~it single-mindedly collects one of the three resource types. As this is a resident mode scenario, the real test is of how well the focal population members coordinate among themselves, and whether their equilibrium is disturbed by the presence of the visitor or not. The visitor may not select the best resource. So this scenario is about testing whether the focal population can handle the introduction of an irrational individual.
\item \emph{Focals are resident, three visitors join who always select the same option as their partner in the previous interaction and do so without being too greedy.} This scenario tests whether the focal population's ability to coordinate is robust to the presence of visitors who try to join into their equilibrium by always selecting the same resource as their partner picked on the previous round. Depending on the focal population's joint strategy, this may lead to scarcity issues with some resource types getting depleted leaving players unable to collect what they need to coordinate successfully.
\item \emph{Focals are resident, three visitors join who always select the same option as their partner in the previous interaction and are greedy in doing so.} This is a more difficult version of SC1.
\item \emph{Visiting a background population where all initially choose option a (without greed) and then switch to either b or c after some time.} One focal player joins seven background players in this visitor mode scenario. This scenario tests flexibility. The focal player needs to adapt accordingly after the background players switch their behavior.
\item \emph{Visiting a background population where all initially choose option b (without greed) and then switch to either a or c after some time.} One focal player joins seven background players in this visitor mode scenario. This scenario tests flexibility. The focal player needs to adapt accordingly after the background players switch their behavior.
\item \emph{Visiting a background population where all initially choose option c (without greed) and then switch to either a or b after some time.} One focal player joins seven background players in this visitor mode scenario. This scenario tests flexibility. The focal player needs to adapt accordingly after the background players switch their behavior. In this case the background players are switching away from the rational choice of c to one of the two irrational choices.
\item \emph{Visiting a background population where all typically choose option a and do not collect it too greedily.} In this visitor mode scenario one focal player joins seven background players. The test here is whether the focal player can ascertain the prevailing convention and join in, even though the prevailing convention is deeply irrational.
\item \emph{Visiting a background population where all typically choose option b and do not collect it too greedily.} In this visitor mode scenario one focal player joins seven background players. The test here is whether the focal player can ascertain the prevailing convention and join in, even though the prevailing convention is  irrational.
\item \emph{Visiting a background population where all typically choose option c and do not collect it too greedily.} In this visitor mode scenario one focal player joins seven background players. The test here is whether the focal player can ascertain the prevailing convention and join in, luckily the convention is the rational one, so this ought to be an easy test.
\end{SC}

\subsubsection[Rationalizable Coordination in the matrix: Repeated]{Rationalizable Coordination in the matrix: Repeated\footnote{For a video of \textit{Rationalizable Coordination in the matrix: Repeated}, see \url{https://youtu.be/3brwR7DtxEI}}}

See the description of \emph{Rationalizable Coordination in the matrix: Arena}. The difference between the two variants is that \emph{Repeated} is a two-player game while \emph{Arena} is an eight-player game.

\noindent{\textbf{\small{Scenarios}}}
\begin{SC}
\item \emph{Meeting any pure strategy player.} Here the focal agent plays with a partner sampled from the background population that was trained to play a pure strategy: either to collect mainly resource a, resource b, or resource c. However, the specific background player is sampled at test time so it could of any of these types. All background bots commit strongly to their choice, aiming to collect at least five resources before interacting. To get a high score the focal agent should scout out which pure strategy its partner is playing and then pick the same pure strategy, collecting as many resources as it can. This may be difficult since the background player sometimes picks irrational choices, nevertheless the focal player must adapt to them anyway.
\item \emph{Meeting an agent who plays the best response to what the focal agent did in the last round.} This scenario ought to be easy since the background player tries to adapt itself to the focal player.
\item \emph{Versus mixture of opponents who often flip to other strategies after some number of interactions.} This scenario tests flexibility since the focal player needs to adapt to players who suddenly change their strategy after some number of interactions. The background player may make irrational choices but the focal player still must adapt to them.
\item \emph{Meeting an agent who almost always chooses resource a.} This scenario tests the focal player's ability to acquiesce to a deeply irrational convention forced by their partner.
\item \emph{Meeting an agent who almost always chooses resource b.} This scenario tests the focal player's ability to acquiesce to an irrational convention forced by their partner.
\item \emph{Meeting an agent who almost always chooses resource c.} This scenario tests the focal player's ability implement the rational convention. It should be easy.
\end{SC}

\subsubsection[Running With Scissors in the matrix: Arena]{Running With Scissors in the matrix: Arena\footnote{For a video of \textit{Running With Scissors in the matrix: Arena}, see \url{https://youtu.be/6BL6JIbS2cE}}}
Same dynamics as \textit{Running with Scissors in the matrix: Repeated} but with eight players. Each pairwise conflict is zero sum so the overall setting is one of competition. As with all zero sum games having more than two players, alliances are possible in principle. But in this case the environment is not very conducive to forming them since rewards are always and only gained in pairwise conflicts and there does not appear to be any way to collude with others.  The matrix for the interaction is:
\[ A_\text{row} = A_\text{col}^T =
\begin{bmatrix}
\hphantom{+}0 & -10 & +10\\
+10 & \hphantom{+}0 & -10\\
-10 & +10 & \hphantom{+}0
\end{bmatrix}
\text{ .} \]

\noindent{\textbf{\small{Scenarios}}}
\begin{SC}
\item \emph{Versus a background population containing bots implementing all three pure strategies.} Here one focal player joins seven from the background population. The background population contains bots who implement all three pure strategies: rock, paper, and scissors. They may either commit to their strategy moderately (aiming to collect three resources before interacting) or more strongly (aiming to collect five). The task for the focal agent is to watch its opponents, see what strategy one of them is implementing, and act accordingly.
\item \emph{Versus gullible bots.} Here one focal player joins seven from the background population. The background population consists entirely of weak bots who were trained to best respond to agents playing pure strategies. They are weak opponents.
\item \emph{Versus mixture of opponents who play rock and some who flip to scissors after two interactions.} Here one focal player joins seven from the background population. The focal player should pay attention to what each prospective partner has collected since $2/3$ of them play rock while $1/3$ play scissors after the first two interactions. Choosing paper to best respond to rock is a bad choice if accidentally paired with an opponent playing scissors.\label{running_with_scissors_itm__arena__sc:flip_r2s}
\item \emph{Versus mixture of opponents who play paper and some who flip to rock after two interactions.} Like \ref{running_with_scissors_itm__arena__sc:flip_r2s} but with bots playing paper and bots switching from paper to rock.
\item \emph{Versus mixture of opponents who play scissors and some who flip to paper after two interactions.} Like \ref{running_with_scissors_itm__arena__sc:flip_r2s} but with bots playing scissors and bots switching from scissors to paper.
\item \emph{Visiting a population of pure paper bots.} Here one focal player joins seven from the background population. All seven background bots play paper so the focal player can get a high score by playing scissors.
\item \emph{Visiting a population of pure rock bots.} Here one focal player joins seven from the background population. All seven background bots play rock so the focal player can get a high score by playing paper.
\item \emph{Visiting a population of pure scissors bots.} Here one focal player joins seven from the background population. All seven background bots play scissors so the focal player can get a high score by playing rock.
\end{SC}

\subsubsection[Running With Scissors in the matrix: One Shot]{Running With Scissors in the matrix: One Shot\footnote{For a video of \textit{Running With Scissors in the matrix: One Shot}, see \url{https://youtu.be/gtemAx4XEcQ}}}
This environment first appeared in \cite{vezhnevets2020options}. Two individuals gather rock, paper or scissor resources in the environment and can challenge others to a 'rock, paper scissor' game, the outcome of which depends on the resources they collected.  It is possible---though not trivial---to observe the policy that one's partner is starting to implement and take countermeasures. This induces a wealth of possible feinting strategies.  The observation window is $5 \times 5$.

\noindent{\textbf{\small{Scenarios}}}
\begin{SC}
\item \emph{Versus mixed strategy opponent.} Here the focal agent must defeat an opponent that was trained to play a pure strategy: either rock, paper, or scissors. However, the specific opponent is sampled at test time so it could be any of those. All opponents commit strongly to their choice, aiming to collect at least three resources before interacting. To defeat them, the focal agent should scout out which pure strategy its opponent is playing and then collect the resources to implement its counter strategy. Since this is a one-shot interaction, success requires the focal agent to pay close attention to which resources are missing since they provide a clue to which strategy their opponent is implementing.
\item \emph{Versus pure rock opponent.} Opponent always plays rock, and commits to it strongly, aiming to collect five resources before interacting. The focal player gets a high score when it picks paper and commits strongly to that choice.\label{running_with_scissors_itm__one_shot__sc:pure_rock}
\item \emph{Versus pure paper opponent.} Same as \ref{running_with_scissors_itm__one_shot__sc:pure_rock} but opponent plays paper so focal player should play scissors. 
\item \emph{Versus pure scissors opponent.} Same as \ref{running_with_scissors_itm__one_shot__sc:pure_rock} but opponent plays scissors so focal player should play rock. 
\end{SC}

\subsubsection[Running With Scissors in the matrix: Repeated]{Running With Scissors in the matrix: Repeated\footnote{For a video of \textit{Running With Scissors in the matrix: Repeated}, see \url{https://youtu.be/rZH9nUKefcU}}}

See the description of \emph{Running With Scissors in the matrix: Arena}. The difference between the two variants is that \emph{Repeated} is a two-player game while \emph{Arena} is an eight-player game.

\noindent{\textbf{\small{Scenarios}}}
\begin{SC}
\item \emph{Versus mixed strategy opponent.} Here the focal agent must defeat an opponent that was trained to play a pure strategy: either rock, paper, or scissors. However, the specific opponent is sampled at test time so it could be any of those. All opponents commit strongly to their choice, aiming to collect at least five resources before interacting. To defeat them, the focal agent should scout out which pure strategy its opponent is playing and then collect the resources to implement its counter strategy. \label{running_with_scissors_itm__repeated__sc:mixed}
\item \emph{Versus opponent who plays the best response to what the focal player did in the last round.} Here the focal agent must defeat an opponent who may change their strategy with each interaction. The opponent will always select the best response to what the focal player selected in the previous interaction. For instance, if the focal player plays rock in one interaction then its opponent will play paper in the next interaction. On the first interaction of each episode it chooses one of the three pure strategies at random. The opponent always commits strongly to its choice, aiming to collect at least five resources before interacting. To win the focal player can trick its opponent into choosing a specific strategy and countering it. This requires changing strategy from interaction to interaction, cycling around the three options.\label{running_with_scissors_itm__repeated__sc:best_response}
\item \emph{Versus opponent who sometimes plays a pure strategy but sometimes plays the best response to what the focal player did in the last round.} Focal player must defeat an opponent sampled from the union of the background populations used in \ref{running_with_scissors_itm__repeated__sc:mixed} and \ref{running_with_scissors_itm__repeated__sc:best_response}. The probability of sampling a pure opponent is $3/4$ while the probability of sampling a best response opponent is $1/4$.
\item \emph{Versus mixture of opponents who often flip to other strategies after two interactions.} Focal player must defeat an opponent that may initially play any pure strategy and, with probability $1/3$, may flip after the second interaction to the best response to the best response to its initial strategy. For example if it starts out playing rock then the best response to that would be paper, so after the second interaction it would switch to playing the best response to paper i.e. scissors. It only weakly commits to its strategy for the first two interactions. That is, it aims to collect only one resource before interacting. After two interactions, at the point when it changes strategy, it also starts committing more strongly to its choice, aiming to collect five resources before each interaction. With probability $2/3$, the opponent instead plays a pure strategy throughout the entire episode. In half of the pure opponent episodes the bot fully commits to its pure strategy, aiming to collect five resources before interacting, while the other half of the time it commits less strongly, aiming to collect only one resource before interacting. Note that opponents may be weakly committed to their strategy for the first two interactions regardless of whether they will ultimately flip strategy or not so it's not possible for the focal agent to observe weak commitment early on as a cue to predict whether or not their opponent will later flip strategies.
\item \emph{Versus mixture of opponents who either flip to another strategy after one interaction and keep it forever or continue to change, always best responding to what the focal player just did.} Two kinds of opponents are possible. Both change their strategy after the first interaction. With probability $3/4$ the opponent will be a bot that flips to a different strategy after the first interaction and then follows it till the end of the episode. It always flips to the best response to the best response to its initial strategy (so if it initially plays rock then it will flip to scissors). With probability $1/4$ the other kind of opponent is sampled. This opponent is identical to the one in \ref{running_with_scissors_itm__repeated__sc:best_response}. Both kinds of opponents always fully commit to their choice, aiming to collect at least five resources before interacting so its not possible to observe the opponent's commitment level to predict which kind they are. To win the focal player must figure out which kind of opponent it is playing against and either best respond by selecting the same choice in all interactions after the first if paired with the first kind of opponent, or apply the cyclic strategy described as the solution to \ref{running_with_scissors_itm__repeated__sc:best_response} if paired with the second kind of opponent.
\item \emph{Versus gullible opponent.} Here the focal agent must defeat an opposing agent that was trained to best respond to agents playing pure strategies. The opponent should attempt to scout out what strategy the focal agent is playing so it can pick the appropriate counter. To defeat it, the focal agent should feint toward one resource and then collect the counter to its counter. So for example, if the focal agent successfully feinted that it would pick rock, inducing its opponent to pick paper, the focal agent should then collect and play scissors. This opponent is fairly weak.
\item \emph{Versus pure rock opponent.} Opponent always plays rock, and commits to it strongly, aiming to collect five resources before interacting. The focal player gets a high score when it picks paper and commits strongly to that choice.\label{running_with_scissors_itm__repeated__sc:pure_rock}
\item \emph{Versus pure paper opponent.} Same as \ref{running_with_scissors_itm__repeated__sc:pure_rock} but opponent plays paper so focal player should play scissors. 
\item \emph{Versus pure scissors opponent.} Same as \ref{running_with_scissors_itm__repeated__sc:pure_rock} but opponent plays scissors so focal player should play rock. 
\end{SC}

\subsubsection[Stag Hunt in the matrix: Arena]{Stag Hunt in the matrix: Arena\footnote{For a video of \textit{Stag Hunt in the matrix: Arena}, see \url{https://youtu.be/agOpo0MZmzs}}}
Individuals collect resources that represent `hare' (red) or `stag' (green) and compare inventories in an encounter. Consequences of this inventory comparison are congruent with the classic Stag Hunt matrix game. This game exposes a tension between reward for the group and risk for the individual. The matrix for the interaction is
\[ A_\text{row} = A_\text{col}^T =
\begin{bmatrix}
4 & 0\\
2 & 2
\end{bmatrix}
\text{ .} \]

\noindent{\textbf{\small{Scenarios}}}
\begin{SC}
\item \emph{Visiting unconditional stag players.} This is a visitor mode scenario. One focal player joins seven background players. In this case the background population implements the stag policy, it's a good idea for the focal player to join them and play stag.
\item \emph{Visiting unconditional hare players.} This is a visitor mode scenario. One focal player joins seven background players. In this case the background population implements the hare policy. Even though the focal player would prefer to play stag, they can't force the others to help them. After determining the others won't change their ways then they should give in and play hare.
\item \emph{Focals are resident and visitors are unconditional stag players.} This is a resident mode scenario where five focals are joined by three background players. The test is mostly about the focals implementing a cooperative joint policy among themselves. They also must be robust to the presence of the three visitors.
\item \emph{Focals are resident and visitors are unconditional hare players.} This is a resident mode scenario where five focals are joined by three background players. The test is mostly about the focals implementing a cooperative joint policy among themselves. They also must be robust to the presence of the three visitors.
\item \emph{Visiting a population of hair-trigger grim reciprocator bots who initially play stag but, if any partner plays hare once, they give up on trying to cooperate and play hare in all future interactions.} This is a visitor mode scenario, the focal player joins seven background players. The focal player would do well by playing stag. 
\item \emph{Visiting a population of two-strikes grim reciprocator bots who initially play stag but, if their partners play hare twice, they give up on trying to cooperate and play hare in all future interactions.} This is like SC4, but this version is a bit more forgiving.
\item \emph{Visiting a mixed population of k-strikes grim reciprocator bots with k values from 1 to 3, they initially play stag but, if their partners play hare k times, they then play hare in all future interactions.} This is a visitor mode scenario where three focal players meet five background players. The focal players can achieve a high score by sticking with playing stag.
\item \emph{Visiting a mixture of pure hare and pure stag players.} This is a visitor mode scenario where three focal players meet five background players. In this case it is useful to scout out specific partners and try to interact preferentially with those who play stag.
\end{SC}

\subsubsection[Stag Hunt in the matrix: Repeated]{Stag Hunt in the matrix: Repeated\footnote{For a video of \textit{Stag Hunt in the matrix: Repeated}, see \url{https://youtu.be/aDp_CArcb1Y}}}

See the description of \emph{Stag Hunt in the matrix: Arena}. The difference between the two variants is that \emph{Repeated} is a two-player game while \emph{Arena} is an eight-player game.

\noindent{\textbf{\small{Scenarios}}}
\begin{SC}
\item \emph{Partner may play either stag or hare.} Here the focal agent plays with a partner sampled from the background population that was trained to play a pure strategy: either to collect mainly stag or mainly hare. However, the specific background player is sampled at test time so it could of either type. All background bots commit strongly to their choice, aiming to collect at least five (or seven) resources before interacting. To get a high score the focal agent should scout out which pure strategy the background bot is playing and then play stag if they play stag and otherwise play hare.
\item \emph{Partner typically plays stag.} Here the background player always picks stag the best course of action is to join them.
\item \emph{Partner typically plays hare.} Here the background player always picks hare the best course of action is also to player hare.
\item \emph{Partner is a hair-trigger grim reciprocator, i.e. one who initially cooperates but, if defected on once, will retaliate by defecting forever after.} The best choice for the focal player is to play stag.
\item \emph{Partner is a two-strikes grim reciprocator, i.e. one who initially cooperates, but if defected on twice, will retaliate by defecting forever after.} This scenario is similar to SC3 but somewhat more forgiving.
\item \emph{Partner is a tit-for-tat conditional cooperator.} The focal player should play stag here.
\item \emph{Partner is a tit-for-tat conditional cooperator who occasionally plays hare instead of stag.} The focal player should play stag here. If employing a conditional cooperation strategy then it's important not to get stuck in a cycle of recrimination as a result of the background player occasionally playing hare. Forgiveness is helpful. 
\item \emph{Partner plays stag for a while then switches to hare.} This scenario tests flexibility.
\item \emph{Partner initially plays hare, but if punished, partner then switches to tit-for-tat conditional cooperation.} In this scenario the background player can be encouraged to switch strategies by punishing it.
\item \emph{Partner initially plays hare, but if punished, partner then switches to noisy tit-for-tat conditional cooperation.} This scenario is like SC6 but the background player occasionally plays hare so forgiveness is important.
\end{SC}


\begin{figure*}[p] 
\includegraphics[height=13mm]{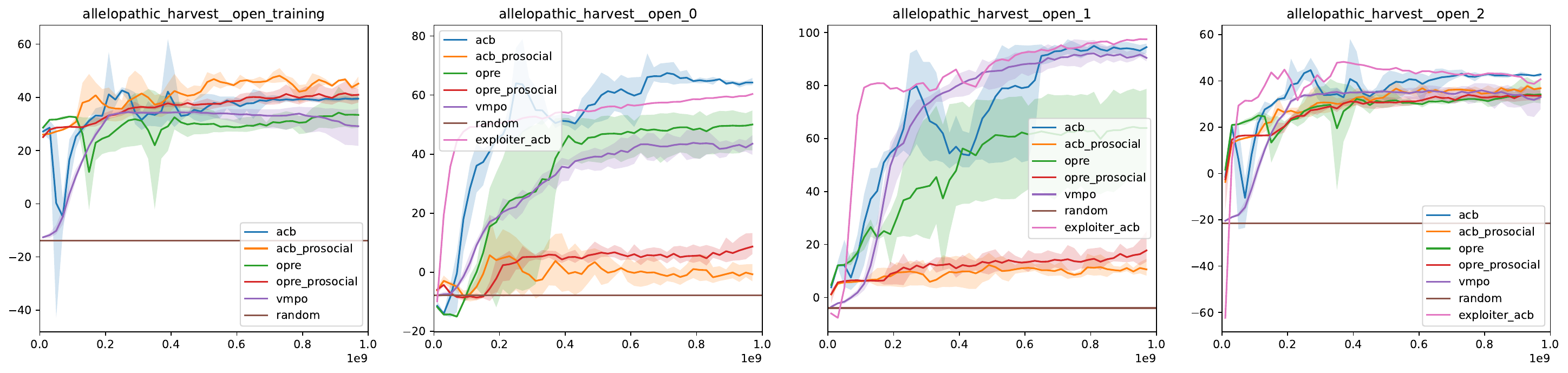}
\hspace{0mm}
\caption{Allelopathic Harvest: Open}\label{fig:allelopathic_harvest__open}
\end{figure*}
\begin{figure*}[p] 
\includegraphics[height=13mm]{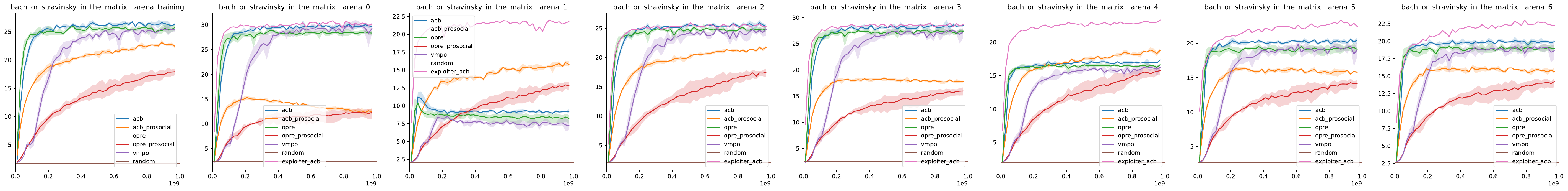}
\hspace{0mm}
\caption{Bach or Stravinsky in the matrix: Arena}\label{fig:bach_or_stravinsky_itm__arena}
\end{figure*}
\begin{figure*}[p] 
\includegraphics[height=13mm]{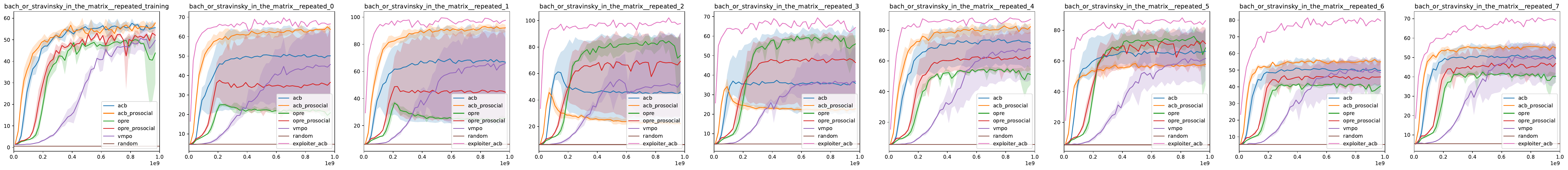}
\hspace{0mm}
\caption{Bach or Stravinsky in the matrix: Repeated}\label{fig:bach_or_stravinsky_itm__repeated}
\end{figure*}
\begin{figure*}[p] 
\includegraphics[height=13mm]{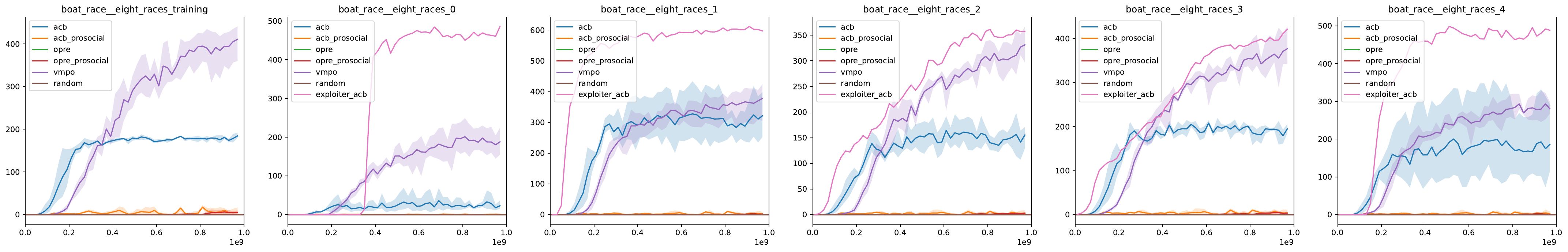}
\hspace{0mm}
\caption{Boat Race: Eight Races}\label{fig:boat_race__eight_races}
\end{figure*}
\begin{figure*}[p] 
\includegraphics[height=13mm]{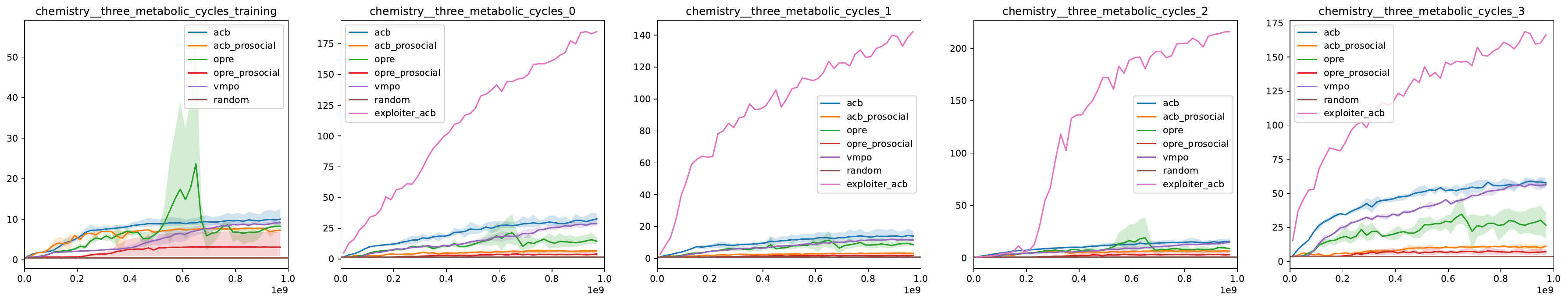}
\hspace{0mm}
\caption{Chemistry: Three Metabolic Cycles}\label{fig:chemistry__three_metabolic_cycles}
\end{figure*}
\begin{figure*}[p] 
\includegraphics[height=13mm]{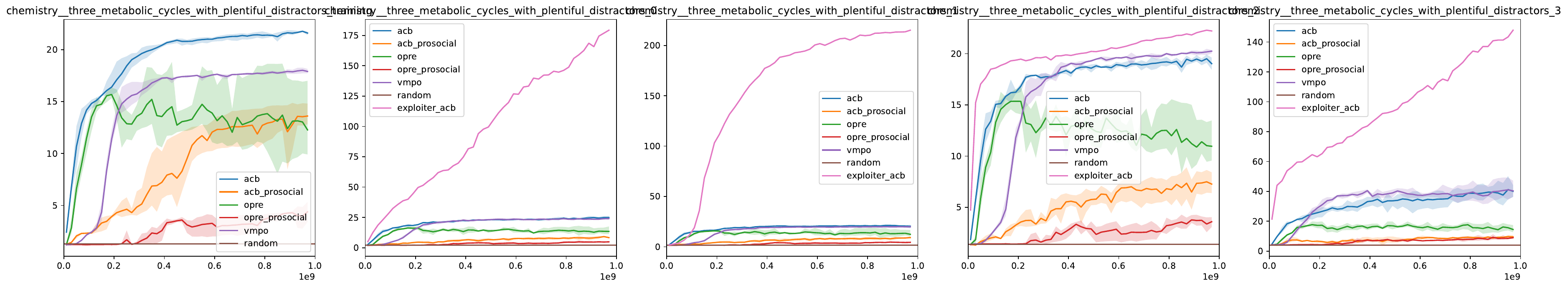}
\hspace{0mm}
\caption{Chemistry: Three Metabolic Cycles with Plentiful Distractors}\label{fig:chemistry__three_metabolic_cycles_with_plentiful_distractors}
\end{figure*}
\begin{figure*}[p] 
\includegraphics[height=13mm]{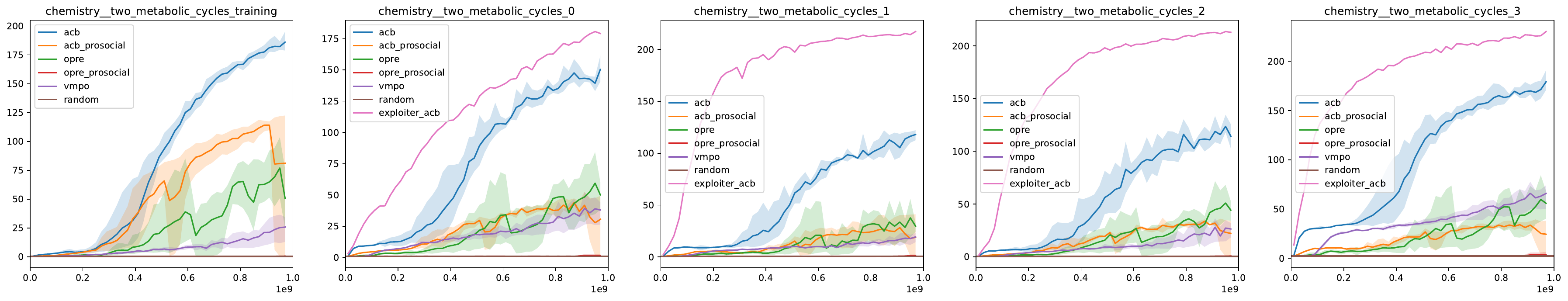}
\hspace{0mm}
\caption{Chemistry: Two Metabolic Cycles}\label{fig:chemistry__two_metabolic_cycles}
\end{figure*}
\begin{figure*}[p] 
\includegraphics[height=13mm]{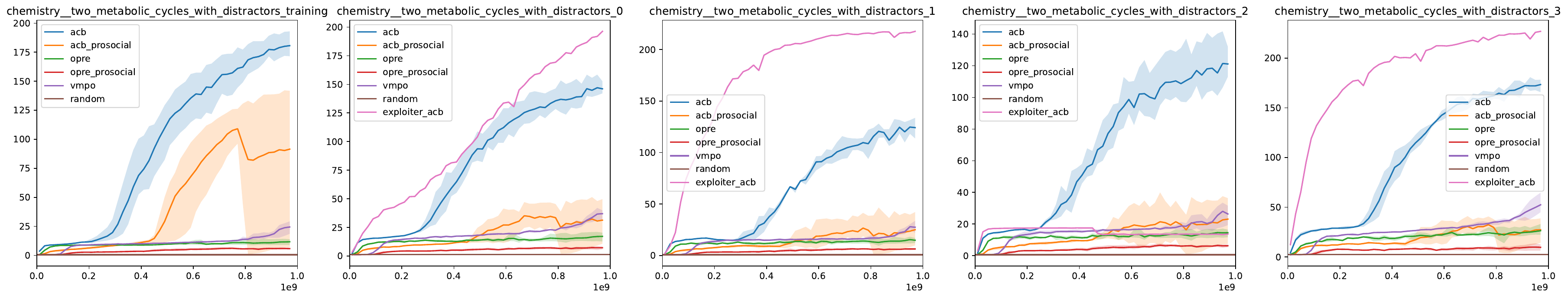}
\hspace{0mm}
\caption{Chemistry: Two Metabolic Cycles with Distractors}\label{fig:chemistry__two_metabolic_cycles_with_distractors}
\end{figure*}
\begin{figure*}[p] 
\includegraphics[height=13mm]{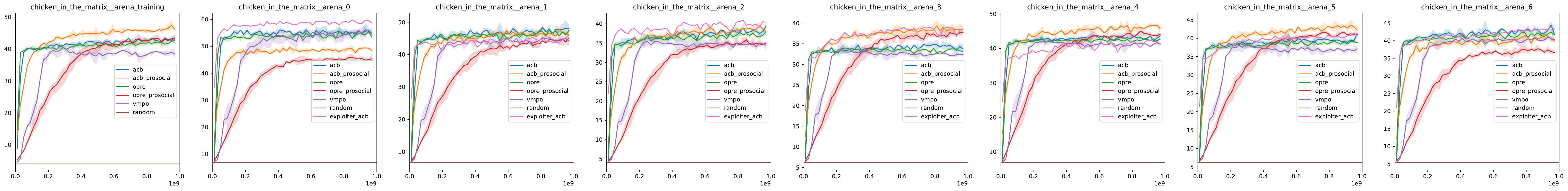}
\hspace{0mm}
\caption{Chicken in the matrix: Arena}\label{fig:chicken_in_the_matrix__arena}
\end{figure*}
\begin{figure*}[p] 
\includegraphics[height=13mm]{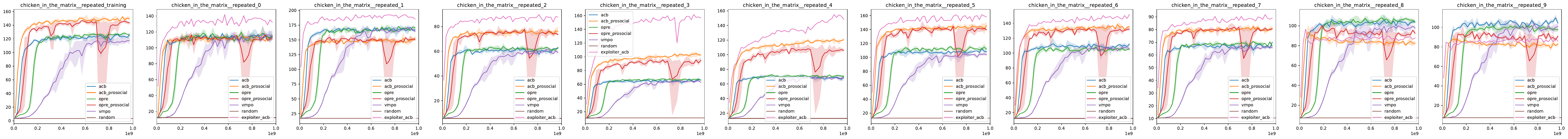}
\hspace{0mm}
\caption{Chicken in the matrix: Repeated}\label{fig:chicken_in_the_matrix__repeated}
\end{figure*}
\begin{figure*}[p] 
\includegraphics[height=13mm]{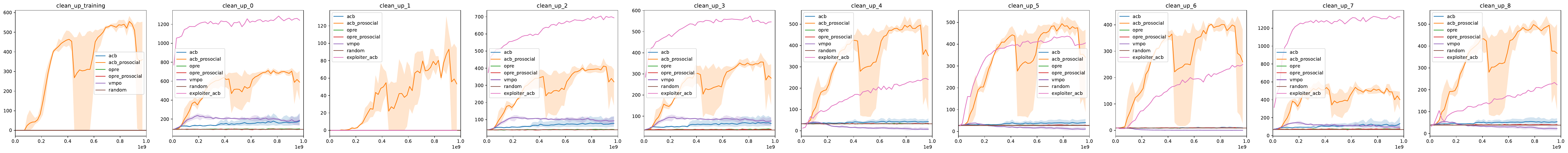}
\hspace{0mm}
\caption{Clean Up}\label{fig:clean_up}
\end{figure*}
\begin{figure*}[p] 
\includegraphics[height=13mm]{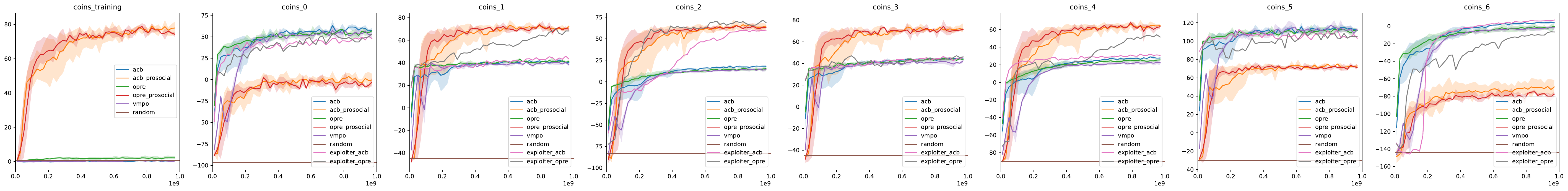}
\hspace{0mm}
\caption{Coins}\label{fig:coins}
\end{figure*}
\begin{figure*}[p] 
\includegraphics[height=13mm]{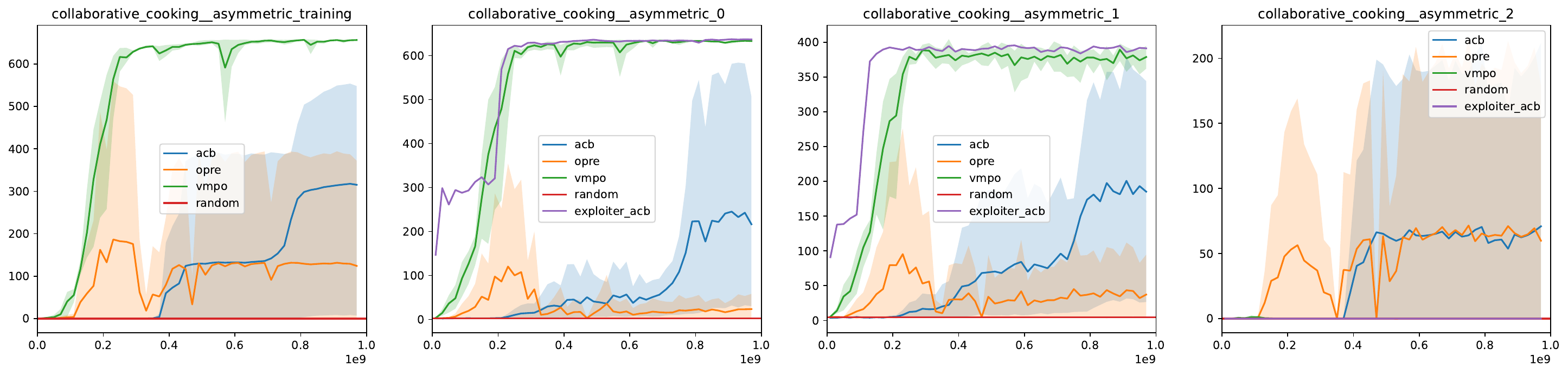}
\hspace{0mm}
\caption{Collaborative Cooking: Asymmetric}\label{fig:collaborative_cooking__asymmetric}
\end{figure*}
\begin{figure*}[p] 
\includegraphics[height=13mm]{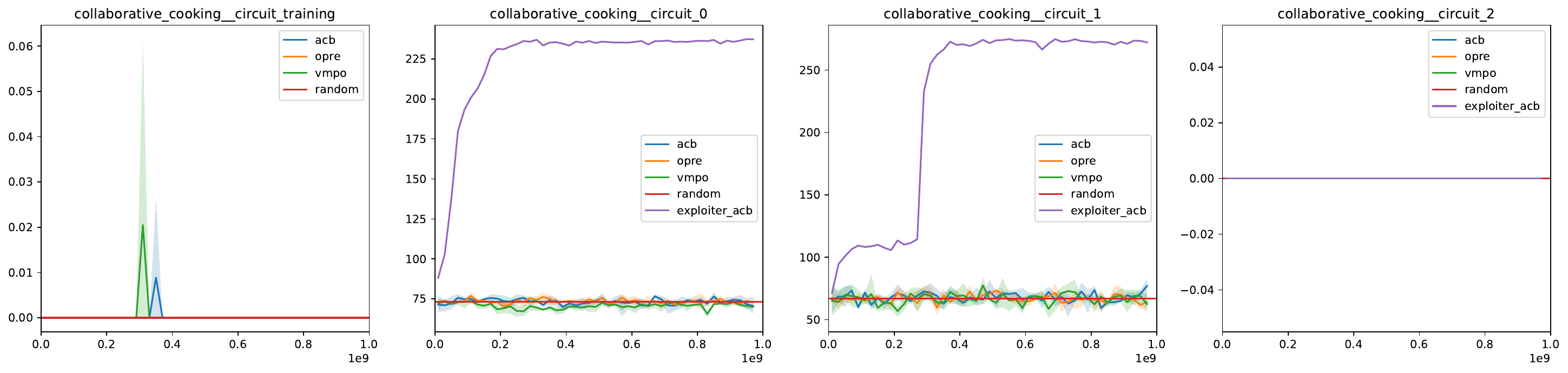}
\hspace{0mm}
\caption{Collaborative Cooking: Circuit}\label{fig:collaborative_cooking__circuit}
\end{figure*}
\begin{figure*}[p] 
\includegraphics[height=13mm]{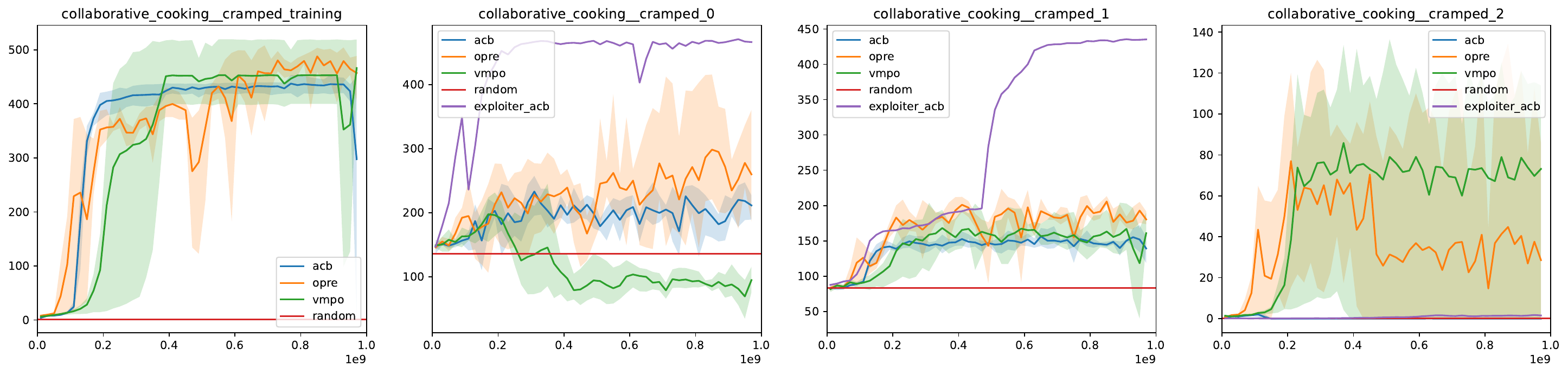}
\hspace{0mm}
\caption{Collaborative Cooking: Cramped}\label{fig:collaborative_cooking__cramped}
\end{figure*}
\begin{figure*}[p] 
\includegraphics[height=13mm]{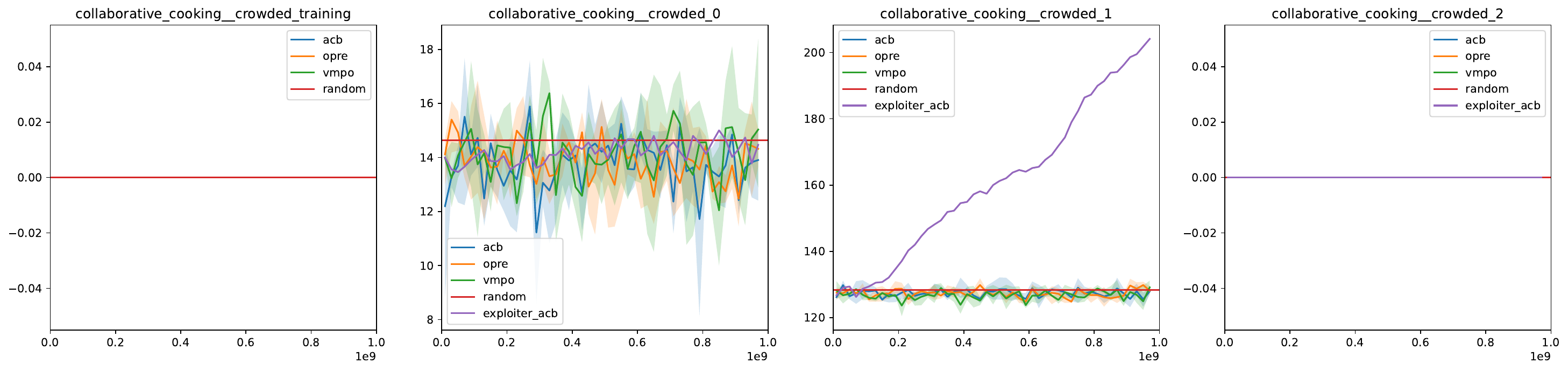}
\hspace{0mm}
\caption{Collaborative Cooking: Crowded}\label{fig:collaborative_cooking__crowded}
\end{figure*}
\begin{figure*}[p] 
\includegraphics[height=13mm]{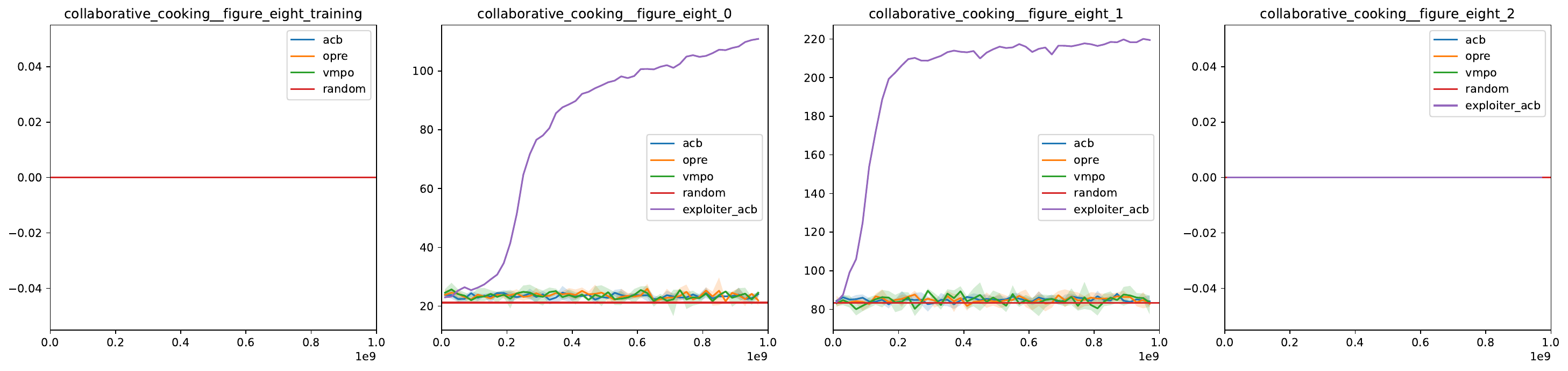}
\hspace{0mm}
\caption{Collaborative Cooking: Figure Eight}\label{fig:collaborative_cooking__figure_eight}
\end{figure*}
\begin{figure*}[p] 
\includegraphics[height=13mm]{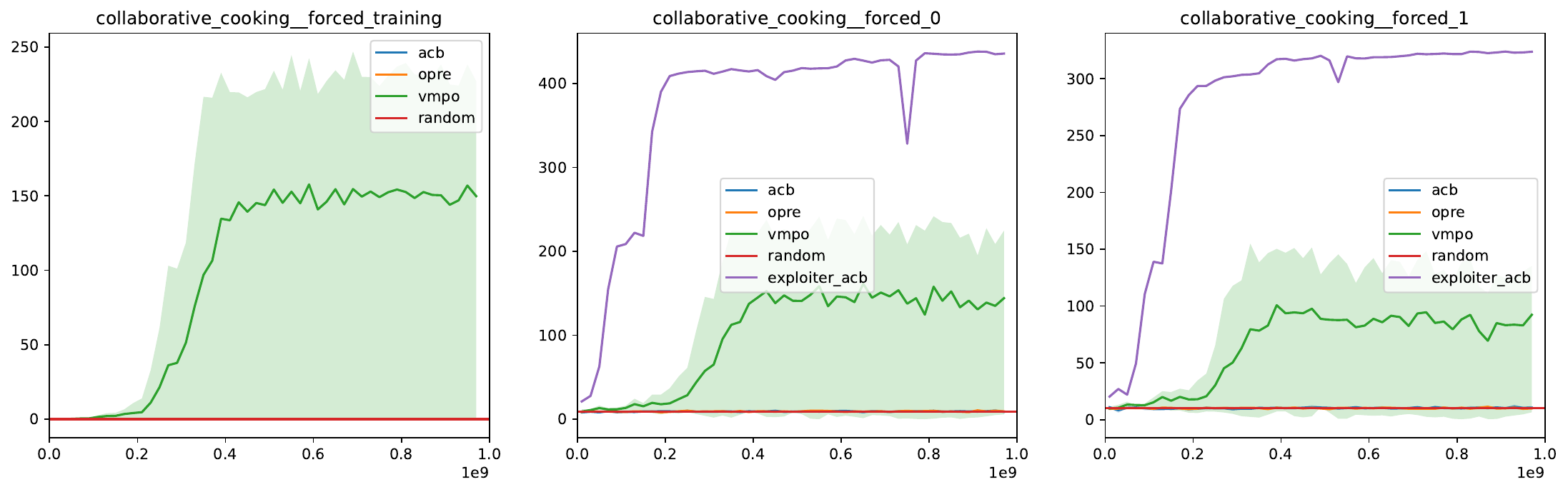}
\hspace{0mm}
\caption{Collaborative Cooking: Forced}\label{fig:collaborative_cooking__forced}
\end{figure*}
\begin{figure*}[p] 
\includegraphics[height=13mm]{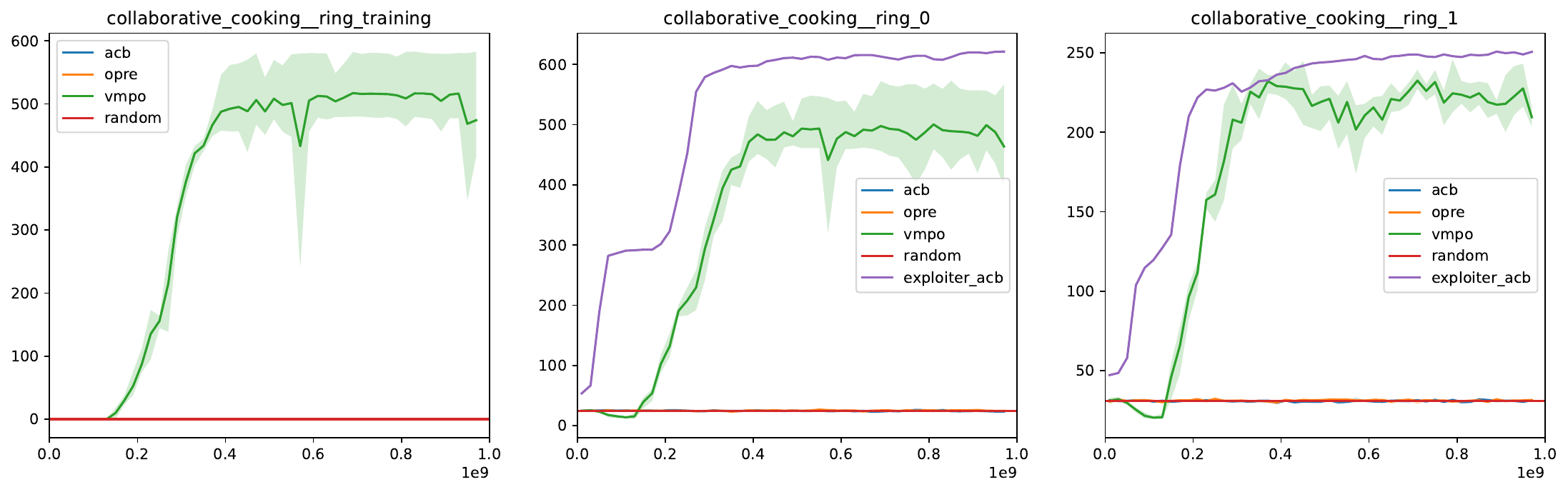}
\hspace{0mm}
\caption{Collaborative Cooking: Ring}\label{fig:collaborative_cooking__ring}
\end{figure*}
\begin{figure*}[p] 
\includegraphics[height=13mm]{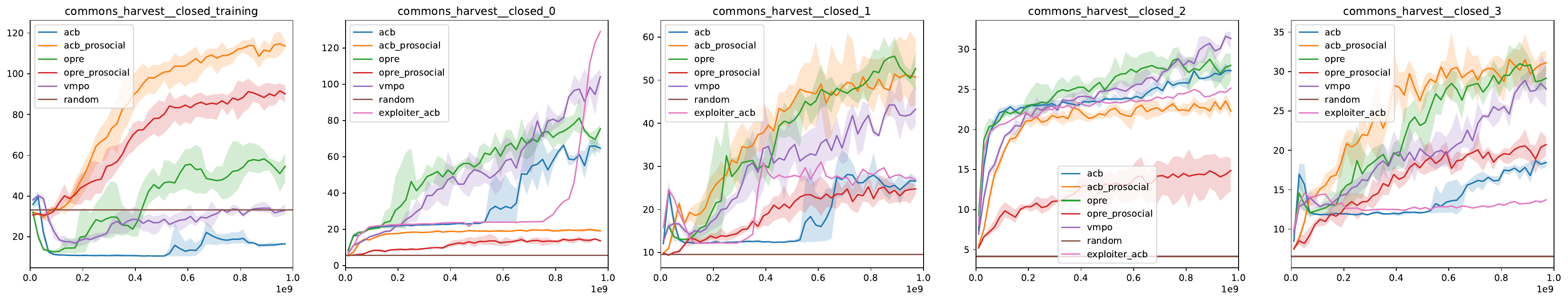}
\hspace{0mm}
\caption{Commons Harvest: Closed}\label{fig:commons_harvest__closed}
\end{figure*}
\begin{figure*}[p] 
\includegraphics[height=13mm]{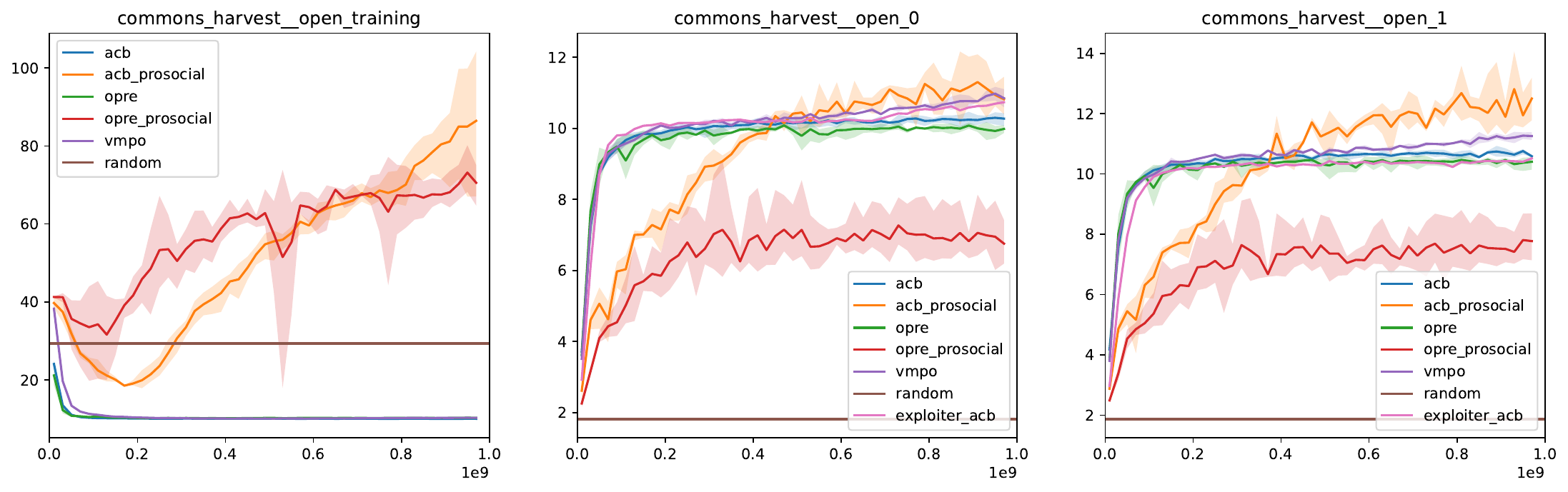}
\hspace{0mm}
\caption{Commons Harvest: Open}\label{fig:commons_harvest__open}
\end{figure*}
\begin{figure*}[p] 
\includegraphics[height=13mm]{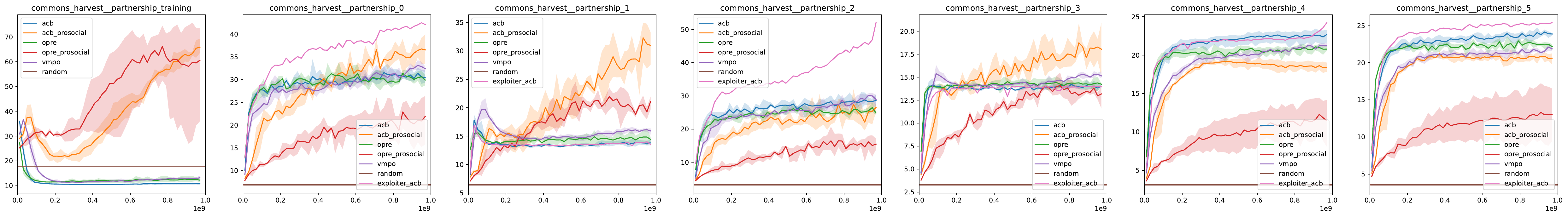}
\hspace{0mm}
\caption{Commons Harvest: Partnership}\label{fig:commons_harvest__partnership}
\end{figure*}
\begin{figure*}[p] 
\includegraphics[height=13mm]{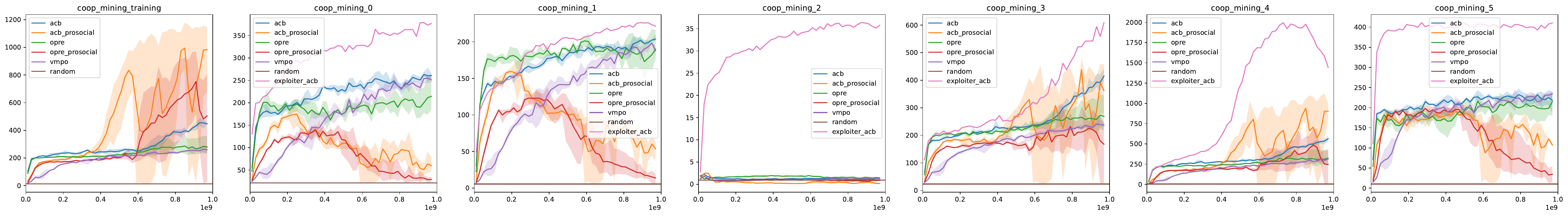}
\hspace{0mm}
\caption{Coop Mining}\label{fig:coop_mining}
\end{figure*}
\begin{figure*}[p] 
\includegraphics[height=13mm]{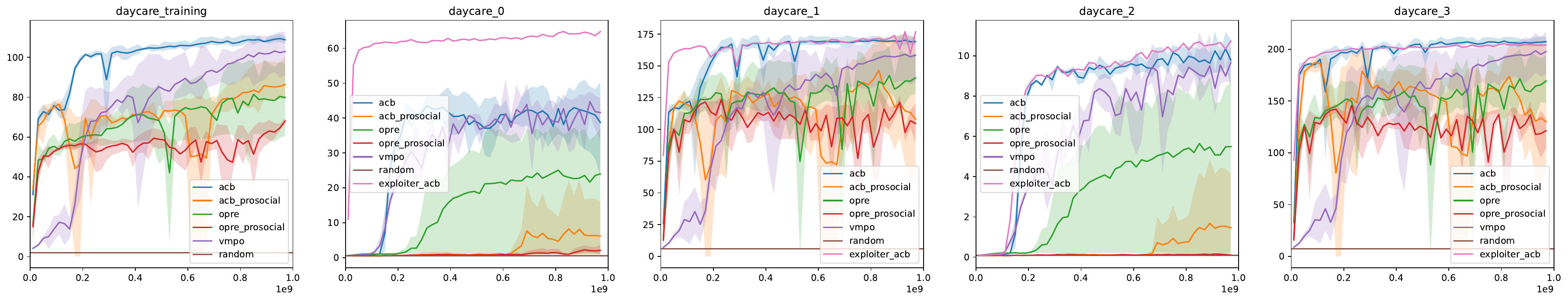}
\hspace{0mm}
\caption{Daycare}\label{fig:daycare}
\end{figure*}
\begin{figure*}[p] 
\includegraphics[height=13mm]{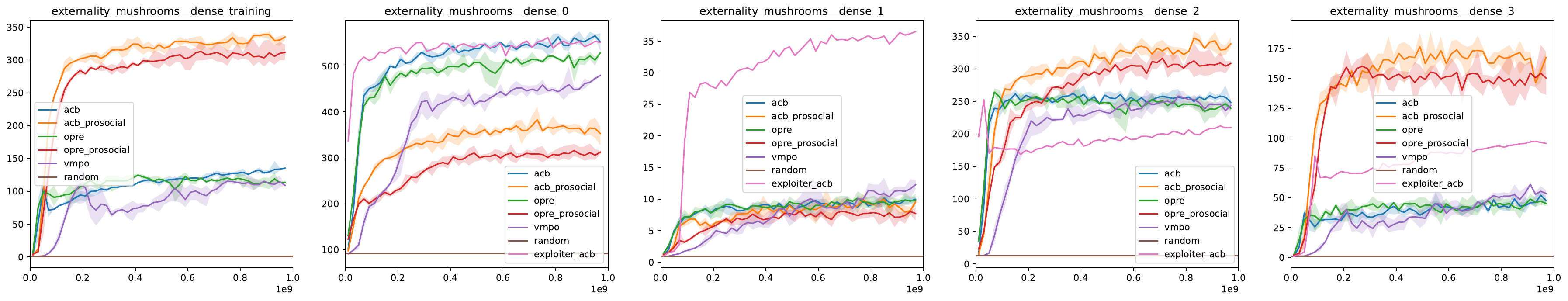}
\hspace{0mm}
\caption{Externality Mushrooms: Dense}\label{fig:externality_mushrooms__dense}
\end{figure*}
\begin{figure*}[p] 
\includegraphics[height=13mm]{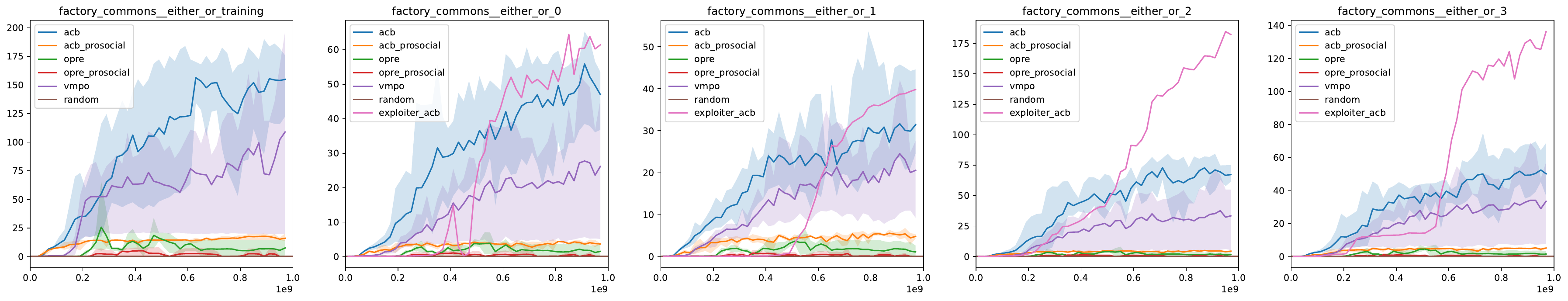}
\hspace{0mm}
\caption{Factory Commons: Either Or}\label{fig:factory_commons__either_or}
\end{figure*}
\begin{figure*}[p] 
\includegraphics[height=13mm]{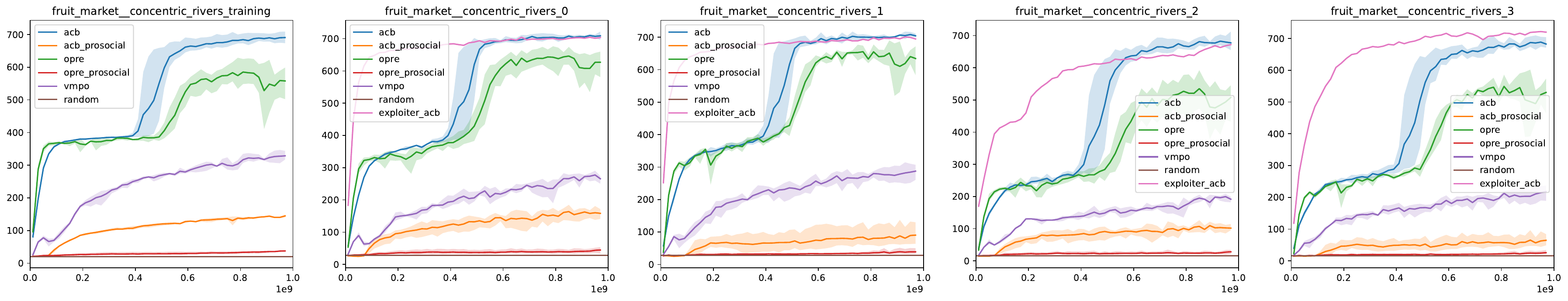}
\hspace{0mm}
\caption{Fruit Market: Concentric Rivers}\label{fig:fruit_market__concentric_rivers}
\end{figure*}
\begin{figure*}[p] 
\includegraphics[height=13mm]{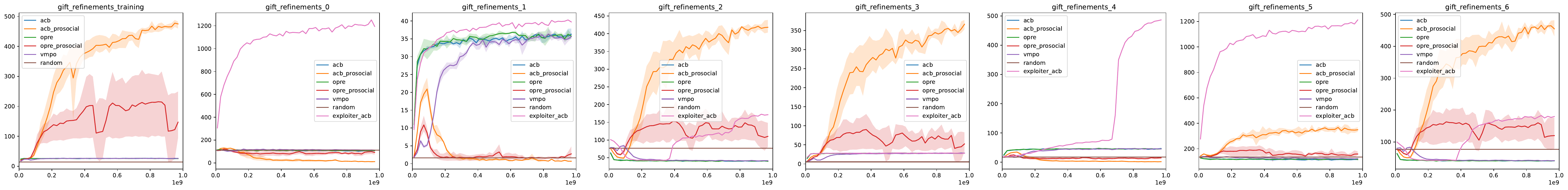}
\hspace{0mm}
\caption{Gift Refinements}\label{fig:gift_refinements}
\end{figure*}
\begin{figure*}[p] 
\includegraphics[height=13mm]{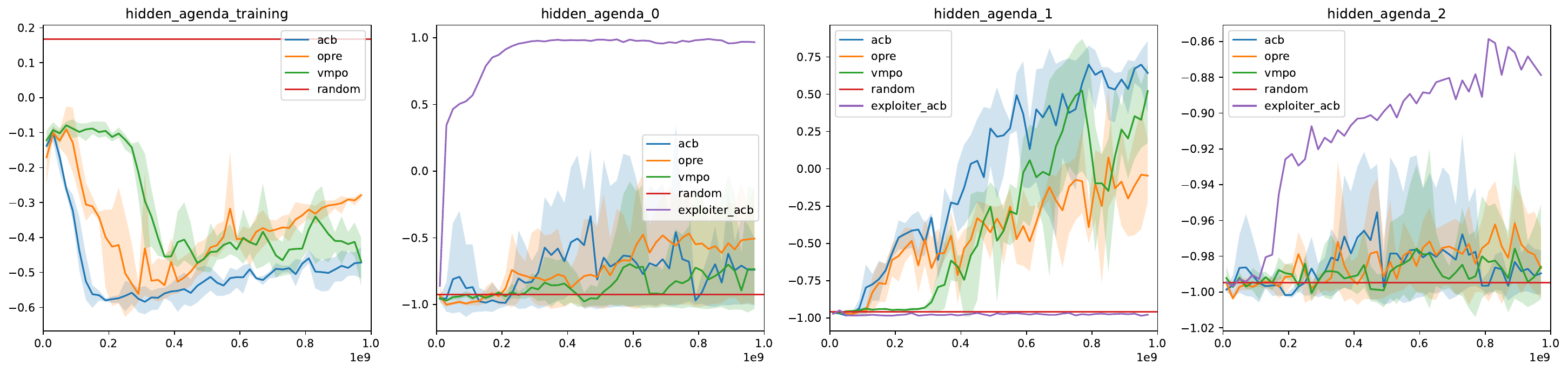}
\hspace{0mm}
\caption{Hidden Agenda}\label{fig:hidden_agenda}
\end{figure*}
\begin{figure*}[p] 
\includegraphics[height=13mm]{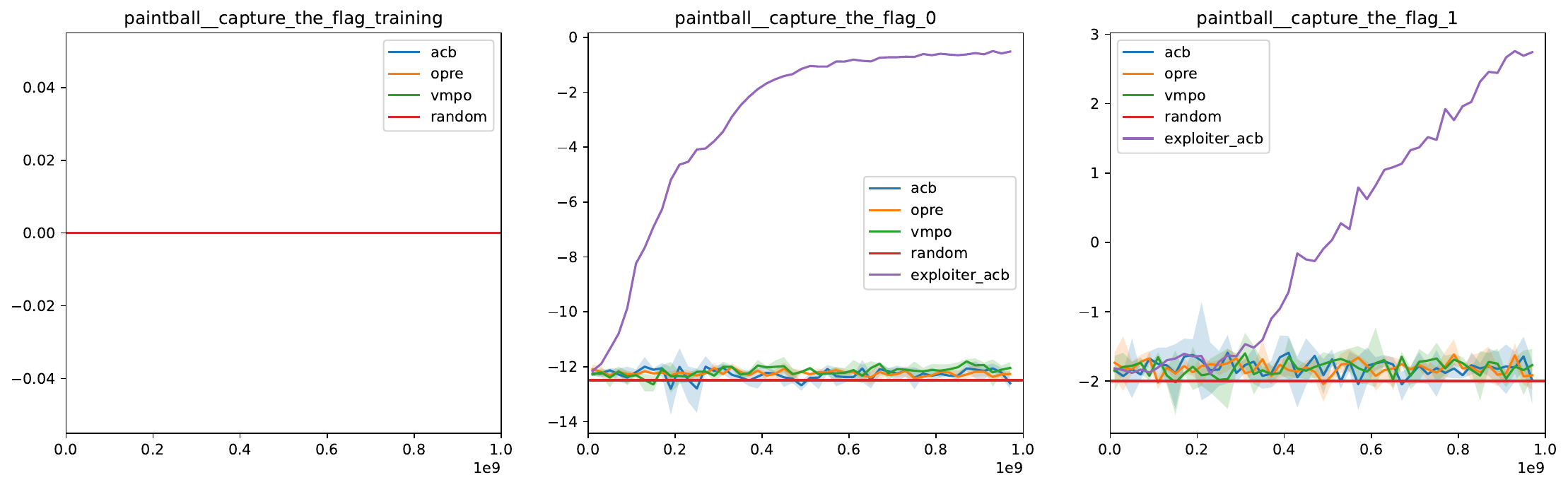}
\hspace{0mm}
\caption{Paintball: Capture the Flag}\label{fig:paintball__capture_the_flag}
\end{figure*}
\begin{figure*}[p] 
\includegraphics[height=13mm]{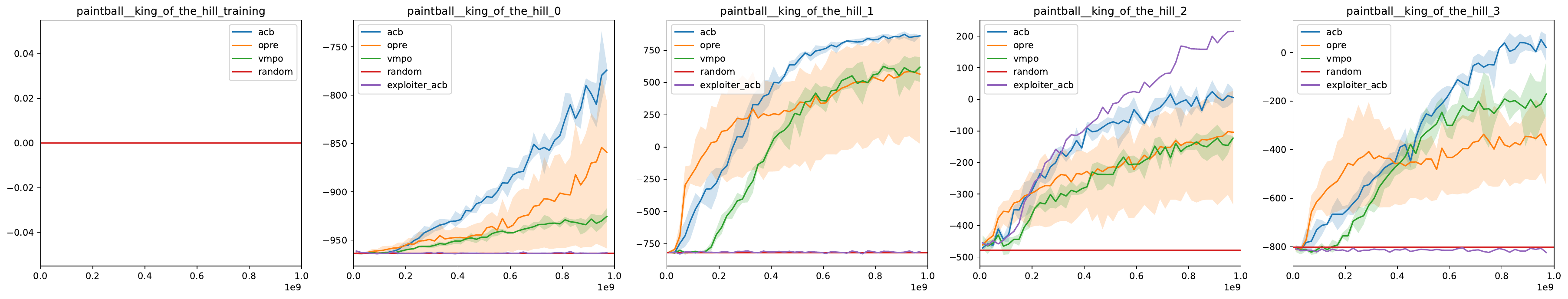}
\hspace{0mm}
\caption{Paintball: King of the Hill}\label{fig:paintball__king_of_the_hill}
\end{figure*}
\begin{figure*}[p] 
\includegraphics[height=13mm]{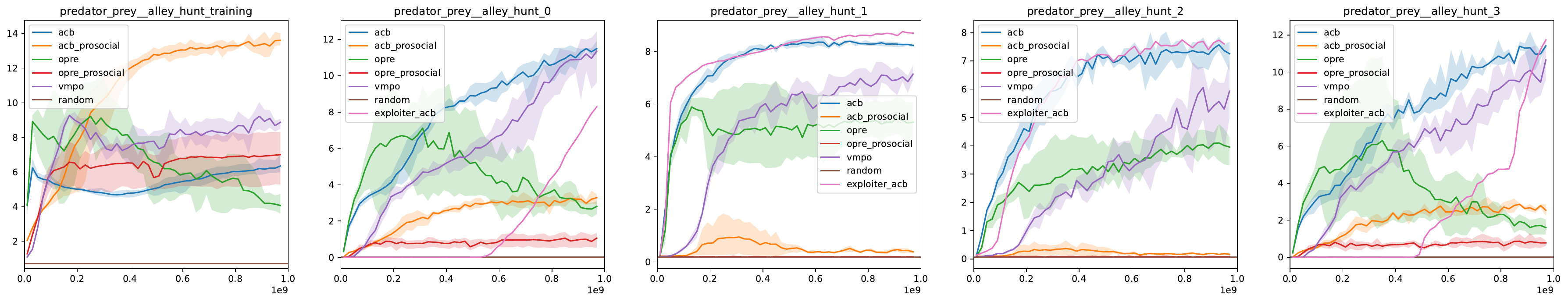}
\hspace{0mm}
\caption{Predator Prey: Alley Hunt}\label{fig:predator_prey__alley_hunt}
\end{figure*}
\begin{figure*}[p] 
\includegraphics[height=13mm]{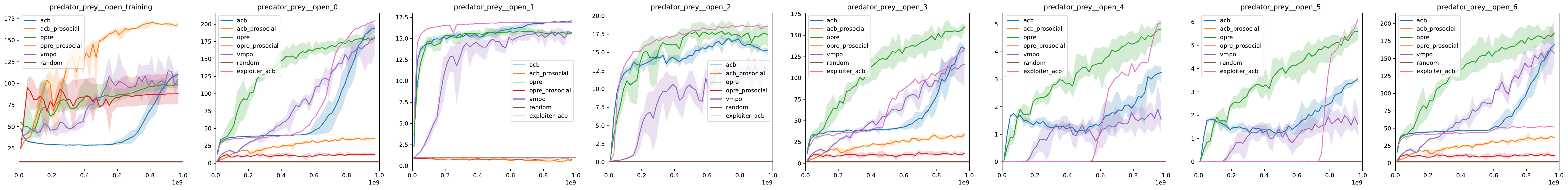}
\hspace{0mm}
\caption{Predator Prey: Open}\label{fig:predator_prey__open}
\end{figure*}
\begin{figure*}[p] 
\includegraphics[height=13mm]{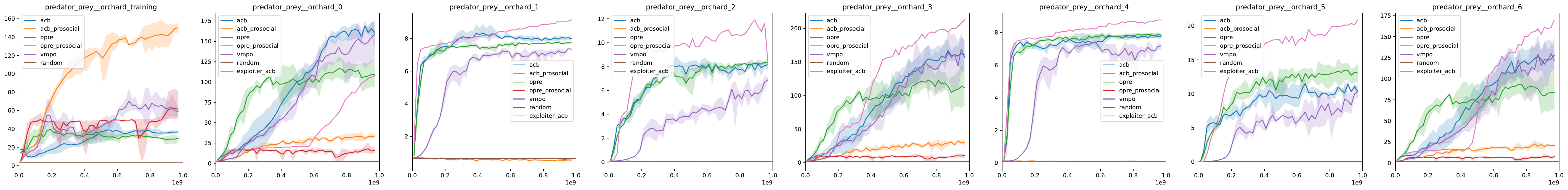}
\hspace{0mm}
\caption{Predator Prey: Orchard}\label{fig:predator_prey__orchard}
\end{figure*}
\begin{figure*}[p] 
\includegraphics[height=13mm]{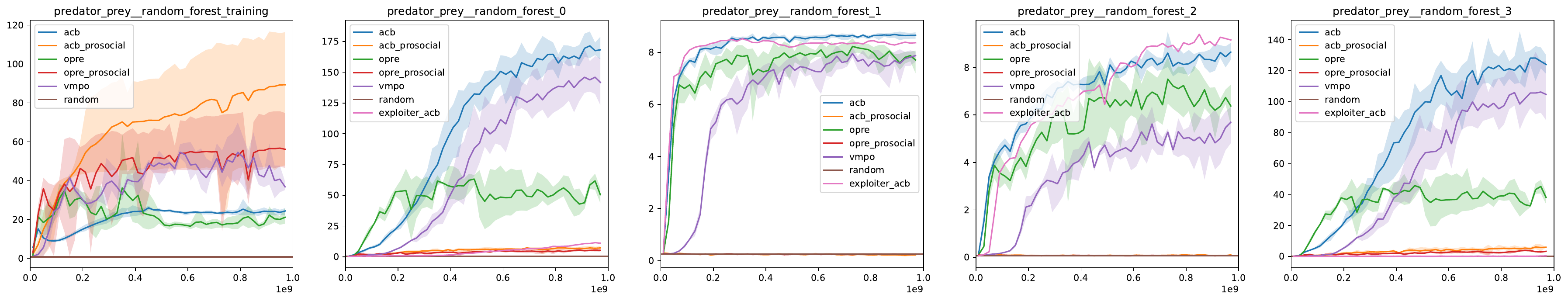}
\hspace{0mm}
\caption{Predator Prey: Random Forest}\label{fig:predator_prey__random_forest}
\end{figure*}
\begin{figure*}[p] 
\includegraphics[height=13mm]{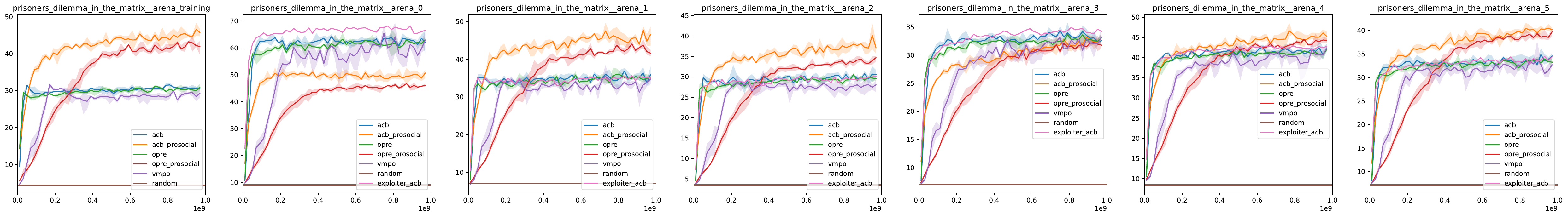}
\hspace{0mm}
\caption{Prisoners Dilemma in the matrix: Arena}\label{fig:prisoners_dilemma_in_the_matrix__arena}
\end{figure*}
\begin{figure*}[p] 
\includegraphics[height=13mm]{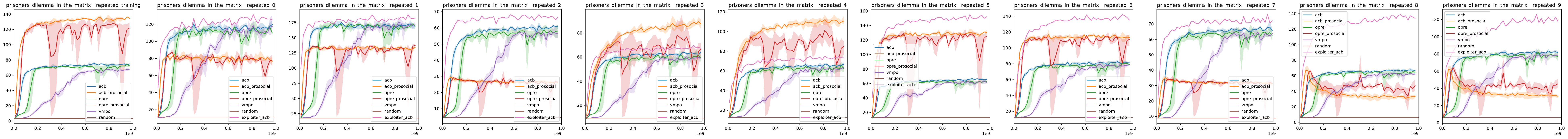}
\hspace{0mm}
\caption{Prisoners Dilemma in the matrix: Repeated}\label{fig:prisoners_dilemma_in_the_matrix__repeated}
\end{figure*}
\begin{figure*}[p] 
\includegraphics[height=13mm]{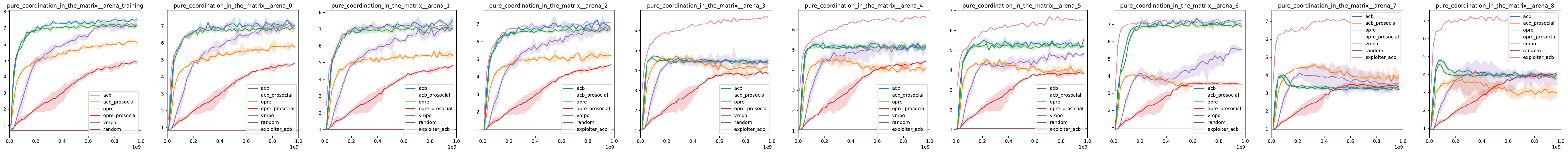}
\hspace{0mm}
\caption{Pure Coordination in the matrix: Arena}\label{fig:pure_coordination_in_the_matrix__arena}
\end{figure*}
\begin{figure*}[p] 
\includegraphics[height=13mm]{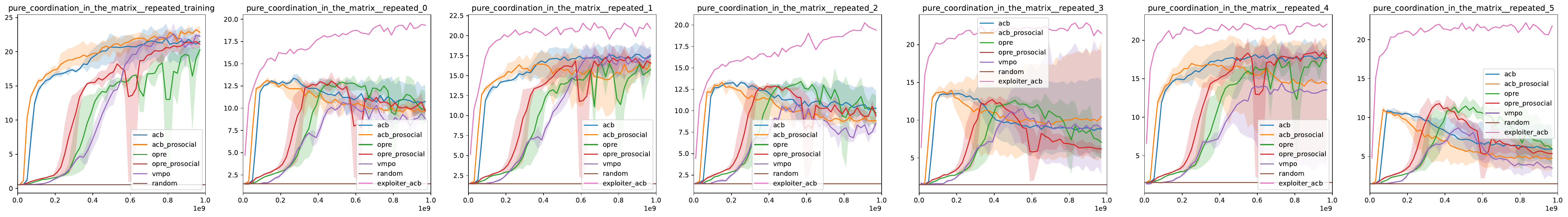}
\hspace{0mm}
\caption{Pure Coordination in the matrix: Repeated}\label{fig:pure_coordination_in_the_matrix__repeated}
\end{figure*}
\begin{figure*}[p] 
\includegraphics[height=13mm]{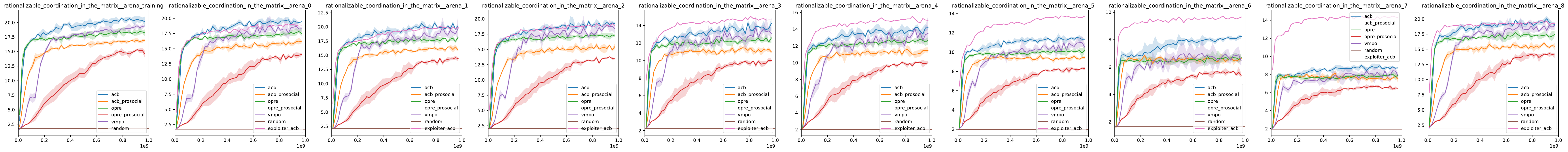}
\hspace{0mm}
\caption{Rationalizable Coordination in the matrix: Arena}\label{fig:rationalizable_coordination_in_the_matrix__arena}
\end{figure*}
\begin{figure*}[p] 
\includegraphics[height=13mm]{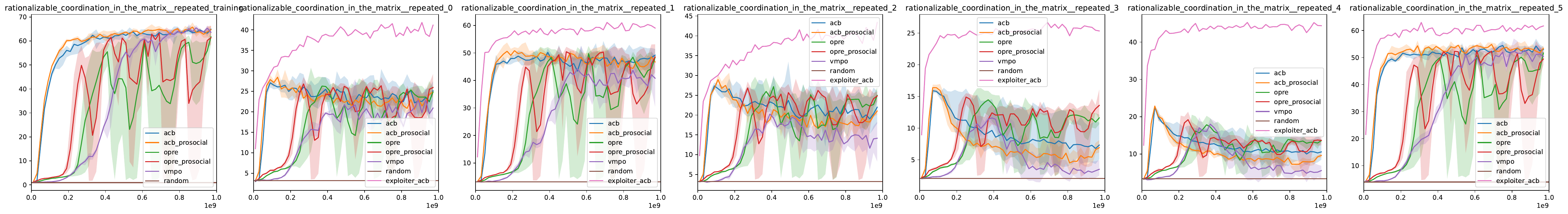}
\hspace{0mm}
\caption{Rationalizable Coordination in the matrix: Repeated}\label{fig:rationalizable_coordination_in_the_matrix__repeated}
\end{figure*}
\begin{figure*}[p] 
\includegraphics[height=13mm]{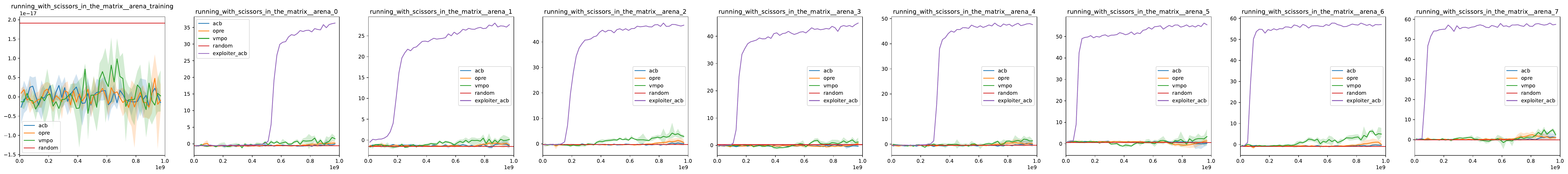}
\hspace{0mm}
\caption{Running With Scissors in the matrix: Arena}\label{fig:running_with_scissors_in_the_matrix__arena}
\end{figure*}
\begin{figure*}[p] 
\includegraphics[height=13mm]{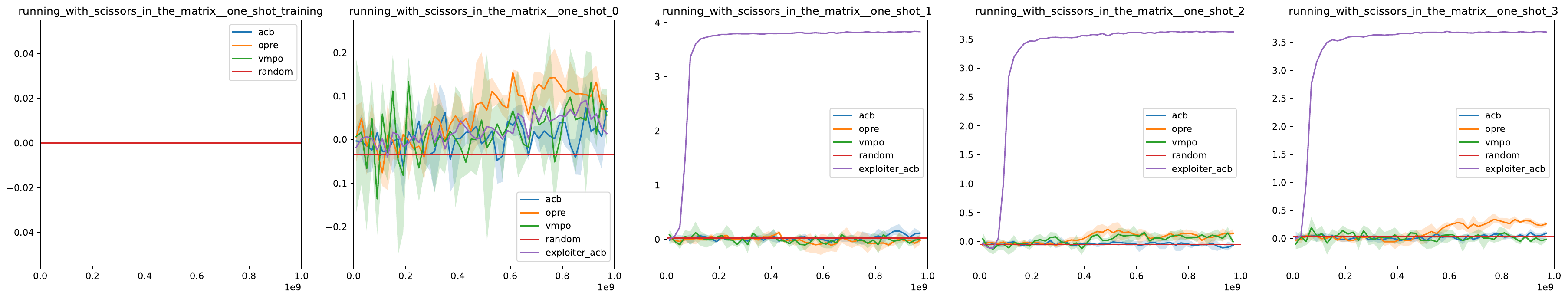}
\hspace{0mm}
\caption{Running With Scissors in the matrix: One Shot}\label{fig:running_with_scissors_in_the_matrix__one_shot}
\end{figure*}
\begin{figure*}[p] 
\includegraphics[height=13mm]{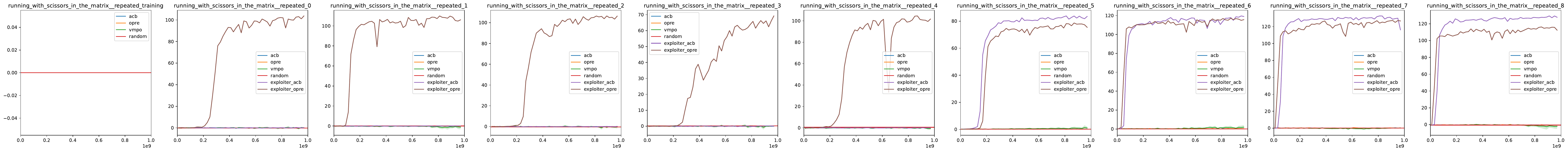}
\hspace{0mm}
\caption{Running With Scissors in the matrix: Repeated}\label{fig:running_with_scissors_in_the_matrix__repeated}
\end{figure*}
\begin{figure*}[p] 
\includegraphics[height=13mm]{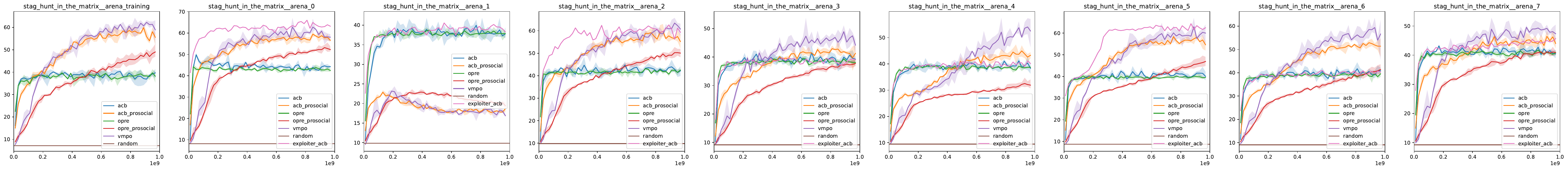}
\hspace{0mm}
\caption{Stag Hunt in the matrix: Arena}\label{fig:stag_hunt_in_the_matrix__arena}
\end{figure*}
\begin{figure*}[p] 
\includegraphics[height=13mm]{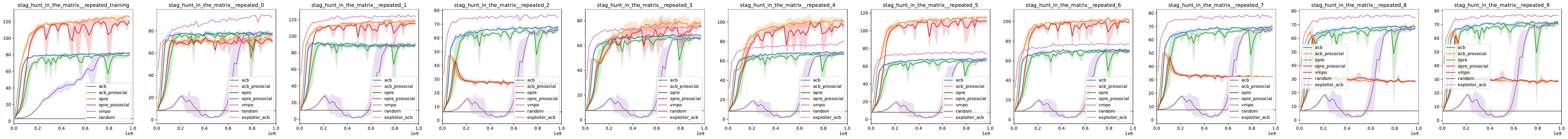}
\hspace{0mm}
\caption{Stag Hunt in the matrix: Repeated}\label{fig:stag_hunt_in_the_matrix__repeated}
\end{figure*}
\begin{figure*}[p] 
\includegraphics[height=13mm]{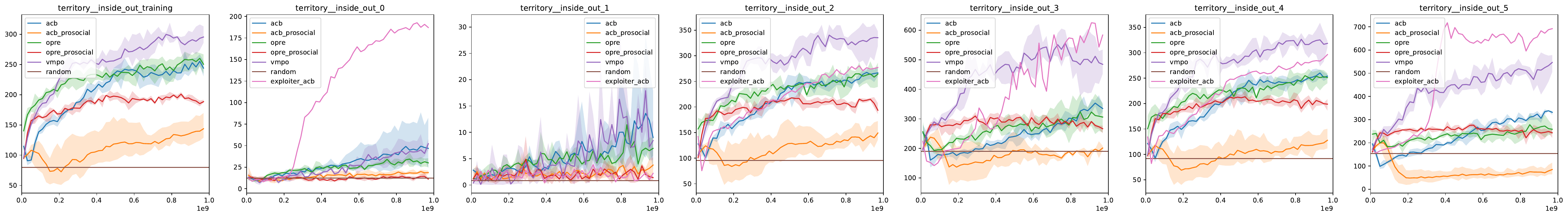}
\hspace{0mm}
\caption{Territory: Inside Out}\label{fig:territory__inside_out}
\end{figure*}
\begin{figure*}[p] 
\includegraphics[height=13mm]{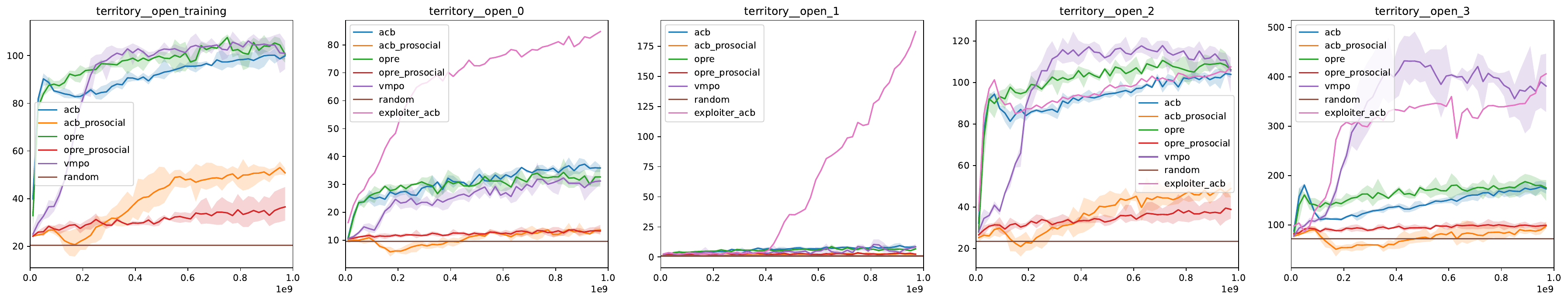}
\hspace{0mm}
\caption{Territory: Open}\label{fig:territory__open}
\end{figure*}
\begin{figure*}[p] 
\includegraphics[height=13mm]{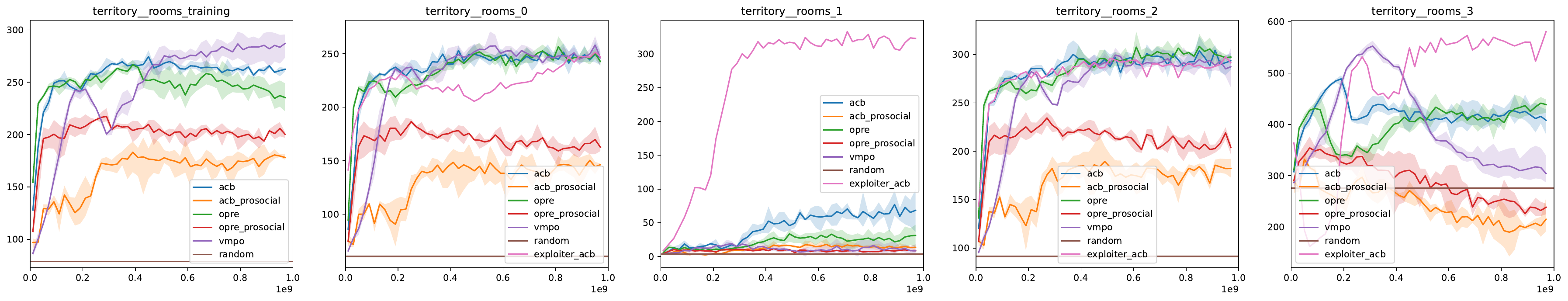}
\hspace{0mm}
\caption{Territory: Rooms}\label{fig:territory__rooms}
\end{figure*}

\section{Reference learning curves}

See Figs.~\ref{fig:allelopathic_harvest__open}--\ref{fig:territory__rooms}.

\FloatBarrier

\bibliography{main}

\begin{thebibliography}{79}
\providecommand{\natexlab}[1]{#1}
\providecommand{\url}[1]{\texttt{#1}}
\expandafter\ifx\csname urlstyle\endcsname\relax
  \providecommand{\doi}[1]{doi: #1}\else
  \providecommand{\doi}{doi: \begingroup \urlstyle{rm}\Url}\fi

\bibitem[Acemoglu(2021)]{acemoglu2021harms}
D.~Acemoglu.
\newblock Harms of {AI}.
\newblock Technical report, National Bureau of Economic Research, 2021.

\bibitem[Amodei et~al.(2016)Amodei, Olah, Steinhardt, Christiano, Schulman, and
  Man{\'e}]{amodei2016concrete}
D.~Amodei, C.~Olah, J.~Steinhardt, P.~Christiano, J.~Schulman, and D.~Man{\'e}.
\newblock Concrete problems in {AI} safety.
\newblock \emph{arXiv preprint arXiv:1606.06565}, 2016.

\bibitem[Axelrod(1984)]{Axelrod84}
R.~Axelrod.
\newblock \emph{The Evolution of Cooperation}.
\newblock Basic Books, 1984.

\bibitem[Baker et~al.(2019)Baker, Kanitscheider, Markov, Wu, Powell, McGrew,
  and Mordatch]{baker2019emergent}
B.~Baker, I.~Kanitscheider, T.~Markov, Y.~Wu, G.~Powell, B.~McGrew, and
  I.~Mordatch.
\newblock Emergent tool use from multi-agent autocurricula.
\newblock In \emph{International Conference on Learning Representations}, 2019.

\bibitem[Bakker et~al.(2021)Bakker, Everett, Weidinger, Gabriel, Isaac, Leibo,
  and Hughes]{bakker2021modelling}
M.~A. Bakker, R.~Everett, L.~Weidinger, I.~Gabriel, W.~S. Isaac, J.~Z. Leibo,
  and E.~Hughes.
\newblock Modelling cooperation in network games with spatio-temporal
  complexity.
\newblock \emph{arXiv preprint arXiv:2102.06911}, 2021.

\bibitem[Balduzzi et~al.(2019)Balduzzi, Garnelo, Bachrach, Czarnecki, Perolat,
  Jaderberg, and Graepel]{balduzzi2019open}
D.~Balduzzi, M.~Garnelo, Y.~Bachrach, W.~Czarnecki, J.~Perolat, M.~Jaderberg,
  and T.~Graepel.
\newblock Open-ended learning in symmetric zero-sum games.
\newblock In \emph{International Conference on Machine Learning}, pages
  434--443, 2019.

\bibitem[Barreto et~al.(2019)Barreto, Borsa, Hou, Comanici, Ayg{\"u}n, Hamel,
  Toyama, Mourad, Silver, Precup, et~al.]{barreto2019option}
A.~Barreto, D.~Borsa, S.~Hou, G.~Comanici, E.~Ayg{\"u}n, P.~Hamel, D.~Toyama,
  S.~Mourad, D.~Silver, D.~Precup, et~al.
\newblock The option keyboard: Combining skills in reinforcement learning.
\newblock \emph{Advances in Neural Information Processing Systems (NeurIPS)},
  32:\penalty0 13052--13062, 2019.

\bibitem[Berner et~al.(2019)Berner, Brockman, Chan, Cheung, Debiak, Dennison,
  Farhi, Fischer, Hashme, Hesse, Józefowicz, Gray, Olsson, Pachocki, Petrov,
  de~Oliveira~Pinto, Raiman, Salimans, Schlatter, Schneider, Sidor, Sutskever,
  Tang, Wolski, and Zhang]{berner2019dota}
C.~Berner, G.~Brockman, B.~Chan, V.~Cheung, P.~Debiak, C.~Dennison, D.~Farhi,
  Q.~Fischer, S.~Hashme, C.~Hesse, R.~Józefowicz, S.~Gray, C.~Olsson,
  J.~Pachocki, M.~Petrov, H.~P. de~Oliveira~Pinto, J.~Raiman, T.~Salimans,
  J.~Schlatter, J.~Schneider, S.~Sidor, I.~Sutskever, J.~Tang, F.~Wolski, and
  S.~Zhang.
\newblock Dota 2 with large scale deep reinforcement learning.
\newblock \emph{arXiv preprint arXiv:1912.06680}, 2019.

\bibitem[Burnell et~al.(2023)Burnell, Schellaert, Burden, Ullman,
  Martinez-Plumed, Tenenbaum, Rutar, Cheke, Sohl-Dickstein, Mitchell, Kiela,
  Shanahan, Voorhees, Cohn, Leibo, and Hernandez-Orallo]{burnell2023rethink}
R.~Burnell, W.~Schellaert, J.~Burden, T.~D. Ullman, F.~Martinez-Plumed, J.~B.
  Tenenbaum, D.~Rutar, L.~G. Cheke, J.~Sohl-Dickstein, M.~Mitchell, D.~Kiela,
  M.~Shanahan, E.~M. Voorhees, A.~G. Cohn, J.~Z. Leibo, and
  J.~Hernandez-Orallo.
\newblock Rethink reporting of evaluation results in ai.
\newblock \emph{Science}, 380\penalty0 (6641):\penalty0 136--138, 2023.

\bibitem[Carroll et~al.(2019)Carroll, Shah, Ho, Griffiths, Seshia, Abbeel, and
  Dragan]{carroll2019utility}
M.~Carroll, R.~Shah, M.~K. Ho, T.~Griffiths, S.~Seshia, P.~Abbeel, and
  A.~Dragan.
\newblock On the utility of learning about humans for human-ai coordination.
\newblock \emph{Advances in neural information processing systems}, 32, 2019.

\bibitem[Claus and Boutilier(1998)]{claus1998dynamics}
C.~Claus and C.~Boutilier.
\newblock The dynamics of reinforcement learning in cooperative multiagent
  systems.
\newblock \emph{AAAI/IAAI}, 1998\penalty0 (746-752):\penalty0 2, 1998.

\bibitem[Cobbe et~al.(2019)Cobbe, Klimov, Hesse, Kim, and
  Schulman]{cobbe2019quantifying}
K.~Cobbe, O.~Klimov, C.~Hesse, T.~Kim, and J.~Schulman.
\newblock Quantifying generalization in reinforcement learning.
\newblock In \emph{International Conference on Machine Learning}, pages
  1282--1289. PMLR, 2019.

\bibitem[Crosby et~al.(2019)Crosby, Beyret, and Halina]{crosby2019animal}
M.~Crosby, B.~Beyret, and M.~Halina.
\newblock The animal-ai olympics.
\newblock \emph{Nature Machine Intelligence}, 1\penalty0 (5):\penalty0
  257--257, 2019.

\bibitem[Dafoe et~al.(2020)Dafoe, Hughes, Bachrach, Collins, McKee, Leibo,
  Larson, and Graepel]{dafoe2020open}
A.~Dafoe, E.~Hughes, Y.~Bachrach, T.~Collins, K.~R. McKee, J.~Z. Leibo,
  K.~Larson, and T.~Graepel.
\newblock Open problems in cooperative {AI}.
\newblock \emph{arXiv preprint arXiv:2012.08630}, 2020.

\bibitem[Du{\'e}{\~n}ez-Guzm{\'a}n et~al.(2021)Du{\'e}{\~n}ez-Guzm{\'a}n,
  McKee, Mao, Coppin, Chiappa, Vezhnevets, Bakker, Bachrach, Sadedin, Isaac,
  et~al.]{duenez2021statistical}
E.~A. Du{\'e}{\~n}ez-Guzm{\'a}n, K.~R. McKee, Y.~Mao, B.~Coppin, S.~Chiappa,
  A.~S. Vezhnevets, M.~A. Bakker, Y.~Bachrach, S.~Sadedin, W.~Isaac, et~al.
\newblock Statistical discrimination in learning agents.
\newblock \emph{arXiv preprint arXiv:2110.11404}, 2021.

\bibitem[Espeholt et~al.(2018)Espeholt, Soyer, Munos, Simonyan, Mnih, Ward,
  Doron, Firoiu, Harley, Dunning, et~al.]{espeholt2018impala}
L.~Espeholt, H.~Soyer, R.~Munos, K.~Simonyan, V.~Mnih, T.~Ward, Y.~Doron,
  V.~Firoiu, T.~Harley, I.~Dunning, et~al.
\newblock Impala: Scalable distributed deep-rl with importance weighted
  actor-learner architectures.
\newblock In \emph{International Conference on Machine Learning}, pages
  1407--1416. PMLR, 2018.

\bibitem[Farebrother et~al.(2018)Farebrother, Machado, and
  Bowling]{farebrother2018generalization}
J.~Farebrother, M.~C. Machado, and M.~Bowling.
\newblock Generalization and regularization in {DQN}.
\newblock \emph{arXiv preprint arXiv:1810.00123}, 2018.

\bibitem[Foerster et~al.(2018)Foerster, Chen, Al-Shedivat, Whiteson, Abbeel,
  and Mordatch]{foerster2018learning}
J.~Foerster, R.~Y. Chen, M.~Al-Shedivat, S.~Whiteson, P.~Abbeel, and
  I.~Mordatch.
\newblock Learning with opponent-learning awareness.
\newblock In \emph{Proceedings of the 17th International Conference on
  Autonomous Agents and MultiAgent Systems}, pages 122--130, 2018.

\bibitem[Hadfield-Menell et~al.(2016)Hadfield-Menell, Russell, Abbeel, and
  Dragan]{hadfield2016cooperative}
D.~Hadfield-Menell, S.~J. Russell, P.~Abbeel, and A.~Dragan.
\newblock Cooperative inverse reinforcement learning.
\newblock \emph{Advances in neural information processing systems}, 29, 2016.

\bibitem[Haidich(2010)]{haidich2010meta}
A.-B. Haidich.
\newblock Meta-analysis in medical research.
\newblock \emph{Hippokratia}, 14\penalty0 (Suppl 1):\penalty0 29, 2010.

\bibitem[Hastie et~al.(2009)Hastie, Tibshirani, Friedman, and
  Friedman]{hastie2009elements}
T.~Hastie, R.~Tibshirani, J.~H. Friedman, and J.~H. Friedman.
\newblock \emph{The elements of statistical learning: data mining, inference,
  and prediction}, volume~2.
\newblock Springer, 2009.

\bibitem[Henrich and Muthukrishna(2021)]{henrich2021origins}
J.~Henrich and M.~Muthukrishna.
\newblock The origins and psychology of human cooperation.
\newblock \emph{Annual Review of Psychology}, 72:\penalty0 207--240, 2021.

\bibitem[Hern{\'a}ndez-Orallo(2017)]{hernandez2017evaluation}
J.~Hern{\'a}ndez-Orallo.
\newblock Evaluation in artificial intelligence: from task-oriented to
  ability-oriented measurement.
\newblock \emph{Artificial Intelligence Review}, 48:\penalty0 397--447, 2017.

\bibitem[Hessel et~al.(2019)Hessel, Soyer, Espeholt, Czarnecki, Schmitt, and
  van Hasselt]{hessel2019multi}
M.~Hessel, H.~Soyer, L.~Espeholt, W.~Czarnecki, S.~Schmitt, and H.~van Hasselt.
\newblock Multi-task deep reinforcement learning with popart.
\newblock In \emph{Proceedings of the AAAI Conference on Artificial
  Intelligence}, volume~33, pages 3796--3803, 2019.

\bibitem[Hessel et~al.(2021)Hessel, Kroiss, Clark, Kemaev, Quan, Keck, Viola,
  and van Hasselt]{hessel2021podracer}
M.~Hessel, M.~Kroiss, A.~Clark, I.~Kemaev, J.~Quan, T.~Keck, F.~Viola, and
  H.~van Hasselt.
\newblock Podracer architectures for scalable reinforcement learning.
\newblock \emph{arXiv preprint arXiv:2104.06272}, 2021.

\bibitem[Hughes et~al.(2018)Hughes, Leibo, Philips, Tuyls,
  Du{\'e}{\~n}ez-Guzm{\'a}n, Casta{\~n}eda, Dunning, Zhu, McKee, Koster, Roff,
  and Graepel]{hughes2018inequity}
E.~Hughes, J.~Z. Leibo, M.~G. Philips, K.~Tuyls, E.~A.
  Du{\'e}{\~n}ez-Guzm{\'a}n, A.~G. Casta{\~n}eda, I.~Dunning, T.~Zhu, K.~R.
  McKee, R.~Koster, H.~Roff, and T.~Graepel.
\newblock Inequity aversion improves cooperation in intertemporal social
  dilemmas.
\newblock In \emph{Advances in Neural Information Processing Systems
  (NeurIPS)}, pages 3330--3340. 2018.

\bibitem[Hume(1739/2012)]{hume1739treatise}
D.~Hume.
\newblock \emph{A treatise of human nature}.
\newblock Courier Corporation, 1739/2012.

\bibitem[Jaderberg et~al.(2016)Jaderberg, Mnih, Czarnecki, Schaul, Leibo,
  Silver, and Kavukcuoglu]{jaderberg2016reinforcement}
M.~Jaderberg, V.~Mnih, W.~M. Czarnecki, T.~Schaul, J.~Z. Leibo, D.~Silver, and
  K.~Kavukcuoglu.
\newblock Reinforcement learning with unsupervised auxiliary tasks.
\newblock \emph{arXiv preprint arXiv:1611.05397}, 2016.

\bibitem[Jaderberg et~al.(2019)Jaderberg, Czarnecki, Dunning, Marris, Lever,
  Castaneda, Beattie, Rabinowitz, Morcos, Ruderman, et~al.]{jaderberg2019human}
M.~Jaderberg, W.~M. Czarnecki, I.~Dunning, L.~Marris, G.~Lever, A.~G.
  Castaneda, C.~Beattie, N.~C. Rabinowitz, A.~S. Morcos, A.~Ruderman, et~al.
\newblock Human-level performance in 3d multiplayer games with population-based
  reinforcement learning.
\newblock \emph{Science}, 364\penalty0 (6443):\penalty0 859--865, 2019.

\bibitem[Janssen et~al.(2010)Janssen, Holahan, Lee, and Ostrom]{janssen2010lab}
M.~A. Janssen, R.~Holahan, A.~Lee, and E.~Ostrom.
\newblock Lab experiments for the study of social-ecological systems.
\newblock \emph{Science}, 328\penalty0 (5978):\penalty0 613--617, 2010.

\bibitem[Johanson et~al.(2022)Johanson, Hughes, Timbers, and
  Leibo]{johanson2022emergent}
M.~B. Johanson, E.~Hughes, F.~Timbers, and J.~Z. Leibo.
\newblock Emergent bartering behaviour in multi-agent reinforcement learning.
\newblock \emph{arXiv preprint arXiv:2205.06760}, 2022.

\bibitem[Koehn and Knowles(2017)]{koehn2017six}
P.~Koehn and R.~Knowles.
\newblock Six challenges for neural machine translation.
\newblock In \emph{First Workshop on Neural Machine Translation}, pages 28--39.
  Association for Computational Linguistics, 2017.

\bibitem[K{\"o}ster et~al.(2020)K{\"o}ster, McKee, Everett, Weidinger, Isaac,
  Hughes, Du{\'e}{\~n}ez-Guzm{\'a}n, Graepel, Botvinick, and
  Leibo]{koster2020model}
R.~K{\"o}ster, K.~R. McKee, R.~Everett, L.~Weidinger, W.~S. Isaac, E.~Hughes,
  E.~A. Du{\'e}{\~n}ez-Guzm{\'a}n, T.~Graepel, M.~Botvinick, and J.~Z. Leibo.
\newblock Model-free conventions in multi-agent reinforcement learning with
  heterogeneous preferences.
\newblock \emph{arXiv preprint arXiv:2010.09054}, 2020.

\bibitem[K{\"o}ster et~al.(2022)K{\"o}ster, Hadfield-Menell, Everett,
  Weidinger, Hadfield, and Leibo]{koster2022spurious}
R.~K{\"o}ster, D.~Hadfield-Menell, R.~Everett, L.~Weidinger, G.~K. Hadfield,
  and J.~Z. Leibo.
\newblock Spurious normativity enhances learning of compliance and enforcement
  behavior in artificial agents.
\newblock \emph{Proceedings of the National Academy of Sciences}, 119\penalty0
  (3):\penalty0 e2106028118, 2022.

\bibitem[Lanctot et~al.(2017)Lanctot, Zambaldi, Gruslys, Lazaridou, Tuyls,
  P{\'e}rolat, Silver, and Graepel]{lanctot2017unified}
M.~Lanctot, V.~Zambaldi, A.~Gruslys, A.~Lazaridou, K.~Tuyls, J.~P{\'e}rolat,
  D.~Silver, and T.~Graepel.
\newblock A unified game-theoretic approach to multiagent reinforcement
  learning.
\newblock In \emph{Advances in neural information processing systems
  (NeurIPS)}, pages 4190--4203, 2017.

\bibitem[Leibo et~al.(2017)Leibo, Zambaldi, Lanctot, Marecki, and
  Graepel]{leibo2017multiagent}
J.~Z. Leibo, V.~Zambaldi, M.~Lanctot, J.~Marecki, and T.~Graepel.
\newblock {Multi-agent Reinforcement Learning in Sequential Social Dilemmas}.
\newblock In \emph{Proceedings of the 16th International Conference on
  Autonomous Agents and Multiagent Systems (AA-MAS 2017)}, Sao Paulo, Brazil,
  2017.

\bibitem[Leibo et~al.(2019)Leibo, Perolat, Hughes, Wheelwright, Marblestone,
  Du{\'e}{\~n}ez-Guzm{\'a}n, Sunehag, Dunning, and
  Graepel]{leibo2019malthusian}
J.~Z. Leibo, J.~Perolat, E.~Hughes, S.~Wheelwright, A.~H. Marblestone,
  E.~Du{\'e}{\~n}ez-Guzm{\'a}n, P.~Sunehag, I.~Dunning, and T.~Graepel.
\newblock Malthusian reinforcement learning.
\newblock In \emph{Proceedings of the 18th International Conference on
  Autonomous Agents and MultiAgent Systems}, pages 1099--1107, 2019.

\bibitem[Leibo et~al.(2021)Leibo, Due{\~n}ez-Guzman, Vezhnevets, Agapiou,
  Sunehag, Koster, Matyas, Beattie, Mordatch, and Graepel]{leibo2021scalable}
J.~Z. Leibo, E.~A. Due{\~n}ez-Guzman, A.~Vezhnevets, J.~P. Agapiou, P.~Sunehag,
  R.~Koster, J.~Matyas, C.~Beattie, I.~Mordatch, and T.~Graepel.
\newblock Scalable evaluation of multi-agent reinforcement learning with
  {M}elting {P}ot.
\newblock In \emph{International Conference on Machine Learning}, pages
  6187--6199. PMLR, 2021.

\bibitem[Lerer and Peysakhovich(2017)]{lerer2017maintaining}
A.~Lerer and A.~Peysakhovich.
\newblock Maintaining cooperation in complex social dilemmas using deep
  reinforcement learning.
\newblock \emph{arXiv preprint arXiv:1707.01068}, 2017.

\bibitem[Lowe et~al.(2017)Lowe, Wu, Tamar, Harb, Abbeel, and
  Mordatch]{lowe2017multi}
R.~Lowe, Y.~Wu, A.~Tamar, J.~Harb, P.~Abbeel, and I.~Mordatch.
\newblock Multi-agent actor-critic for mixed cooperative-competitive
  environments.
\newblock \emph{arXiv preprint arXiv:1706.02275}, 2017.

\bibitem[Luce and Raiffa(1957/1989)]{luce1957games}
R.~D. Luce and H.~Raiffa.
\newblock \emph{Games and decisions: Introduction and critical survey}.
\newblock Courier Corporation, 1957/1989.

\bibitem[Mart{\'\i}nez-Plumed et~al.(2019)Mart{\'\i}nez-Plumed, Prud{\^e}ncio,
  Mart{\'\i}nez-Us{\'o}, and Hern{\'a}ndez-Orallo]{martinez2019item}
F.~Mart{\'\i}nez-Plumed, R.~B. Prud{\^e}ncio, A.~Mart{\'\i}nez-Us{\'o}, and
  J.~Hern{\'a}ndez-Orallo.
\newblock Item response theory in {AI}: Analysing machine learning classifiers
  at the instance level.
\newblock \emph{Artificial intelligence}, 271:\penalty0 18--42, 2019.

\bibitem[Mart{\i}nez-Plumed et~al.(2020)Mart{\i}nez-Plumed,
  Hern{\'a}ndez-Orallo, and G{\'o}mez]{martinez2020tracking}
F.~Mart{\i}nez-Plumed, J.~Hern{\'a}ndez-Orallo, and E.~G{\'o}mez.
\newblock Tracking the impact and evolution of ai: The aicollaboratory.
\newblock 2020.

\bibitem[McKee et~al.(2020)McKee, Gemp, McWilliams, Du{\`e}{\~n}ez-Guzm{\'a}n,
  Hughes, and Leibo]{mckee2020social}
K.~R. McKee, I.~Gemp, B.~McWilliams, E.~A. Du{\`e}{\~n}ez-Guzm{\'a}n,
  E.~Hughes, and J.~Z. Leibo.
\newblock Social diversity and social preferences in mixed-motive reinforcement
  learning.
\newblock In \emph{Proceedings of the 19th International Conference on
  Autonomous Agents and MultiAgent Systems}, pages 869--877, 2020.

\bibitem[Mehrabi et~al.(2021)Mehrabi, Morstatter, Saxena, Lerman, and
  Galstyan]{mehrabi2021survey}
N.~Mehrabi, F.~Morstatter, N.~Saxena, K.~Lerman, and A.~Galstyan.
\newblock A survey on bias and fairness in machine learning.
\newblock \emph{ACM Computing Surveys (CSUR)}, 54\penalty0 (6):\penalty0 1--35,
  2021.

\bibitem[Mnih et~al.(2016)Mnih, Badia, Mirza, Graves, Lillicrap, Harley,
  Silver, and Kavukcuoglu]{mnih2016asynchronous}
V.~Mnih, A.~P. Badia, M.~Mirza, A.~Graves, T.~Lillicrap, T.~Harley, D.~Silver,
  and K.~Kavukcuoglu.
\newblock Asynchronous methods for deep reinforcement learning.
\newblock In \emph{International conference on machine learning}, pages
  1928--1937. PMLR, 2016.

\bibitem[Obermeyer et~al.(2019)Obermeyer, Powers, Vogeli, and
  Mullainathan]{obermeyer2019dissecting}
Z.~Obermeyer, B.~Powers, C.~Vogeli, and S.~Mullainathan.
\newblock Dissecting racial bias in an algorithm used to manage the health of
  populations.
\newblock \emph{Science}, 366\penalty0 (6464):\penalty0 447--453, 2019.

\bibitem[Oord et~al.(2018)Oord, Li, and Vinyals]{oord2018representation}
A.~v.~d. Oord, Y.~Li, and O.~Vinyals.
\newblock Representation learning with contrastive predictive coding.
\newblock \emph{arXiv preprint arXiv:1807.03748}, 2018.

\bibitem[Ostrom(2009)]{ostrom2009understanding}
E.~Ostrom.
\newblock \emph{Understanding institutional diversity}.
\newblock Princeton university press, 2009.

\bibitem[Ott et~al.(2022)Ott, Barbosa-Silva, Blagec, Brauner, and
  Samwald]{ott2022mapping}
S.~Ott, A.~Barbosa-Silva, K.~Blagec, J.~Brauner, and M.~Samwald.
\newblock Mapping global dynamics of benchmark creation and saturation in
  artificial intelligence.
\newblock \emph{Nature Communications}, 13\penalty0 (1):\penalty0 1--11, 2022.

\bibitem[Pascanu et~al.(2013)Pascanu, Mikolov, and
  Bengio]{pascanu2013difficulty}
R.~Pascanu, T.~Mikolov, and Y.~Bengio.
\newblock On the difficulty of training recurrent neural networks.
\newblock In \emph{International conference on machine learning}, pages
  1310--1318. PMLR, 2013.

\bibitem[Perolat et~al.(2017)Perolat, Leibo, Zambaldi, Beattie, Tuyls, and
  Graepel]{perolat2017}
J.~Perolat, J.~Z. Leibo, V.~Zambaldi, C.~Beattie, K.~Tuyls, and T.~Graepel.
\newblock A multi-agent reinforcement learning model of common-pool resource
  appropriation.
\newblock In \emph{Advances in Neural Information Processing Systems
  (NeurIPS)}, pages 3643--3652, 2017.

\bibitem[Peysakhovich and Lerer(2017)]{peysakhovich2017prosocial}
A.~Peysakhovich and A.~Lerer.
\newblock Prosocial learning agents solve generalized stag hunts better than
  selfish ones.
\newblock \emph{arXiv preprint arXiv:1709.02865}, 2017.

\bibitem[Rahwan et~al.(2019)Rahwan, Cebrian, Obradovich, Bongard, Bonnefon,
  Breazeal, Crandall, Christakis, Couzin, Jackson, et~al.]{rahwan2019machine}
I.~Rahwan, M.~Cebrian, N.~Obradovich, J.~Bongard, J.-F. Bonnefon, C.~Breazeal,
  J.~W. Crandall, N.~A. Christakis, I.~D. Couzin, M.~O. Jackson, et~al.
\newblock Machine behaviour.
\newblock \emph{Nature}, 568\penalty0 (7753):\penalty0 477--486, 2019.

\bibitem[Rashid et~al.(2018)Rashid, Samvelyan, Schroeder, Farquhar, Foerster,
  and Whiteson]{rashid2018qmix}
T.~Rashid, M.~Samvelyan, C.~Schroeder, G.~Farquhar, J.~Foerster, and
  S.~Whiteson.
\newblock Qmix: Monotonic value function factorisation for deep multi-agent
  reinforcement learning.
\newblock In \emph{International Conference on Machine Learning}, pages
  4295--4304. PMLR, 2018.

\bibitem[Roelofs et~al.(2019)Roelofs, Shankar, Recht, Fridovich-Keil, Hardt,
  Miller, and Schmidt]{roelofs2019meta}
R.~Roelofs, V.~Shankar, B.~Recht, S.~Fridovich-Keil, M.~Hardt, J.~Miller, and
  L.~Schmidt.
\newblock A meta-analysis of overfitting in machine learning.
\newblock \emph{Advances in Neural Information Processing Systems}, 32, 2019.

\bibitem[Russakovsky et~al.(2015)Russakovsky, Deng, Su, Krause, Satheesh, Ma,
  Huang, Karpathy, Khosla, Bernstein, et~al.]{russakovsky2015imagenet}
O.~Russakovsky, J.~Deng, H.~Su, J.~Krause, S.~Satheesh, S.~Ma, Z.~Huang,
  A.~Karpathy, A.~Khosla, M.~Bernstein, et~al.
\newblock Imagenet large scale visual recognition challenge.
\newblock \emph{International journal of computer vision}, 115\penalty0
  (3):\penalty0 211--252, 2015.

\bibitem[Russell(2019)]{russell2019human}
S.~Russell.
\newblock \emph{Human compatible: Artificial intelligence and the problem of
  control}.
\newblock Penguin, 2019.

\bibitem[Russell et~al.(2015)Russell, Dewey, and Tegmark]{russell2015research}
S.~Russell, D.~Dewey, and M.~Tegmark.
\newblock Research priorities for robust and beneficial artificial
  intelligence.
\newblock \emph{Ai Magazine}, 36\penalty0 (4):\penalty0 105--114, 2015.

\bibitem[Silver et~al.(2016)Silver, Huang, Maddison, Guez, Sifre, Van
  Den~Driessche, Schrittwieser, Antonoglou, Panneershelvam, Lanctot,
  et~al.]{silver2016mastering}
D.~Silver, A.~Huang, C.~J. Maddison, A.~Guez, L.~Sifre, G.~Van Den~Driessche,
  J.~Schrittwieser, I.~Antonoglou, V.~Panneershelvam, M.~Lanctot, et~al.
\newblock Mastering the game of {Go} with deep neural networks and tree search.
\newblock \emph{Nature}, 529\penalty0 (7587):\penalty0 484, 2016.

\bibitem[Silver et~al.(2017)Silver, Schrittwieser, Simonyan, Antonoglou, Huang,
  Guez, Hubert, Baker, Lai, Bolton, et~al.]{silver2017mastering}
D.~Silver, J.~Schrittwieser, K.~Simonyan, I.~Antonoglou, A.~Huang, A.~Guez,
  T.~Hubert, L.~Baker, M.~Lai, A.~Bolton, et~al.
\newblock Mastering the game of go without human knowledge.
\newblock \emph{Nature}, 550\penalty0 (7676):\penalty0 354--359, 2017.

\bibitem[Silver et~al.(2018)Silver, Hubert, Schrittwieser, Antonoglou, Lai,
  Guez, Lanctot, Sifre, Kumaran, Graepel, et~al.]{silver2018general}
D.~Silver, T.~Hubert, J.~Schrittwieser, I.~Antonoglou, M.~Lai, A.~Guez,
  M.~Lanctot, L.~Sifre, D.~Kumaran, T.~Graepel, et~al.
\newblock A general reinforcement learning algorithm that masters chess, shogi,
  and go through self-play.
\newblock \emph{Science}, 362\penalty0 (6419):\penalty0 1140--1144, 2018.

\bibitem[Soares and Fallenstein(2014)]{soares2014aligning}
N.~Soares and B.~Fallenstein.
\newblock Aligning superintelligence with human interests: A technical research
  agenda.
\newblock \emph{Machine Intelligence Research Institute (MIRI) technical
  report}, 8, 2014.

\bibitem[Song et~al.(2020)Song, Abdolmaleki, Springenberg, Clark, Soyer, Rae,
  Noury, Ahuja, Liu, Tirumala, Heess, Belov, Riedmiller, and
  Botvinick]{Song2020VMPO}
H.~F. Song, A.~Abdolmaleki, J.~T. Springenberg, A.~Clark, H.~Soyer, J.~W. Rae,
  S.~Noury, A.~Ahuja, S.~Liu, D.~Tirumala, N.~Heess, D.~Belov, M.~Riedmiller,
  and M.~M. Botvinick.
\newblock V-mpo: On-policy maximum a posteriori policy optimization for
  discrete and continuous control.
\newblock In \emph{International Conference on Learning Representations}, 2020.

\bibitem[Strouse et~al.(2021)Strouse, McKee, Botvinick, Hughes, and
  Everett]{strouse2021collaborating}
D.~Strouse, K.~McKee, M.~Botvinick, E.~Hughes, and R.~Everett.
\newblock Collaborating with humans without human data.
\newblock \emph{Advances in Neural Information Processing Systems},
  34:\penalty0 14502--14515, 2021.

\bibitem[Sunehag et~al.(2018)Sunehag, Lever, Gruslys, Czarnecki, Zambaldi,
  Jaderberg, Lanctot, Sonnerat, Leibo, Tuyls, et~al.]{sunehag2018value}
P.~Sunehag, G.~Lever, A.~Gruslys, W.~M. Czarnecki, V.~Zambaldi, M.~Jaderberg,
  M.~Lanctot, N.~Sonnerat, J.~Z. Leibo, K.~Tuyls, et~al.
\newblock Value-decomposition networks for cooperative multi-agent learning
  based on team reward.
\newblock In \emph{Proceedings of the 17th International Conference on
  Autonomous Agents and MultiAgent Systems}, pages 2085--2087, 2018.

\bibitem[Sutton and Barto(2018)]{sutton2018reinforcement}
R.~S. Sutton and A.~G. Barto.
\newblock \emph{Reinforcement learning: An introduction}.
\newblock MIT press, 2018.

\bibitem[Sutton et~al.(1999)Sutton, Precup, and Singh]{sutton1999between}
R.~S. Sutton, D.~Precup, and S.~Singh.
\newblock Between mdps and semi-mdps: A framework for temporal abstraction in
  reinforcement learning.
\newblock \emph{Artificial intelligence}, 112\penalty0 (1-2):\penalty0
  181--211, 1999.

\bibitem[Sutton et~al.(2011)Sutton, Modayil, Delp, Degris, Pilarski, White, and
  Precup]{sutton2011horde}
R.~S. Sutton, J.~Modayil, M.~Delp, T.~Degris, P.~M. Pilarski, A.~White, and
  D.~Precup.
\newblock Horde: A scalable real-time architecture for learning knowledge from
  unsupervised sensorimotor interaction.
\newblock In \emph{The 10th International Conference on Autonomous Agents and
  Multiagent Systems-Volume 2}, pages 761--768, 2011.

\bibitem[Tesfatsion(2021)]{tesfatsion2021}
L.~Tesfatsion.
\newblock Agent-based computational economics: Overview and brief history.
\newblock Working Paper 21004, Department of Economics, Iowa State University,
  2021.

\bibitem[Vezhnevets et~al.(2020)Vezhnevets, Wu, Eckstein, Leblond, and
  Leibo]{vezhnevets2020options}
A.~Vezhnevets, Y.~Wu, M.~Eckstein, R.~Leblond, and J.~Z. Leibo.
\newblock Options as responses: Grounding behavioural hierarchies in
  multi-agent reinforcement learning.
\newblock In \emph{International Conference on Machine Learning}, pages
  9733--9742. PMLR, 2020.

\bibitem[Vezhnevets et~al.(2017)Vezhnevets, Osindero, Schaul, Heess, Jaderberg,
  Silver, and Kavukcuoglu]{vezhnevets2017feudal}
A.~S. Vezhnevets, S.~Osindero, T.~Schaul, N.~Heess, M.~Jaderberg, D.~Silver,
  and K.~Kavukcuoglu.
\newblock Feudal networks for hierarchical reinforcement learning.
\newblock In \emph{International Conference on Machine Learning}, pages
  3540--3549. PMLR, 2017.

\bibitem[Vinitsky et~al.(2023)Vinitsky, K{\"o}ster, Agapiou,
  Du{\'e}{\~n}ez-Guzm{\'a}n, Vezhnevets, and Leibo]{vinitsky2023learning}
E.~Vinitsky, R.~K{\"o}ster, J.~P. Agapiou, E.~A. Du{\'e}{\~n}ez-Guzm{\'a}n,
  A.~S. Vezhnevets, and J.~Z. Leibo.
\newblock A learning agent that acquires social norms from public sanctions in
  decentralized multi-agent settings.
\newblock \emph{Collective Intelligence}, 2\penalty0 (2):\penalty0
  26339137231162025, 2023.

\bibitem[Vinyals et~al.(2019)Vinyals, Babuschkin, Czarnecki, Mathieu, Dudzik,
  Chung, Choi, Powell, Ewalds, Georgiev, et~al.]{vinyals2019grandmaster}
O.~Vinyals, I.~Babuschkin, W.~M. Czarnecki, M.~Mathieu, A.~Dudzik, J.~Chung,
  D.~H. Choi, R.~Powell, T.~Ewalds, P.~Georgiev, et~al.
\newblock Grandmaster level in starcraft {II} using multi-agent reinforcement
  learning.
\newblock \emph{Nature}, 575\penalty0 (7782):\penalty0 350--354, 2019.

\bibitem[Wang et~al.(2020)Wang, Wu, Evans, Tenenbaum, Parkes, and
  Kleiman-Weiner]{wang2020too}
R.~E. Wang, S.~A. Wu, J.~A. Evans, J.~B. Tenenbaum, D.~C. Parkes, and
  M.~Kleiman-Weiner.
\newblock Too many cooks: Coordinating multi-agent collaboration through
  inverse planning.
\newblock \emph{arXiv preprint arXiv:2003.11778}, 2020.

\bibitem[Weibull(1997)]{weibull1997evolutionary}
J.~W. Weibull.
\newblock \emph{Evolutionary game theory}.
\newblock MIT press, 1997.

\bibitem[Weidinger et~al.(2021)Weidinger, Mellor, Rauh, Griffin, Uesato, Huang,
  Cheng, Glaese, Balle, Kasirzadeh, et~al.]{weidinger2021ethical}
L.~Weidinger, J.~Mellor, M.~Rauh, C.~Griffin, J.~Uesato, P.-S. Huang, M.~Cheng,
  M.~Glaese, B.~Balle, A.~Kasirzadeh, et~al.
\newblock Ethical and social risks of harm from language models.
\newblock \emph{arXiv preprint arXiv:2112.04359}, 2021.

\bibitem[Ye et~al.(2020)Ye, Chen, Zhang, Chen, Yuan, Liu, Chen, Liu, Qiu, Yu,
  et~al.]{ye2020towards}
D.~Ye, G.~Chen, W.~Zhang, S.~Chen, B.~Yuan, B.~Liu, J.~Chen, Z.~Liu, F.~Qiu,
  H.~Yu, et~al.
\newblock Towards playing full moba games with deep reinforcement learning.
\newblock \emph{Advances in Neural Information Processing Systems (NeurIPS)},
  33, 2020.

\bibitem[Zhang et~al.(2018)Zhang, Ballas, and Pineau]{zhang2018dissection}
A.~Zhang, N.~Ballas, and J.~Pineau.
\newblock A dissection of overfitting and generalization in continuous
  reinforcement learning.
\newblock \emph{arXiv preprint arXiv:1806.07937}, 2018.

\end{thebibliography}

\end{document}